\documentclass{aa} 
\usepackage{rotating}
\usepackage[colorlinks=true, citecolor=blue, urlcolor=blue]{hyperref}
\usepackage{float}
\usepackage{graphicx}
\usepackage{tabularx}
\usepackage{txfonts}
\usepackage{amssymb}
\usepackage{lscape}
\usepackage[bf,nooneline]{subfigure}
\bibpunct{(}{)}{;}{a}{}{,}

\begin{document}

\title{Ultramassive dense early-type galaxies: velocity dispersion and
number density evolution since z=1.6}

   \author{A. Gargiulo\inst{1}\thanks{adriana@lambrate.inaf.it}, P. Saracco\inst{2}, S. Tamburri\inst{2,3}, I. Lonoce\inst{2,3}, 
F. Ciocca\inst{2,3} }

   \institute{INAF – Istituto di Astrofisica Spaziale e Fisica Cosmica Milano, via Bassini 15, 20133 Milano, Italy\\ 
      \and INAF - Osservatorio Astronomico di Brera, Via Brera 28, 20121 Milano, Italy\\
         \and Dipartimento di Scienza e Alta Tecnologia, Universit\`{a} degli Studi dell'Insubria, via Valleggio 11, 22100 Como, Italy}

   \date{Received 2015 May 20; accepted 2016 May 4}

\abstract{}{}{}{}{}
 
  \abstract{}
{In this paper we investigate the stellar mass assembly history of  ultramassive 
(M$_{\star}\gtrsim$ 10$^{11}$M$_{\odot}$) dense ($\Sigma$ = M$_{\star}$/2$\pi$R$_{e}^{2}>$ 2500 M$_{\odot}$pc$^{-2}$)
early-type galaxies (ETGs, elliptical and spheroidal galaxies visually selected) over the last 9 Gyr.}
{We have traced the evolution of the comoving number density $\rho$ of ultramassive dense ETGs 
and have compared their structural (effective radius R$_{e}$ and stellar mass M$_{\star}$) and dynamical 
(velocity dispersion $\sigma_{e}$) parameters 
over the redshift range $0<z<1.6$.
We have derived the number density $\rho$ at 1.6 $<z<$ 1 from the MUNICS and GOODS-South surveys, while we have taken advantage of the 
COSMOS spectroscopic survey to probe the intermediate redshift range [0.2 - 1.0].
The number density of ultramassive dense local ETGs has been derived from the SDSS sample taking 
into account all the selection bias affecting the spectroscopic sample.
For the comparison of the dynamical and structural parameters, we have collected a sample
of 11 ultramassive dense ETGs at 1.2 $<z<$ 1.6  for which velocity dispersion measurements 
are available. 
For 4 of them (plus one at z = 1.91) we present previously unpublished estimates 
of velocity dispersion, based on optical VLT-FORS2 spectra.  
We probe the intermediate redshift range (0.2 $ \lesssim$ z $\lesssim $ 0.9) 
using the sample of ETGs by \citet{saglia10}, and \citet{zahid15} and the local universe
using the sample of ETGs by \citet{thomas10} 
}
{We find that the comoving number density of ultramassive dense ETGs evolves with z as 
$\rho(z)\propto(1 + z)^{0.3\pm0.8}$ implying a decrease of $\sim$ 25$\%$ of the population 
of ultramassive dense ETGs since z = 1.6. 
By comparing in the [R$_{e}$, M$_{\star}$, $\sigma_{e}$]  plane the structural and dynamical properties of 
high-z ultramassive dense ETGs over the range 0 $\lesssim z <$ 1.6  
we find that all the ETGs of the high-z sample have a counterpart with similar properties in the local universe. 
This implies either that the majority ($\sim$70$\%$) of ultramassive dense ETGs has 
already completed its assembly and its shaping at $<z>$ = 1.4, or that,
if a significant fraction of them evolves in size, new ultramassive dense ETGs must
form at z $<$ 1.5 to maintain their number density almost constant. The difficulty into identify  
good progenitors for these new dense ETGs at z $\lesssim$ 1.5, 
and the stellar populations properties of local ultramassive dense ETGs point toward the first hypothesis.
In this case, the ultramassive dense galaxies missing in local universe 
can have joined, in the last 9 Gyr, the ``non dense'' ETGs population through minor mergers, thus
contributing to mean size growth. In any case, the comparison between their number density and the number density of the whole population 
of ultramassive ETGs relegates their contribution to the mean size evolution to a secondary process. }{}

\titlerunning {Number density evolution of ultra-massive dense early type
galaxies since z=1.6}
   \authorrunning {Gargiulo et al.}

   \maketitle
%
\section{Introduction}

Most of the studies focusing on the structural parameters of massive passive galaxies both at high and at low 
redshift show that this population of galaxies has increased the mean effective radius over the cosmic time 
\citep[e.g.][]{daddi05, trujillo06, trujillo07, longhetti07, zirm07, toft07,
vandokkum08, franx08, cimatti08, buitrago08, vanderwel08, bernardi09, 
damjanov09, bezanson09, mancini09, mancini10, cassata11, damjanov11, bruce12, szomoru12, newman12, cassata13, vanderwel14, belli14}. 
For example, \citet{vanderwel14} find that the mean effective radius $<R_{e}>$ of
passive galaxies selected with a colour-colour criterion 
increases as $<R_{e}>$ $\propto$ (1+z)$^{-1.48}$ over the redshift range 0 $< z <$ 3.
Similar results, among the others, have been reached also by \citet{cassata13}, on samples 
of passive (specific star formation rate sSFR $<$ 10$^{-11}$ yr$^{-1}$) galaxies, 
and by \citet{cimatti12} for galaxies with red colours and elliptical morphology.
Nonetheless, other analysis contrast these results showing a mild or an even negligible evolution
of the mean size with cosmic time \citep[][]{saracco11, stott11, jorgensen13, saracco14, jorgensen14, poggianti13a}.
Most of these works study sample of galaxies selected on the basis of their elliptical morphology.

In fact, the reason for these different results is not easily detectable. A major point 
to be considered is that 
a selection based on the passivity (or on 
Sersic index $n>$ 2.5, see, e.g., Buitrago et al. \citeyear{buitrago13}) 
includes in the sample a significant fraction ($>30\%$) of disk galaxies  
\citep[e.g.][]{vanderwel11, ilbert10, cassata11, tamburri14}
and, concurrently, does not include all the galaxies visually classified as ellipticals.
At the same time, a selection based on a cut in sSFR constant with time (usually 
sSFR $<$ 10$^{-11}$ yr$^{-1}$) selects predominantly red spheroids at $z\sim2$
but includes a large contamination of disk/spiral galaxies in the local universe \citep{szomoru11}.
Actually, \citet{keating15} show that the properties of high-z 
ETGs are highly sensitive to the definitions used in selecting the samples \citep[see also][]{moresco13}.

On top of this debated issue on the amount of evolution in size, 
its possible origin is also investigated. 
Is the increase of the mean size
due to the size-growth of individual galaxies across the time (resulting from an inside-out stellar matter accretion
due to minor dry merging) or rather to the addition 
of new and larger galaxies over the cosmic time
\citep[e.g.][]{valentinuzzi10, carollo13}? 
A promising approach to shed light on the origin of the size evolution  
is to investigate how the number density of dense galaxies evolves over the cosmic time 
\citep{saracco10, bezanson12, cassata13, poggianti13, carollo13, damjanov14, damjanov15}. 
The analysis by different authors slightly differ in the selection of the sample (e.g. passivity, visual classification, colour, Sersic index)
 in the definition of ``dense'' galaxies (e.g selected on the basis of their mean stellar mass density 
$\Sigma$ = M$_{\star}$/2$\pi$R$_{e}^{2}$ or of the sigma deviation from the local size-mass relation (SMR)), 
and in the treatment of the progenitor bias.
\citet{saracco10} find that the number density of dense (1$\sigma$ below the local SMR) massive 
(M$_{\star} >$ 3$\times$10$^{10}$M$_{\odot}$) ellipticals (visually selected) at 0.9$<z<$1.9 
is comparable with that of equally dense ones in local clusters.  
In the same direction, \citet{poggianti13} studying the number density 
of massive (M$_{\star} \gtrsim$ 10$^{10}$M$_{\odot}$) compact (Log (M$_{\star}$/ R$_{e}^{-1.5}$) $\geq$ 10.3 M$_{\odot}$kpc$^{-1.5}$) 
local old galaxies find that it is just a factor 2 lower than that of compact passive galaxies observed at z = 2.
On the opposite side, \citet{vanderwel14} find that 
the number density of dense 
(R$_{e}$/(M$_{\star}$/10$^{11}$M$_{\odot}$) $<$ 1.5 kpc) massive (M$_{\star} >$ 5$\times$10$^{10}$M$_{\odot}$)
passive galaxies decreases by a factor $\sim$ 10 
from z$\sim$2 at z$\sim$0.2.

Actually, the individual size growth is expected to be more efficient for the most massive (M$_{\star} >$ 10$^{11}$) galaxies.
In fact, these systems are expected to host the center of the most massive halo, and to evolve mainly 
through minor mergers \citep{hopkins09}.
However, given the fact that at each redshift the most massive galaxies
are extremely rare \citep[see, e.g.,][]{muzzin13}, 
very few works have investigated the number density evolution of their dense subpopulation over the cosmic time.
\citet{carollo13} find that the number density of passive (sSFR $<$ 10$^{-11}$ yr$^{-1}$) $and$ elliptical 
galaxies with M$_{\star}>$10$^{11}$M$_{\odot}$ and R$_{e} <$ 2kpc decreases of $\sim$ 30$\%$ in the 
5 Gyr between 0.2$<z<$1. An even mild evolution is claimed by \citet{damjanov15}, who find 
for the passive (colour selected), dense (Log (M$_{\star}$/ R$_{e}^{-1.5}$) $\geq$ 10.3M$_{\odot}$kpc$^{-1.5}$)
galaxies with M$_{\star}>$ 8$\times$10$^{10}$M$_{\odot}$ an almost
constant number density  in the range 0.2$ < z <$0.8.

What is still missing is a study of the number density of the ``pure'' 
elliptical (elliptical and spheroidal
galaxies, visually selected, hereafter ETGs) massive  (M$_{\star}>$10$^{11}$M$_{\odot}$) dense 
population over the time, and in particular at 
z $>$ 0.8-1, the range in which most of the evolution of spheroidal galaxies occurs.

In this paper, we want to probe the mass accretion history of these most massive and dense
(M$_{\star} >$ 10$^{11}$M$_{\odot}$, $\Sigma >$ 2500 M$_{\odot}$pc$^{-2}$)
ETGs over the last $\sim$ 9 Gyr, by tracing the evolution of 
their comoving number density and by comparing their 
structural and dynamical parameters from redshift 1.6 to $\sim$ 0.
One of the aspects of this work which differs from the previous analysis is the selection of the samples 
based on the visual classification (or on methodology which can mimic the visual classification) over the whole redshift range. 
In fact, the star formation histories of the galaxies are often more complex than we model them, being characterized by 
bursty star formation histories against smooth and gentle declining ones. 
As a consequence of this, in a phase of low star formation activity, an active galaxy can just ``temporarily'' 
show red colours and/or low sSFR. Moreover, galaxies tend to become passive with time. 
Hence, the population of passive galaxies collects a different mix of morphological types at different redshift. 
This leads to a non homogeneous comparison of their properties at different time \citep[see, e.g.,][]{huertascompany13}.
In this context, the spheroidal shape of a galaxy is a more stable property over the time and  
allows us to select the same population over the cosmic time.
In addition, we have selected dense galaxies on the basis of their $\Sigma$ rather then 
using a constant cut in R$_{e}$ or on the basis of their deviation with respect to the local SMR. The main 
reason for this choice is that the mean stellar mass density, up to z $\sim$ 1.5, 
is better correlated to, e.g., the age, the sSFR of the galaxy
\citep{franx08, wake12}
and hence constitutes a more meaningful parameter.

We have organized the analysis in two parts. 
 
The first one (see Sec. 2), is focused on the evolution of the number density 
of ultramassive dense ETGs over the redshift range 0$< z <$1.6
to investigate whether the ultramassive dense high-z ETGs
are numerically consistent with those observed at z $\sim$ 0, or on the contrary they 
numerically decrease along the time suggesting an individual growth of the galaxies.
For this purpose, we have referred to SDSS DR4 survey to constrain the number density of ultramassive dense ETGs at $z\sim0.1$
paying particular attention to the bias affecting the catalog (see Sec. 2.3),
and to COSMOS field to cover the range $0.2 \lesssim z \lesssim 1$ (see Sec. 2.2). 
For the highest redshift bin (1.2 $\lesssim z <$ 1.6) 
we have considered two different surveys of similar area ($\sim$ 150 arcmin$^{2}$), the GOODS-South \citep{giavalisco04} and the MUNICS \citep{drory01}, 
in order to average over the cosmic variance (see Sec. 2.1). 

The second part of the paper is focused on the comparison
of the structural (i.e. R$_{e}$, M$_{\star}$) and dynamical (i.e. velocity dispersion $\sigma$)
parameters of ETGs over the last 9 Gyr (see Sec. 3).
The aim of this analysis is to asses whether high-z ultramassive dense ETGs have  
counterparts at lower z characterized by the same structural and dynamical properties 
or rather some of them are not present at later epochs.
To perform this comparison we have selected data from literature (see Sec. 3.1). 
Clearly, the request of available velocity dispersion drastically reduces the 
statistic of the sample at  1.2 $ < z <$ 1.6 to 11 ETGs.
For 4 of them (plus one at z = 1.9) 
we present here the new measurements of velocity 
dispersions (see Sec. 3.1). 
In Sec. 4 we summarize the work and present our results and conclusions.

Throughout the paper we adopt standard cosmology with H$_0$ = 70 kms$^-1$ Mpc$^-1$, $\Omega_m$ = 0.3
and $\Omega_\lambda$ = 0.7, the stellar mass we report are derived using the Chabrier 
Initial Mass Function (IMF) \citep{chabrier03} and magnitude are in the AB system when not otherwise specified.

\section{The number density of ultramassive dense ETGs}

In this section we constrain the stellar mass
accretion history of  ultramassive ETGs investigating the evolution of their number density $\rho$
over the last $\sim$ 9 Gyr. 
Since we cover the redshift range from 1.6 to 0 using different surveys (with, e.g., different magnitude cut and completeness)
in Tab. \ref{sum} we summarize their main characteristics, and the references for the main quantities involved in this part of analysis 
(i.e., stellar mass, and R$_{e}$).
As a reminder, we highlight that the analysis presented in this paper is 
aimed to probe the evolution of ``ETGs'', i.e. elliptical and spheroidal galaxies selected
on the basis of the visual classification. When this classification is not feasible, we have explicitly discussed it in the text.
\begin{table*}\footnotesize
\begin{center}
\caption{The main characteristics of the surveys used to derived the number density evolution from z = 1.6 to z $\sim$ 0.
\textit{Column 1}: Name of the survey, 
\textit{Column 2}: Selection criteria, \textit{Column 3}: redshift range probed by the survey in our analysis, 
\textit{Column 4}: Level of completeness for galaxies with M$_{\star} >$ 10$^{11}$M$_{\odot}$ in the redshift range probed, \textit{Column 5 (6)}: 
Reference adopted for stellar mass (R$_{e}$) estimates, \textit{Column 7}: Filter of the images used to derived the R$_{e}$.
In all the surveys, stellar masses have been derived through the fit of the spectral energy distribution, and effective radius fitting
a Sersic profile to the images.}
\begin{tabular}{ccccccc}
\hline
\hline
 Name    &  Selection   &    redshift range   &   M$_{\star} >$ 10$^{11}$M$_{\odot}$  &  M$_{\star}$ reference & R$_{e}$ reference & R$_{e}$ filter \\  
         &        criteria              &           probed                 &  completeness           &                           &          &         \\
\hline
\hline
  MUNICS &  K < 20.4 $\&$  & 1.2 $\lesssim z <$ 1.6 & 100$\%$  & Sec. 3.1 & \citet{longhetti07} & F160W/HST\\
         &  R - K $\lesssim$ 3.6$^{*}$ &                     &                             &           &                   &            \\ 
  GOODS-South & Ks < 22.5 & 0.65 $ < z <$ 1.6 & 100$\%$  & \citet{tamburri14} & \citet{tamburri14} & F850LP/HST\\ 
  COSMOS & F814W < 22.5 & 0.2 $ < z <$ 1.0 & 100$\%$ up to z $\sim$ 0.8 & \citet{ilbert13} & \citet{sargent07} & F814W/HST\\
         &              &                   & $\sim$80$\%$ at 0.8 $< z <$ 1.0 &                  &                   &          \\
  SDSS & $r <$ 16.8 & 0.063 $< z <$ 0.1 & 100$\%$   & MPA - JHU$^{**}$ & \citet{blanton05} & $r$\\
\hline
\hline
 \end{tabular}
 \label{sum}\\
\end{center}
$^{*}$ the color cut remove from the sample the less dense systems, 
so does not compromise the selection of dense ETGs. $^{**}$http://www.mpa-garching.mpg.de/SDSS/DR7/
\end{table*}

\subsection{The number density of ultramassive dense  ETGs at $1 \lesssim z < 1.6$}

For the comoving number density of ultramassive dense ETGs at z$>$ 1, we have referred 
 to two samples of ETGs, one selected in the MUNICS-S2F1 field, and published in \citet{saracco05} and \citet{longhetti07}, 
and the other selected in the GOODS South-field and described in \citet{tamburri14}. 

\textit{MUNICS sample -} The MUNICS sample consists of 7 ultramassive dense ETGs spectroscopically confirmed at z $\sim$ 1.5.
These galaxies were observed as part of the spectroscopic follow up of the complete sample of 19 extremely 
red objects (EROs, K $\lesssim$ 20.4, R-K $\lesssim$ 3.6) in the S2F1 field \citep{saracco05}. 
We have checked whether the cut both in magnitude and colour of the MUNICS sample
affect our selection of ultramassive dense ETGs.
Using the sample of ETGs in the GOODS-South field presented in \citet{tamburri14} (see below), 
we have verified that a sample of ETGs brighter than K = 20.4 is 100$\%$ complete up to z$\sim$ 1.6.
For what concerns the colour cut, it tends to deprive the massive sample of the less dense systems (to which we are not interested in), 
since the redder the colour of a galaxies, the higher its $\Sigma$ \citep{franx08, saracco11}.
All the galaxies of the MUNICS sample, indeed, have $\Sigma >$ 2500M$_{\odot}$pc${-2}$.
The effective radii of the galaxies were derived in the F160W filter with GALFIT
\citep{longhetti07} and the stellar masses were estimated fitting their spectral energy distribution (SED)
with the Bruzual and Charlot models using the Chabrier IMF, 
while visual classification was performed on HST/NIC2 - F160W images (for more details see Sec. 3).
We have revised the number density estimated by \citet{saracco05} considering only the 6 ETGs at 1.2 $< z <$ 1.6 
in the field S2F1 ($\sim$ 160 arcmin$^{2}$),
thus, excluding the ETG S2F1-443 at z = 1.91 (see Sec. 3).
We have found a comoving number density of ultramassive dense ETGs 
of 3.8($\pm$ 1.7)$\times$10$^{-5}$ gal/Mpc$^{-3}$ at 1.2 $ < z <$1.6, which 
becomes 5.6($\pm$ 1.9)$\times$10$^{-5}$ gal/Mpc$^{-3}$ considering the spectroscopic
incompleteness (a factor $\sim$ 1.4). The errors on number densities, throughout the paper,
are based on Poisson statistic.

\textit{GOODS - South sample - } Given the small area of the S2F1-MUNICS field, for comparison, and to average the cosmic variance, 
we have also derived the number density of 
ultramassive dense ETGs in the GOODS-South field
covering an area (143 arcmin$^{2}$) comparable to the S2F1 field.
To this end, we have used the sample of galaxies
visually classified as ETGs on ACS/F850LP images and studied by \citet{tamburri14}.
The sample contains all the ETGs in the GOODS-South field brighter than K$_{s}$ = 22 and for masses 
M$_{\star} >$ 10$^{11}$M$_{\odot}$ is 100$\%$ complete up to z $\sim$ 2 \citep{tamburri14}.
Their stellar masses were derived assuming Bruzual and Charlot models and both
the Salpeter IMF and the Chabrier IMF. We refer to the latter in our analysis.
We have found a number density of ultramassive dense ETGs of 
$1.8(\pm1.3)\times10^{-5}$ gal/Mpc$^3$ in the redshift range $1.3 \lesssim z < 1.6$
and of $2.9(\pm1.6)\times10^{-5}$ gal/Mpc$^3$ in the redshift range $0.95<z<1.25$.
It is worth noting that the structural parameters in the Tamburri's et al.
sample have been derived on the ACS-F850LP images with GALFIT.
Hence the selection of ETGs could be biased against dense ETGs due to their
larger apparent size ($\sim$30$\%$, Gargiulo et al. \citeyear{gargiulo12}) at this wavelength.
However, even correcting  the effective radius of
massive ETGs in the GOODS sample for this difference, the number density of dense ETGs does not
change. In Table 1 and in Fig. \ref{evolunumdens} we report the measures.

\subsection{The number density of ultramassive dense  ETGs at intermediate redshift}

In order to derive the comoving number density of ultramassive dense ETGs at intermediate redshift,
we took advantage of the COSMOS survey.
We have considered the catalogue of \citet{davies15} including
all the available spectroscopic redshifts in the COSMOS area,
the catalogue of structural parameters based on the HST-ACS
images in the F814W filter of \citet{scarlata07} and the
catalogue including the stellar masses of \citet{ilbert13}
based on the multiwavelength UltraVISTA photometry.
The cross match of these three catalogues produced a sample
of 110171 galaxies in the magnitude range $19<F814W<24.8$ over the
1.6 deg$^{2}$ of the HST-ACS COSMOS field.
We have then selected all the galaxies (25262) brighter than
F814W$<22.5$ mag for which structural parameters and morphological 
classification
are robust \citep{sargent07, scarlata07}.
From this sample we have finally extracted our parent sample of 1412 
galaxies with stellar mass M$_*>10^{11}$ M$_\odot$ and early-type 
morphology. 
For the morphological classification we refer to the ZEST (Zurich Estimator of Structural Types) 
estimates by \citet{scarlata07}. ZEST is an automated tool for the galaxies classification
which uses the Principal Component Analysis to store the full information provided
by the entire set of five basic non-parametric diagnostics of galaxy structure 
(symmetry, concentration, Gini coefficient, 2nd-order moment of
the brightest 20$\%$ of galaxy pixels, and the ellipticity) plus the Sersic index.
The software subdivided the galaxies in three main classes: early-types (type = 1), 
disk galaxies (type = 2) and irregular galaxies (type = 3).
The robustness of the software was tested both on $\sim$ 56000 COSMOS galaxies and on local galaxies visually classified, 
with excellent results.  
For our analysis, thus, we have selected as COSMOS ETGs all the galaxies with ZEST-type = 1. 
Among these, 551 have a secure spectroscopic redshift (parameter Z\_USE $\le2$
in Davies et al. 2014 catalogue), and 145 at 0.2 $< z <$ 1.0, have $\Sigma \geq$ 2500 M$_{\odot}$pc$^{-2}$.
As further check on morphology, we have visually classified on the 
HST/F814W-ACS images \citep{koekemoer07} the 145 
ultramassive dense COSMOS galaxies with ZEST-type = 1. 
Actually, we have found that all but one are ETGs.

The selection at F814W$<22.5$ mag implies that for masses
M$_*>10^{11}$ M$_\odot$ the sample is 100$\%$ complete up to $z\sim0.8$
(80$\%$ complete to $z\sim1$).
The stellar masses of \citet{ilbert13} are based on the
Bruzual and Charlot (2003) models and the Chabrier IMF, consistently with
our assumptions.
The structural parameters of \citet{sargent07} are based on
a Sersic profile fitting as in our analysis even if performed
with GIM2D instead of GALFIT.
However, \citet{damjanov15} show that the two methods,
GIM2D and GALFIT,  are in excellent agreement when based on
a single Sersic component, as in our case.
In order to derive the number densities of massive ETGs,
we have applied a correction for the spectroscopic incompleteness that we have
defined as the ratio between the magnitude distribution (in bin of 0.1 
mag) of the parent sample of elliptical galaxies and the sample of ellipticals with
spectroscopic redshift.
In table \ref{numdens} we report the comoving number density of ultramassive 
dense ETGs from z = 0.2 to z = 1.0, estimated in bin of redshift of 0.2 (see also Fig. \ref{evolunumdens}).
It is worth noting that at z$>$0.6 the rest-frame wavelength sampled by the 
F814W filter is significantly different from the one sampled by the filter F160W at z=1.5.
However, as done for the GOODS-south data, we verified that the number
densities derived from the COSMOS data at z$>$0.6 do not change even
assuming a correction of 30$\%$ of the effective radius of galaxies.

\begin{table}\footnotesize
\begin{center}
\caption{The number density $\rho$ of ultramassive dense 
(M$_{\star} > $ 10$^{11}$ M$_{\odot}$, $\Sigma >$ 2500 M$_{\odot}$pc$^{-2}$) 
ETGs from z = 1.6 to $\sim$ 0. $\textit{Column 1}$: redshift range; 
$\textit{Column 2}$: number of massive dense ETGs; 
$\textit{Column 3}$: number density of ultramassive dense ETGs; $\textit{Column 4}$: 
error on the number density quoted; $\textit{Column 5}$: field. 
The number density derived both in the SDSS and in the MUNICS field is a lower limits.}
\begin{tabular}{ccccc}
\hline
\hline
 $ z $    &  N  &   $\rho$               &     $\sigma_{\rho}$  & Field\\  
          &     &   10$^{-5}$ Mpc$^{-3}$ &  10$^{-5}$ Mpc$^{-3}$ & \\
\hline
\hline
  0.066 - 0.1 & 124  & 0.75   &  0.07 & SDSS \\
  0.2 - 0.4   & 13  & 2.6   &  0.9 & COSMOS \\
  0.4 - 0.6   & 14  & 1.9   &  0.5 & COSMOS \\
  0.6 - 0.8   & 39  & 3.6   &  0.7 & COSMOS \\
  0.8 - 1.0   & 79  & 4.7   &  0.8 & COSMOS \\
  0.95 - 1.25 & 3  & 2.9    &  1.6  & GOODS-South \\
  1.3 - 1.6   & 2  & 1.8    &  1.3  & GOODS-South \\
  1.26 - 1.6  & 6  &3.81   &  1.7  & MUNICS\\

\hline
\hline
 \end{tabular}
 \label{numdens}\\
\end{center}
\end{table}

\subsection{The number density of ultramassive dense ETGs in local universe}
\begin{figure}[h!]
\begin{center}
	\includegraphics[angle=0,width=9cm]{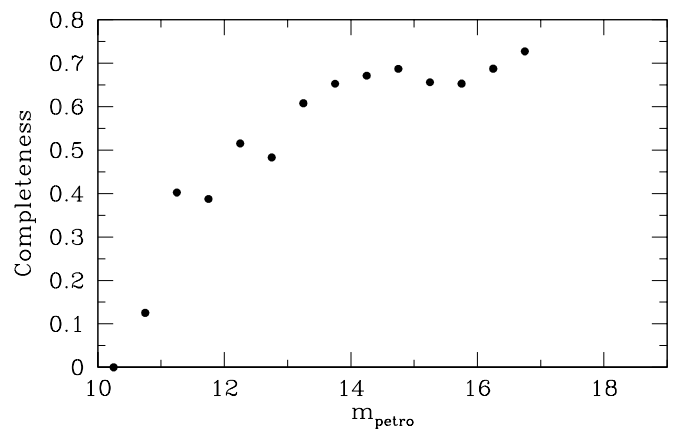} \\
	\caption{The completeness of the spectroscopic SDSS galaxy catalogue with 
respect to the Main Galaxy Sample as a function of galaxy petrosian magnitude.}
	\label{incompleteness}
\end{center}
\end{figure}

As representative sample of local ETGs we have chosen the dataset analyzed by 
\citet{thomas10}\footnote{http://www.icg.port.ac.uk/$\sim$thomasd/moses.html}. The authors 
have collected a sample of 16502 ETGs extracted from the magnitude
limited sample (r $<$ 16.8) of 48023 galaxies selected from the SDSS Data Release 4 \citep{adelman06} 
in the redshift range 0.05$\leq z \leq$0.1.
Contrary to the other available samples, this sample is the only one in which ETGs
are selected through a visual inspection of SDSS multiband images 
as we did for high-z ETGs.
The sample is 100$\%$ complete at M$_{\star} >$10$^{11}$M$_{\odot}$.
We have associated to each ETG its stellar mass taken from MPA-JHU catalogue\footnote{http://www.mpa-garching.mpg.de/SDSS/DR7/}
and obtained through the fit of the multiband photometry (Chabrier IMF), as for the higher redshift galaxies. 
 For the effective radius we have adopted as reference the NYU-VAG 
catalogue \citep{blanton05}. In appendix A we discuss about the robustness and limits of these estimates.

We bear in mind that the galaxy spectroscopic SDSS survey is biased against dense galaxies
\citep[e.g.][]{strauss02, poggianti13a, taylor10b}.
Briefly, the ``main galaxy sample`` (MGS) of galaxy target for the spectroscopic follow up 
is composed by all the galaxies detected at $>$ 5$\sigma$ in the r band with Petrosian magnitude
m$_{pet,r}$ < 17.77 and r-band Petrosian half-light surface brightness $>$ 24.5 mag arcsec$^{-2}$.
On these first cuts, three further selection criteria have been imposed
to reject stars, to avoid the cross talk and to exclude bright objects that can saturate the CCD.
\citet{taylor10b} have investigated how these cuts affect the distribution of local galaxies in the size mass plane and have shown 
that 
at z$<$ 0.05, galaxies with masses and sizes similar to those of z $\sim$2 galaxies are not targeted for the spectroscopic
follow-up. Nonetheless, this incompleteness decreases at increasing redshift:  
at z $\gtrsim$ 0.06, on average, the
completeness of the spectroscopic survey for compact massive galaxies is $\sim$ 80$\%$. 
For this reason, we have selected from the Thomas et al.'s catalogue, all
the ultramassive dense ETGs  at 0.063 $ < z <$ 0.1 (156 galaxies). 
Based on the evidence of Fig. \ref{checkstime} we have removed from this sample
those classified as ultramassive outliers in Appendix A (22 galaxies), 
i.e. galaxies that should not be included in the local ultramassive sample
since, most probably, their stellar masses derived through the SED fitting are overestimated. 
We have visually checked this final subsample of 124 galaxies (156 - 22), and have found that no galaxy has
a close companion, or is in the halo of a bright star, or is involved in a merger, factors that 
can alter the estimate of the effective radius. However, we have noted that $\sim$15 $\%$ of 
the sample (20 galaxies) has a not clear elliptical or S0 morphology, despite they are included in the Thomas et al.'s ETGs sample.
We have checked the structural parameters of these galaxies, and they have a mean Sersic index $n$ $\simeq$ 3.5 
(for all the galaxies $n$ $>$ 2.5) consistent with the light 
profile of a spheroidal galaxy, but an axis ratio b/a $<$ 0.4. 
Since the Blanton's catalogue does not provide the axis ratio, we have used the one obtained 
from the SDSS pipeline, fitting the real galaxies with a de Vaucouleur profile. Actually, the b/a should be not 
dependent of the surface brightness law used to fit the profile.
We have left these elongated objects in the final sample, since the Thomas et al. sample
is widely adopted, and studied, but in Sec. 2.4 we show that this choice does 
not impact our results.

Beside this source of incompleteness due to the 
targets selection, we have to consider that
not all the the spectroscopic target were actually observed (for example because two fibres cannot be placed 
closer than 55`` on a given plate). This source of incompleteness should be more important for the brightest 
 galaxies and should be taken into account in the derivation of the comoving number density.
To quantify it we have extracted the MGS from DR4 database 
restricting the query to 
only the sciencePrimary objects (in order to reject the multiple observations, 648852 galaxies).
From this sample we have rejected galaxies classified as quasar remaining with 636980 galaxies.
Among these objects
384564 ($\sim$60$\%$) have been spectroscopically observed
and $\sim$88$\%$ of them has secure redshift (i.e. zwarning = 0).
We have defined the completeness in bin of magnitude as the ratio of galaxies with secure redshift,
over the number of galaxies in the MGS (see Fig. 1).
Taking in consideration this, we have found a number density for local ultramassive dense ETGs of 
0.75($\pm$0.07)$\times$10$^{-5}$ gal/Mpc$^{-3}$.
We have repeated the measure of $\rho$ assuming 
as stellar masses for local ETGs those derived by \citet{kauffmann03} with spectral indices - see Appendix A - 
and we  have found $\rho$ =
0.7($\pm$0.06)$\times$10$^{-5}$ gal/Mpc$^{-3}$ .
We have corrected the number density for the incompleteness of the 
spectroscopic sample with respect to the MGS but
we remind that  the original selection cuts of the MGS can exclude part of massive dense ETGs 
especially in the lowest redshift bin. The number density we quote is therefore a lower limit.
\begin{figure*}[t]
\begin{minipage}[t]{1\linewidth}
  \centering
  \includegraphics[angle=0,width=17cm]{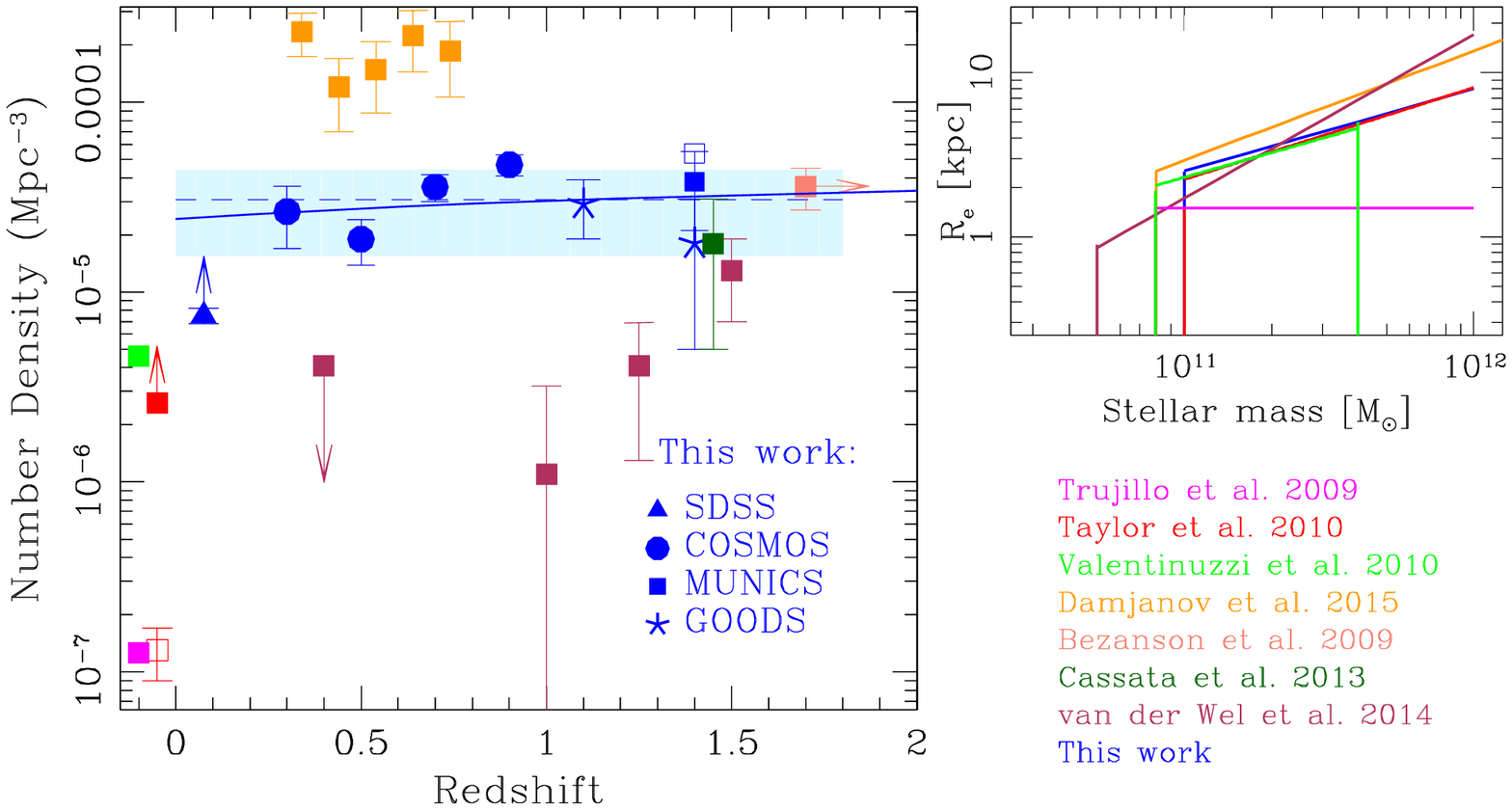} 
\end{minipage}
\begin{minipage}[b]{\linewidth}
    \centering
\footnotesize
\begin{tabular}[b]{ccccc}
\hline 
\hline
Reference & z & Stellar mass range & Compactness definition & Supplementary galaxy selections \\
 \hline
\hline
 & & & & \\
This work & 0 $ \lesssim z \lesssim$ 1.6 & M$_{\star} >$ 10$^{11}$ & $\Sigma \geq$ 2500 M$_{\odot}$pc$^{-2}$  & Elliptical morphology \\
          &                              &                         &                                          & (Visual classification) \\
Trujillo et al. (2009) & 0 $ < z <$ 0.2 &  M$_{\star} >$ 8$\times$10$^{10}$ & R$_{e} <$ 1.5 kpc &  -- \\
         &                              &                         &                                          &   \\
Taylor et al. (2010) & 0.066 $ < z <$ 0.12 & M$_{\star} >$ 10$^{11}$ & log(R$_{e,z}$/kpc) $<$  & (u - r) > 2.5 \\
                     &                     &                                  &0.56$\times$(log(M$_{\star}$/M$_{\odot}$) - 9.84) - 0.3 & \\
Valentinuzzi et al. (2010) & 0.04 $ < z <$ 0.07 & 8$\times$10$^{10} <$ M$_{\star} <$ 4$\times$10$^{11}$M$_{\odot}$ & $\Sigma >$ 3000M$_{\odot}$pc$^{-2}$ & -- \\
          &                              &                         &                                          &  \\
Damjanov et al. (2015) & 0.2 $ < z <$ 0.8 & M$_{\star} >$ 8$\times$10$^{10}$ & log(R$_{e}$/kpc) $<$  & NUVrJ colours\\
                        &                 &                                  &         0.568$\times$log(M$_{\star}$/M$_{\odot}$) - 5.74                      &     \\

Bezanson et al. (2009) & z = 2.5 & M$_{\star} >$ 10$^{11}$ & -- & Quiescent  \\
          &                              &                         &                                          &  \\
Cassata et al. (2013) & 1.2  $ < z <$ 1.6 & M$_{\star} >$ 10$^{11}$ & $\Sigma \geq$ 2500 M$_{\odot}$pc$^{-2}$ & sSFR $<$ 10$^{-11}$yr$^{-1}$ \\
                      &                   &                          &                                         &  + elliptical morphology \\
van der Wel et al. (2014) & 0 $ \lesssim z \lesssim$ 1.6 & M$_{\star} >$ 5$\times$10$^{10}$ & R$_{e}$/(M$_{\star}$/10$^{11}$M$_{\odot}$)$^{0.75}$ $<$ 1.5 kpc & UVJ red colours \\
\hline
\hline
\end{tabular}
\end{minipage}
	\caption{\textit{Top - left}: 
The evolution of the comoving number density of ultramassive dense ETGs (M$_{\star} >$ 10$^{11}$M$_{\odot}$, 
$\Sigma >$ 2500 M$_{\odot}$pc$^{-2}$)
from redshift 1.6 to redshift $\sim$ 0 (blue symbols).  
The open blue square indicates the number density of ultramassive dense MUNICS ETGs at $<z>$ =1.4 corrected 
for the spectroscopic incompleteness of the MUNICS survey (see text).
Solid blue line is the best fit relation to our data in the range 0.066 $< z <$ 1.6 ($\rho$ $\propto$ (1 + z)$^{0.3}$), 
while blue dotted line indicates the mean value over the redshift range  0.066 $< z <$ 1.6 
(dashed area marks the 1$\sigma$ deviation around this value).
Orange squares are the number density for ultramassive compact galaxies by \citet{damjanov15},
magenta one by \citet{trujillo09}, light green one by \citet{valentinuzzi10}, purple ones by \citet{vanderwel14}, 
salmon one by \citet{bezanson09}, dark green one by \citet{cassata13}, and open red square by \citet{taylor10}.
The points by Taylor et al. (open red square) and Cassata et al. (dark green square) 
are derived selecting from their sample only the galaxies with M$_{\star} >$ 10$^{11}$M$_{\odot}$ (see Tab. in the figure). 
Filled red square is the number density by Taylor et al. obtained using the circularized Sersic R$_{e}$.
\textit{Top - right}: The cuts in R$_{e}$ and M$_{\star}$  of our and others samples in the size mass plane 
(colours are the same in the top - left panel). \textit{Bottom table}: A quick comparison of the 
main characteristics of our sample with respect to those by other authors we compare. 
\textit{Column 1}: reference paper, \textit{Column 2}: redshift range, \textit{Column 3}: stellar mass limits (for Taylor et al.
and Cassata et al. is reported the one we use), \textit{Column 4}: compactness definition 
, \textit{Column 5}: other supplementary 
selections applied to the samples.}
	\label{evolunumdens}
\end{figure*}

\subsection{The evolution of the number density of ultramassive dense ETGs over the last 9 Gyr: results }

Figure \ref{evolunumdens} shows the evolution of number density for the ultramassive dense ETGs in the last $\sim$ 9 Gyr.
Once the possible sources of non-homogeneity are taken into account the mean value of the number density 
of ultramassive dense ETGs, within the uncertainties, is $\sim$3.0$\pm$1.5$\times$10$^{-5}$Mpc$^{-3}$. 
We observe that at z$\sim$0.8, in the COSMOS field, there is a significant overdensity \citep[see e.g.][and reference therein]{damjanov15} 
and that, correspondingly, this is the point that most deviates in Fig. 2. 
We have fitted the number densities estimated in the redshift range 0.06 $\lesssim z< $1.6
with a power law, treating the SDSS and MUNICS point as lower limits,
and have found $\rho$(z) $\propto$ (1 + z)$^{0.3\pm0.8}$,
showing that the number density of ultramassive dense ETGs may decrease from z=1.6 to z=0, approximately by a factor 1.3.
We have repeated the fit excluding from local sample those galaxies with ambiguous morphology (see Sec. 2.3), 
but the slope does not change, since in the fit this point is considered a lower limit.

\subsection{Comparison with previous works}

The number density estimates of dense ETGs are clearly dependent on many factors, such as
the stellar mass limit,
the definition of compact galaxies, and the criteria adopted to select the sample (e.g. visual classification, colour, passivity).
This implies that any comparison of our results with those by other authors is not straightforward.
In the following we try to compare  our 
estimates of number density with those by other authors which adopted selection criteria as similar as possible to ours.
Given the many factors concurring in the final estimate of $\rho$, to better visualize these comparisons, 
in Fig. \ref{evolunumdens} we report, in 
the size mass plane, the
stellar mass and effective radius limits of each sample. 
At the same time, we add a table that can 
immediately show the characteristics of all the samples studied.

\subsubsection{Local universe}

For what concerns the local universe, there is a clear tension among the results already published, 
since the number densities derived from the SDSS database \citep{trujillo09, taylor10} turn out to be
at least 1-2 order of magnitude lower than that derived from other database (e.g. WINGS/PM2GC) 
\citep[][see Fig. \ref{evolunumdens}]{valentinuzzi10, poggianti13}. 
Interestingly, Fig. \ref{evolunumdens} shows that our estimate of local number density, 
although derived on SDSS database,
is much more in agreement with the finding of \citet{valentinuzzi10} than with that by \citet{trujillo09} and \citet{taylor10}.
In the following, we investigate the possible origins of the observed discrepancy. 

In fact, \citet{taylor10} search for counterparts of high-z  compact massive galaxies
in the DR7 SDSS spectroscopic database. 
To this aim, they select from the sample of local galaxies at 0.066 $< z <$ 0.12, with M$_{\star} >$ 10$^{10.7}$ M$_{\odot}$,  
and log(R$_{e,z}$/kpc) $<$ 0.56$\times$(log(M$_{\star}$/M$_{\odot}$) - 9.84) - 0.3  (i.e. with 
radius in the SDSS-$z$ band at least twice 
smaller than the local SMR by \citet{shen03}), those with $^{0.1}$(u -r) > 2.5\footnote{the superscript 0.1 
indicates rest-frame photometry redshifted to z = 0.1, \citep[see, e.g.,][]{blanton07}}, 
in order to select the descendant candidates of high-z dense galaxies.
They reject from their sample all the galaxies with possible problems with their size measurement 
(due to confusion with, e.g., other galaxies, or with extended halos and/or with spikes of bright stars) or mass estimate,
for a total final sample of 63 galaxies. Among these galaxies, 11 have stellar mass $>$ 10$^{11}$M$_{\odot}$
and yield a number density of 1.3($\pm$0.4)10$^{-7}$gal/Mpc$^{3}$ over the entire volume of the SDSS-DR7 spectroscopic survey  
(see FIG. \ref{evolunumdens} open red square). 
This value, not corrected for the spectroscopic incompleteness of SDSS-DR7 survey, is $\sim$ two orders of magnitude 
lower than the one we have found. Actually, this huge observed difference 
cannot be ascribed only to the completeness function. In fact, it is really difficult to pinpoint
the origin of such a difference between our and their result, since the two samples are differently selected.

In fact, the cut they apply in R$_{e}$ is very similar to our cut in $\Sigma$ (see size mass plane in Fig. 2), 
however, differently from us they select only red galaxies.  
To test the impact of the colour selection, we have
computed the $^{0.1}$(u -r) colour 
for the galaxies of our sample and have found that all but one have $^{0.1}$(u -r) > 2.5, so the colour
selection cannot be the source of the 
observed difference in the number density.
We have then investigated the impact on the number density of the different R$_{e}$ used. 
Actually, we use the $circularized$ R$_{e}$ obtained fitting the 
surface brightness profile with  a Sersic law (hereafter R$_{e,Ser}$), instead
of the semi-major axis SM$_{e}$, used by Taylor et al., obtained from SDSS pipeline fitting the SDSS images 
with a de Vaucouleur profile (hereafter SM$_{e,dV}$).
For this reason, we have repeated their analysis looking for differences in the
$\rho$ estimates varying the definition of R$_{e}$.
To do this, we have selected from the primary observations of the DR7 spectroscopic sample the 226312 galaxies 
with secure redshift (i.e. zwarning = 0) and spectroscopic redshift 0.066 $< z_{spec} < $ 0.12.
Among them, we have collected those with M$_{\star}>$10$^{11}$M$_{\odot}$, 
log(SM$_{e,dV}$/kpc) $<$ 0.56$\times$(log(M$_{\star}$/M$_{\odot}$) - 9.84) - 0.3, 
and $^{0.1}$(u -r) > 2.5. 
For the SM$_{e,dV}$, consistent with Taylor et al., we retrive the catalogue from the SDSS database, while
 for the stellar masses 
derived through the SED fitting, we refer to the MPA - JHU catalogue (see Sec. 2.3).
We have visually inspected this subsample, and we have rejected all the galaxies with 
problems related with the size measurement (e.g. merging galaxies, or close pairs or galaxies
in the halo of a bright stars), leaving us with 49 galaxies. None of them is a quasar.
To check the reliability of the stellar mass estimates we have applied the same procedure of Appendix A.
Among the 49 galaxies of our selection, 38 have a stellar mass estimate derived also from the spectral indeces
(stellar masses from spectral indeces were derived only for galaxies in the DR4 database),
 and for 10  of them ($\sim$ 26$\%$) the two estimates are consistent. These ten galaxies are all included in the subsample of 11 
massive galaxies of Taylor and collaborators. Assuming that the fraction of galaxies with consistent estimates 
of M$_{\star}$ is constant, we should expect  $\sim$ 12 galaxies (49$\times$0.26) with 
M$_{\star}>$10$^{11}$M$_{\odot}$, log(SM$_{e,dV}$/kpc) $<$ 0.56$\times$(log(M$_{\star}$/M$_{\odot}$) - 9.84) - 0.3, 
$^{0.1}$(u -r) > 2.5 in DR7 release, in agreement with Taylor and collaborators.
When circularized SDSS de Vaucouleur radii (R$_{e,dV}$) are instead used, we have found 569 galaxies with 
M$_{\star}>$10$^{11}$M$_{\odot}$, log(R$_{e,dV}$/kpc) $<$ 0.56$\times$(log(M$_{\star}$/M$_{\odot}$) - 9.84) - 0.3, 
$^{0.1}$(u -r) > 2.5, among which 186 have good  R$_{e}$ 
estimates and stellar mass, against the 12 found when semi-major axes are used.
Finally, when we use the Blanton's R$_{e,Ser}$ (circularized and obtained fitting a Sersic profile), 
we find 687 galaxies with 
M$_{\star}>$10$^{11}$M$_{\odot}$, log(R$_{e,Ser}$/kpc) $<$ 0.56$\times$(log(M$_{\star}$/M$_{\odot}$) - 9.84) - 0.3, and
$^{0.1}$(u -r) > 2.5, which after all the checks and the visual inspection, result in a final sample of $\sim$ 224 galaxies,
a factor $>$ 20 higher than the one obtained using the semi-major axis.
The higher number of galaxies we have found using R$_{e,Ser}$ with respect to the one 
obtained using circularized de Vaucouleur radius 
is consistent with the expectations. Compact galaxies are known to 
generally have Sersic index $n <$ 4, thus, the value of R$_{e}$ obtained fitting a de Vaucouleur profile should be higher with 
respect to that obtained fitting a Sersic law. This implies that the number of compact objects, defined using 
a ``Sersic'' - R$_{e}$ should be greater than that obtained using a ``de Vaucouleur'' - R$_{e}$ , as we find.
For more details on the comparison of the local samples selected using  
circularized R$_{e}$ and semi-major axis (SM$_{e}$), or de Vaucouleurs and Sersic R$_{e}$
see Appendix B.
Thus, these checks show that the much lower number density found by Taylor et al. with respect to our estimate is due 
to the fact that they use the semi-major axis. 
Once the circularized Sersic R$_{e}$ are used, the number density (not corrected 
for spectroscopic incompleteness) increases up to 
2.6($\pm$0.2)10$^{-6}$ gal/Mpc$^{-3}$, qualitatively in agreement with our estimate.

\citet{trujillo09}, from the analysis of the SDSS DR6 database, claim that only the 0.03$\%$
of galaxies at z$<$ 0.2 with M$_{\star} >$ 8$\times$10$^{10}$M$_{\odot}$ have 0.05'' $<$ R$_{e} <$1.5 kpc.
We have found that among the 99783 galaxies in the SDSS DR4 spectroscopic sample 
with z $<$ 0.2, zwarning = 0 and M$_{\star} >$ 8$\times$10$^{10}$M$_{\odot}$, 
only 113 (i.e. $\sim$ 0.11$\%$ of the massive sample) 
 have 0.05'' < R$_{e} < $ 1.5 kpc.
The visual inspection rejects 76 galaxies since e.g. 
in the halo of a bright star, or a merger galaxy. This leave a sample of 37 galaxies, i.e. the 0.037$\%$ (vs 0.03$\%$ by Trujillo) 
of the original sample of local galaxies with M$_{\star} >$8$\times$10$^{10}$M$_{\odot}$. 
In this case, thus, the difference we have found among our estimate and the one by Trujillo et al. is to totally ascribed to the 
different selection criteria. The authors probe a region of the size-mass plane 
included in our region (see size mass plane in Fig. \ref{evolunumdens}) and known to be highly incomplete in the 
SDSS spectroscopic survey. Both factors lead to a very low number density. 

\citet{valentinuzzi10} studying the sample of 78 clusters included in the WINGS survey, 
have derived the number density of local (0.04 < z <0.07) galaxies with 
8$\times$10$^{10} <$ M$_{\star} <$ 4$\times$10$^{11}$M$_{\odot}$ and $\Sigma >$ 3000M$_{\odot}$pc$^{-2}$. 
Assuming no dense massive galaxies outside the clusters, they found a hard lower limit to the number density of 
these systems of 0.46$\times$10$^{-5}$ gal/Mpc$^{-3}$. Using the DR7, we have found 181 galaxies in the 
same z/M$_{\star}$/$\Sigma_{g}$ range, where $\Sigma_{g}$ is the mean stellar mass density derived 
using the effective radius measured by Blanton on SDSS $g$-band images, 
i.e. the SDSS band that gets closer to the v-band used by \citet{valentinuzzi10}.
Among the 181 selected galaxies, 120 have reliable structural parameters,
and $\sim$ 64$\%$ of them has a robust stellar mass estimate.
Taking into account the effect of the incompleteness, we have found that the 
number density is $\rho$ = 5.21$\times$10$^{-5}$ gal/Mpc$^{3}$, thus a factor 10 higher than the lower limit of 
\citet{valentinuzzi10}.
Our estimate places in the right direction with respect to that of Valentinuzzi,  but we are not 
in the position to make a more quantitative comparison.
We avoid to compare our results with those by other analysis \citep[e.g.][]{poggianti13} which, although 
treating the evolution of the number density  of compact galaxies, adopt a much lower cut in stellar mass with respect to our
(e.g. $\sim$ 1-5$\times$10$^{10}$M$_{\odot}$).

\subsubsection{Intermediate redshift range}

\citet{carollo13} derive the number density of M$_{\star}>$10$^{11}$M$_{\odot}$ quenched 
(sSFR $<$ 10$^{11}$yr$^{-1}$) and elliptical galaxies 
with R$_{e}<$ 2.5kpc over the redshift range 0.2 $< z <$ 1  and find that it decreases by a factor $\sim$ 30$\%$.
Their analysis is focused on a sample of galaxies, on average, denser than the one studied here, 
and differently from us they do not consider all the elliptical galaxies, but the subsample of passive ones.
Despite this, their trend is in fair agreement with our findings. 
In fact, we find a decrease
of $\rho$ of $\sim$ 20$\%$ in the redshift range 0.2 $ < z <$ 1.0.

\citet{damjanov15} present the number density of quiescent (colour - colour selected) galaxies with 
Log(M$_{\star}$/R$_{e}^{1.5}$)$>$ 10.3M$_{\odot}$kpc$^{-1.5}$  and M$_{\star} >$ 8$\times$10$^{10}$ in the COSMOS field
and find that it is roughly constant in the redshift range 0.2 $< z <$ 0.8. 
Their sample includes our one, thus, as expected their number density values are higher than those we find, but 
once again, their trend of $\rho$ with z is in fair agreement with us. Actually, from z = 0.8 to z = 0.2 we find that $\rho$
decreases by $\sim$ 10 $\%$.

\subsubsection{High redshift}

\citet{bezanson09}, integrating the Schechter mass function given in \citet{marchesini09} find that the
number density of the galaxies with M$_{\star}>$10$^{11}$M$_{\odot}$ is 7.2$^{+1.1}_{-1.1}$ $\times$ 10$^{-5}$ Mpc$^{-3}$ at z = 2.5.
Assuming that $\sim$ half of the galaxies with M$_{\star}>$10$^{11}$M$_{\odot}$ are compact, they 
find that the number density of compact massive galaxies at z = 2.5 is 3.6$^{+0.9}_{-0.9}$ $\times$ 10$^{-5}$ Mpc$^{-3}$.
We note that the authors assume the \citet{kroupa01} IMF, which provides stellar masses a factor $\sim$ 1.2 higher than that 
obtained assuming the Chabrier IMF, so their value with respect to ours is an upper limit. Really interestingly 
the estimate by \citet{bezanson09} at z = 2.5 is in agreement with the mean value we find at z $<$ 1.6 and with our best-fit.

\citet{cassata13} derive the number density of compact (i.e. 1$\sigma$ below the local SMR) 
and ultracompact (more than 0.4 dex smaller
than  local  counterparts  of  the  same  mass) passive (sSFR $<$ 10$^{-11}$yr$^{-1}$) galaxies  
with elliptical morphology and M$_{\star}>$ 10$^{10}$M$_{\odot}$ 
in the $\sim$ 120 arcmin$^{2}$ of the GOODS-South field covered by the 
first four CANDELS observations \citep{grogin11, koekemoer11} and by the Early Release Science Program 2 \citep{windhorst11}.
Using their published sample of 107 galaxies, we have selected the only two galaxies
with M$_{\star}>$10$^{11}$M$_{\odot}$ and $\Sigma>$ 2500M$_{\odot}$pc$^{-2}$ at 
1.2 $\leqslant z \leqslant$1.6, which account for a number density of 1.47($\pm$1.04)10$^{-5}$ gal/Mpc$^{3}$.
This value is slightly lower than the one we have found in the GOODS - South field (see Tab. 1),
nonetheless we bear in mind that the authors selected passive $and$ elliptical galaxies
while we apply a pure visual classification.
\citet{tamburri14}, comparing the sample of passive galaxies with that of elliptical ones, visually classified, in the GOODS - South field, 
have shown that about $\sim$26$\%$ of ETGs have sSFR $>=$ 10$^{-11}$yr$^{-1}$, i.e. are not passive. 
Taking in consideration this, we have corrected the number density by a factor 1.26, obtaining 1.85($\pm$1.30)$\times$10$^{-5}$ gal/Mpc$^{3}$, in
agreement with our estimate on the GOODS - South field.
We note that the agreement between our number density estimate and that from Cassata's sample, 
although on the same field, is not straightforward.
Actually, all the quantities involved in our estimate 
(e.g. stellar mass, effective radius) are derived independently by us.
However, despite the different approaches and procedures, the number
densities are consistent, reinforcing the robustness of the result.

In the 3D-HST survey, \citet{vanderwel14} derive the number density of 
compact (R$_{e}$/(M$_{\star}$/10$^{11}$M$_{\odot}$)$^{0.75}$ $<$ 1.5 kpc) 
passive (on the basis of their UVJ colours) galaxies with M$_{\star} >$ 5$\times$10$^{10}$M$_{\odot}$
in the redshit range 0 $< z <$ 3. 
In contrast with us, the authors find a decrease by a factor $\gtrsim$ 10 from z= 1.6 
to z = 0. We have to premise that a reliable comparison with their work
is unfeasible given the much different criteria used to select the samples,
so we cannot go deeper in the orgin of the drop that can be due to, e.g., the different stellar mass limit 
(M$_{\star} >$ 5$\times$10$^{10}$M$_{\odot}$ vs M$_{\star} >$ 10$^{11}$M$_{\odot}$)
and/or to a peculiarity of the population of red galaxies (in principle, we have no 
reason to suppose that the formation and evolutionary picture of ultramassive dense ETGs is similar to 
that of less massive and red galaxies). 
However, we find jarring that the values they find are 
a factor $>$ 5 lower than ours, despite their more relaxed cut 
in stellar mass. In the following we simply try to understand
the origin of this discepancy.
The criterion they adopted to select compact galaxies is very similar to our (see Fig. \ref{evolunumdens}),
as well as the reference restframe wavelength for the R$_{e}$ ($r$-band for us and 5000\,$\AA$ for them).  
Thus, the reasons for the huge discrepancy we observe cannot reside in these
features.
There are two main differences between their and our
approach. 
The first one is that they used the semi-major axes SM$_{e}$, instead of 
the R$_{e}$.
The second one is that they did not directly measure the
semi-major axis on images sampling the 5000\,$\AA$ rest-frame
at the various redshift considered, but they derived it
from the SM$_{e,F125W}$ (i.e the semi-major axis measured on the F125W-band 
images) assuming a variation with 
$\lambda$ ($\Delta$ $\log$ SM$_{e}$/$\Delta$ $\log$ $\lambda$) = -0.25 
(see their Eq. 2).
In order to investigate whether one of these aspects (or both) are responsible of the 
observed discrepancy,
we have used our sample of ETGs selected in the GOODS-South field by \citet{tamburri14} and have derived the number densities 
of ultramassive dense ETGs following 
the recipe by \citet{vanderwel14}. 
In details, we have associated at each galaxy of our sample
the structural parameters derived by van der Wel in both the F125W and F160W filter. 
From them we have estimated the R$_{e,5000\AA der}$, and the SM$_{e, 5000\AA der}$
following their Eq. 2. 
In the redshift range 0.95 $ < z <$ 1.25 in the GOODS South field we have found 3 galaxies 
with M$_{\star}$ $\geq$ 10$^{11}$M$_{\odot}$ and 
$\Sigma \geq$ 2500 M$_{\odot}$pc$^{-2}$ (see Table 2), while 
using R$_{e,5000\AA der}$ and SM$_{e,5000\AA der}$, there are no ETGs. 
At 1.3 $ < z <$ 1.6 we find 2 galaxies. We have found the same number of objects
using R$_{e,5000\AA der}$, and just 1 using the semi-major axis SM$_{e,5000\AA der}$.
Thus, applying the same procedure by van der Wel et al. to derive the number density of ultramassive dense ETGs
we find values of $\rho$ lower or comparable (and no more higher) than the ones presented in their paper, 
thus much more consistent with the expectations considering the fact that our cut in stellar mass is more restrictive 
(for a more exsaustive analysis on the effect of the extrapolation of the SM$_{e}$ using the van der Wel et al.'s Eq. 2, see Appendix C.). 
However, we should take in mind that in this comparison both the cosmic variance, and also the different selection of the sample 
(elliptical vs. passive) have a strong effect on the number density estimates.
All these checks highlight the huge impact of the criteria and quantities used in the analysis on the comparison of the results.
A fair comparison is really hard to address, and this stresses the impelling necessity of
analysis on homogeneous samples over the cosmic time to draw firm conclusions.

\section{The evolution of structural and dynamical parameters of ultramassive dense ETGs}

The mild evolution in the number density shows that in local universe, the ultramassive dense ETGs are at most  
$\sim$25$\%$ less numerous than at $<z>$ = 1.4. 
Thus, if a significant fraction of ultramassive dense ETGs evolves in size, new ones have to appear 
at lower redshift to maintain their number density almost constant. However, the observed mild number density 
evolution opens also to the possibility that most ($\sim$ 75$\%$) of the ultramassive dense 
ETGs passively evolve from $<z>$ = 1.4 to z = 0. In this case, local and high-z 
ultramassive dense ETGs should share also the same structural and dynamical properties. 
In the following we compare the structural (R$_{e}$ and M$_{\star}$) 
and dynamical properties ($\sigma_{e}$) of the ultramassive dense ETGs over the last 9 Gyr.
We highlight that in this second part of analysis, we investigate the same class of objects studied in Sec. 2, 
so ETGs visually classified with M$_{\star} >$ 10$^{11}$M$_{\odot}$ and $\Sigma >$ 2500 M$_{\odot}$pc$^{-2}$. 
However, the request of available velocity dispersion, does not allow us to use the same samples presented in Sec. 2.
In Sec. 3.1 we present the samples used for this part of analysis.

\subsection{The comparison samples}

In order to compare the structural and dynamical properties of ultramassive dense ETGs over the last 9 Gyr, 
 we still refer, as local reference, to the sample of 124 ETGs presented in Section 2.3.
For what concerns the stellar mass and the effective radius used, we resend to Sec. 2.3.
For the velocity dispersions we have referred to DR7 $\sigma$ 
since in DR7 there have been improvements to the algorithms which photometrically calibrate the spectra, 
and all spectra have been re-reduced (see \citet{thomas13} for more tests on the reliability of DR7 $\sigma$ estimates).

For the intermediate and high-z sample, we cannot refer exactly to the same samples used to estimate
the number density in Sec. 2 , since not all the ETGs have available velocity dispersions. 
In order to probe the intermediate redshift range (0.2 $ \lesssim z \lesssim$ 0.9), we refer to the sample of field and cluster
ETGs published by \citet{saglia10}. 
The authors selected the
galaxies of their sample on the basis of their spectra (requesting the total  absence  or  the  presence  of
only weak [OII] lines), and for those with HST images they provided also the morphological classification (T type). 
We have selected from their sample, the galaxies with elliptical morphology (T type $\leqslant$ -4), and
with M$_{\star}>$10$^{11}$M$_{\odot}$ and $\Sigma >$2500$_{\odot}$pc$^{-2}$.
The stellar mass provided by the authors are derived adopting the Salpeter IMF. We have converted these estimates 
to Chabrier IMF scaling the masses by 0.23 dex. 
For the majority of the galaxies, the structural parameters were derived on the F814W-band images, 
while for a small fraction, I-band VLT images were used. 
We have integrated the Saglia et al.'s sample, with that by \citet{zahid15}, in which passive 
galaxies are selected on COSMOS field using a NUVrJ color-color cut. 
Using the coordinates of the galaxies available at the author's  
website\footnote{https://www.cfa.harvard.edu/$\sim$hzahid/Data$\_$files/ApJ97868$\_$table1.ascii},
we have visually checked their morphology on HST/ACS images, founding that all but one passive galaxies are ellipticals. 

For the high-z sample, we have collected all the high-z ETGs
at 1.2 $< z <$ 1.6  with $\Sigma>$ 2500M$_{\odot}$pc$^{-2}$, M$_{\star}$ $>$ 10$^{11}$ M$_{\odot}$,
available kinematics
(i.e. velocity dispersion) and effective radius derived in the F160W band (11 objects). 
The morphological classification has been performed on the basis of a visual inspection of the galaxies carried out
on the HST images in the F160W filter.
Among the 11 galaxies of the total high-z sample 7 galaxies have been taken from the literature (see below) 
and for the remaining four of them (plus one at z=1.9, hereafter ``our sample of high-z ETGs'') we present 
the unpublished VLT-FORS2 spectra together with their velocity dispersion measurements (Section 3.2).  
For what concerns the 7 galaxies from literature:
\begin{itemize}
\item $Four$ ETGs are taken from \citet{bezanson13}. From the eight galaxies of their sample we have removed 
the galaxy U53937 being less massive than 10$^{11}$M$_{\odot}$,
U55531 since its image shows that is not an elliptical, and C13412 and C22260 having $\Sigma <$ 2500M$_{\odot}$pc$^{-2}$.
\item $Two$ ETGs are taken from \citet{belli14}. From the 56 quiescent galaxies of their sample, 
we have excluded 23 passive galaxies having
z$<$1.2; of the remaining 33, 24 have been rejected having M$_{\star} <$ 10$^{11}$M$_{\odot}$, three for not being ETGs 
(4906, 34609 and 34265) and four (1244914, 2823, 34879, 7310) to have $\Sigma <$ 2500M$_{\odot}$pc$^{-2}$.
\item $One$ ETG is taken from \citet{longhetti14}. This galaxy belongs to MUNICS sample of 6 ETGs described in Sec. 2.1 and used 
to estimate the number density.
\end{itemize}
The sample is summarized in Table \ref{table2}.
We do not include the 3 ETGs out of 17 galaxies 
presented in \citet{newman10} (ID GN3, GN4$^{b}$, GN5$^{b}$) with available $\sigma$, redshift z$>$1.26 
and M$_{\star} >$ 10$^{11}$M$_{\odot}$, 
since their effective radii were derived fitting a de Vaucouleur and not a Sersic
profile.

\begin{table*}\footnotesize
\caption{Sample of high-z ETGs collected from literature. $Column 1$: ID, $Column 2$: RA, $Column 3$: Dec, $Column 4$; redshift 
$Column 5$: circularized effective radius, 
$Column 6$: Sersic index, $Column 7$: axis ratio, $Column 8$: filter adopted to derive the surface brightness parameters, 
 $Column 9$: measured velocity dispersion, $Column 10$: velocity dispersion within R$_{e}$, 
$Column 11$: Stellar masses. 
References for the velocity dispersion values are 
listed near the ID value. When a data of the galaxy comes from a paper different from the one where the $\sigma$ 
is published, the new reference is specified near its value.}

\begin{tabular}{ccccccccccc}
\hline
\hline
Object              & RA     &    Dec    & z     & R$_{e}$             & n       &  b/a & Camera-Filter  & $\sigma$           & $\sigma_{e}$         & logM$_{\star}$  \\
                    & h:m:s  &    d:m:s  &       & (kpc)               &         &      &                &  (km/s)            & (km/s)               & (M$_{\odot}$)  \\
\hline
\hline
A17300$^{1}$        & 14:18:39.78   & 52:41:59.51 & 1.423   & 2.9$\pm$0.51$^{2}$  & 5.3       & -- & HST/WFC3-F160W & 265$\pm$7          &  272$\pm$7           & 11.26          \\
A21129$^{1}$        & 14:19:10.56   & 52:46:26.79 & 1.583   & 1.5$\pm$0.27$^{2}$  & 5.0       & -- & HST/WFC3-F160W & 260$\pm$9          &  278$\pm$10          & 11.18           \\
C21434$^{1}$        & 10:00:36.10   & 2:32:13.77  & 1.522   & 1.9$\pm$0.33$^{2}$  & 3.1       & -- & HST/WFC3-F160W & 218$\pm$16         &  230$\pm$17          & 11.20          \\
C20866$^{1}$        & 9:59:18.64    & 2:31:39.03  & 1.522   & 2.4$\pm$0.42$^{2}$  & 3.0       & -- & HST/WFC3-F160W & 272$\pm$23         &  282$\pm$23          & 11.30         \\
21750$^{3}$         & 215.03490     & 52.9829     & 1.242   & 2.59$\pm$0.26       & 5.2       & 0.57 & HST/WFC3-F160W &         --           &  259$\pm$16          & 11.03          \\
7310$^{3}$          & 150.05791     & 2.2904      & 1.255   & 4.34$\pm$0.43       & 3.8       & 0.87 & HST/WFC3-F160W &         --           &  167$\pm$15          & 11.13          \\
S2F1-142$^{4}$      & 03:06:36.51   & 00:03:01.0  & 1.386   & 3.04$\pm$0.12       & 3.5 & 0.74 & HST/NIC2-F160W & 340$^{120}_{-60}$   & 347$^{120}_{-60}$          & 11.54           \\ 
\hline
\hline  
 \end{tabular}
 \label{table2}\\
$^{1}$ \citet{bezanson13},$^{2}$ \citet{vandesande13},$^{3}$ \citet{belli14},$^{4}$ \citet{longhetti14}\\
\end{table*}

\subsection{Our sample of high-z ETGs: spectroscopic observations and velocity dispersion measurements}

\begin{figure*}
  \begin{center}
   \begin{tabular}{cc}
	\includegraphics[angle=0,width=2.7cm]{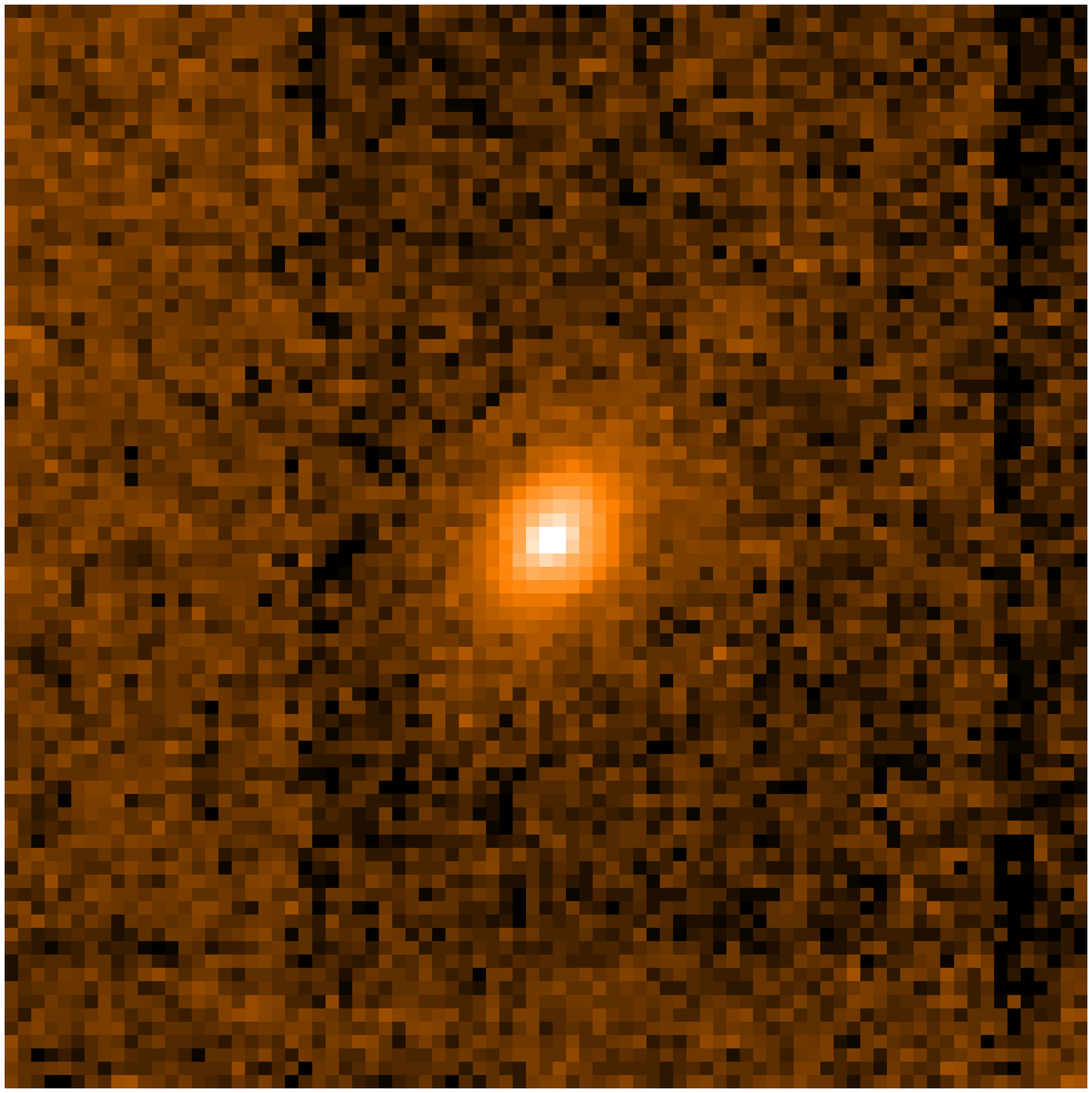}  & \includegraphics[angle=0,width=9.0cm]{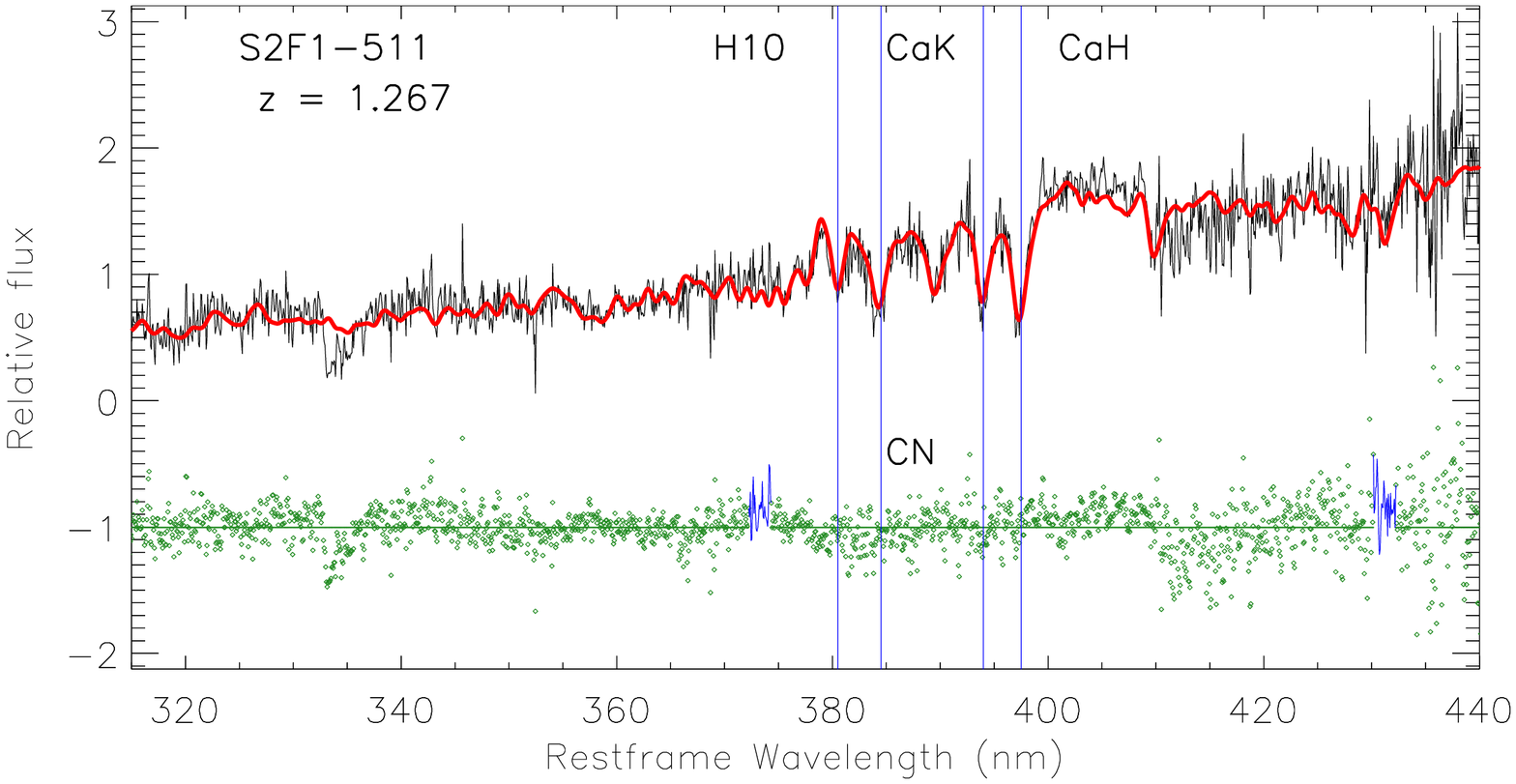} \\
	\includegraphics[angle=0,width=2.7cm]{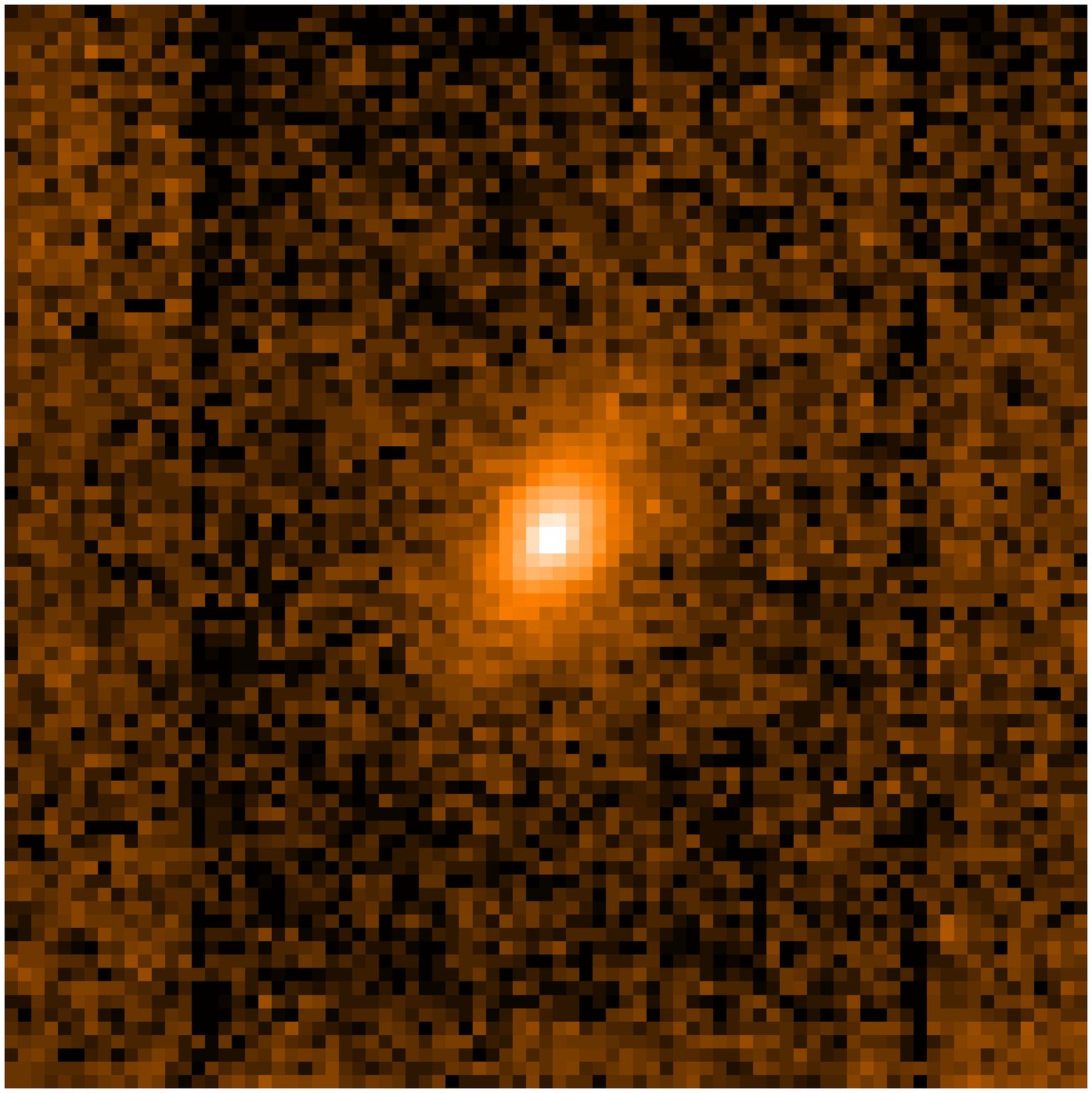}  & \includegraphics[angle=0,width=9.0cm]{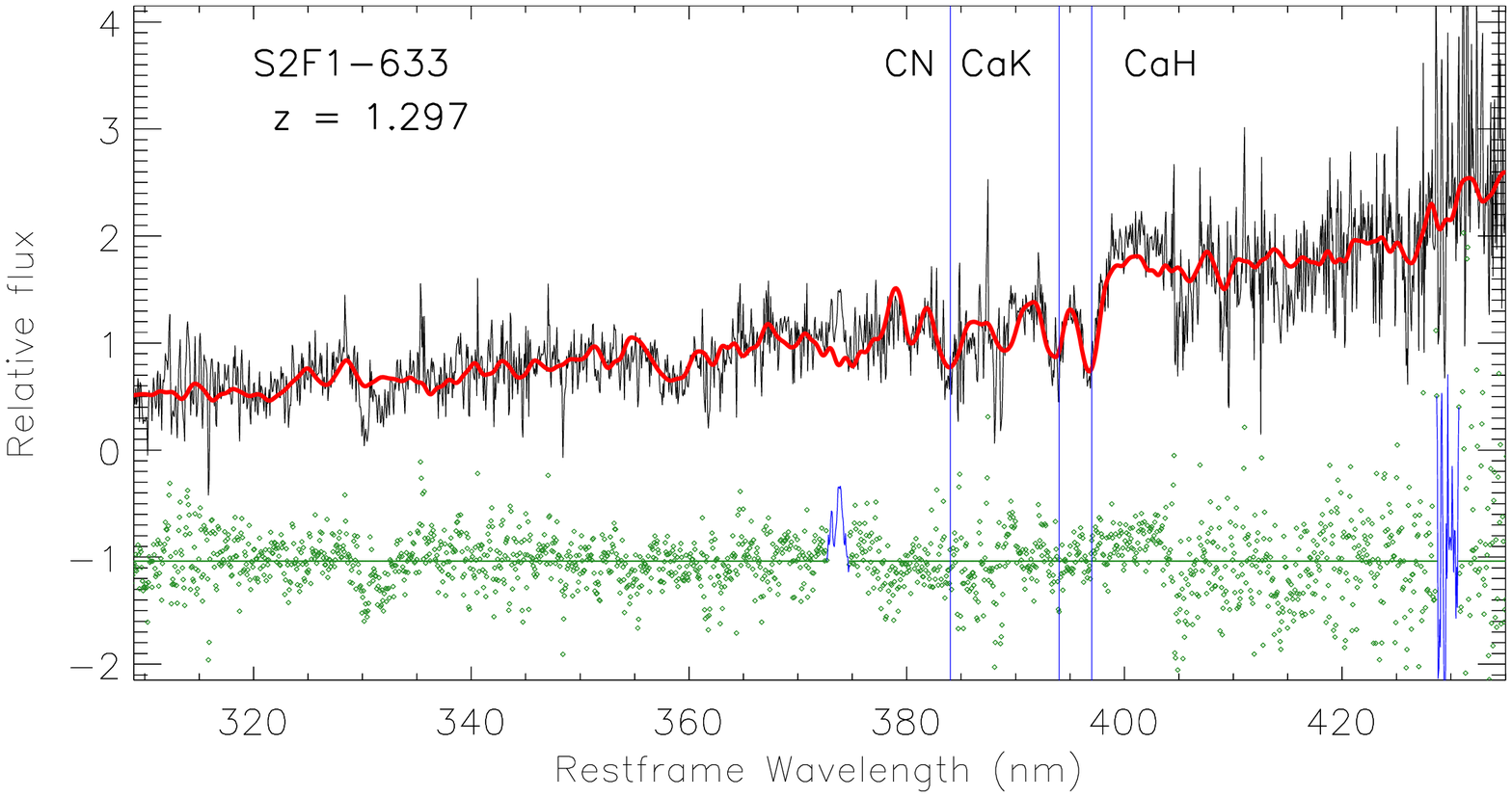} \\
	\includegraphics[angle=0,width=2.7cm]{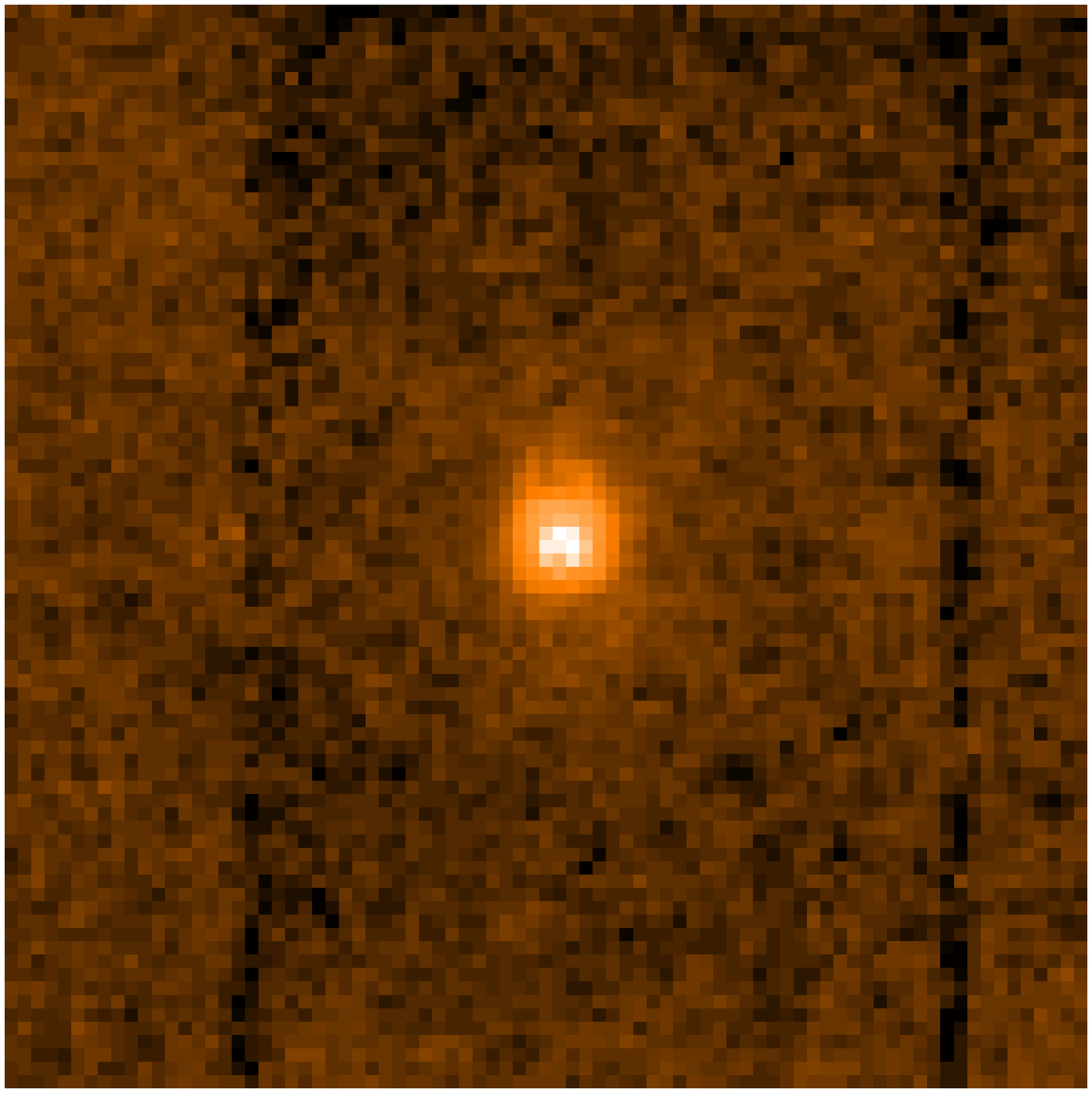}  & \includegraphics[angle=0,width=9.0cm]{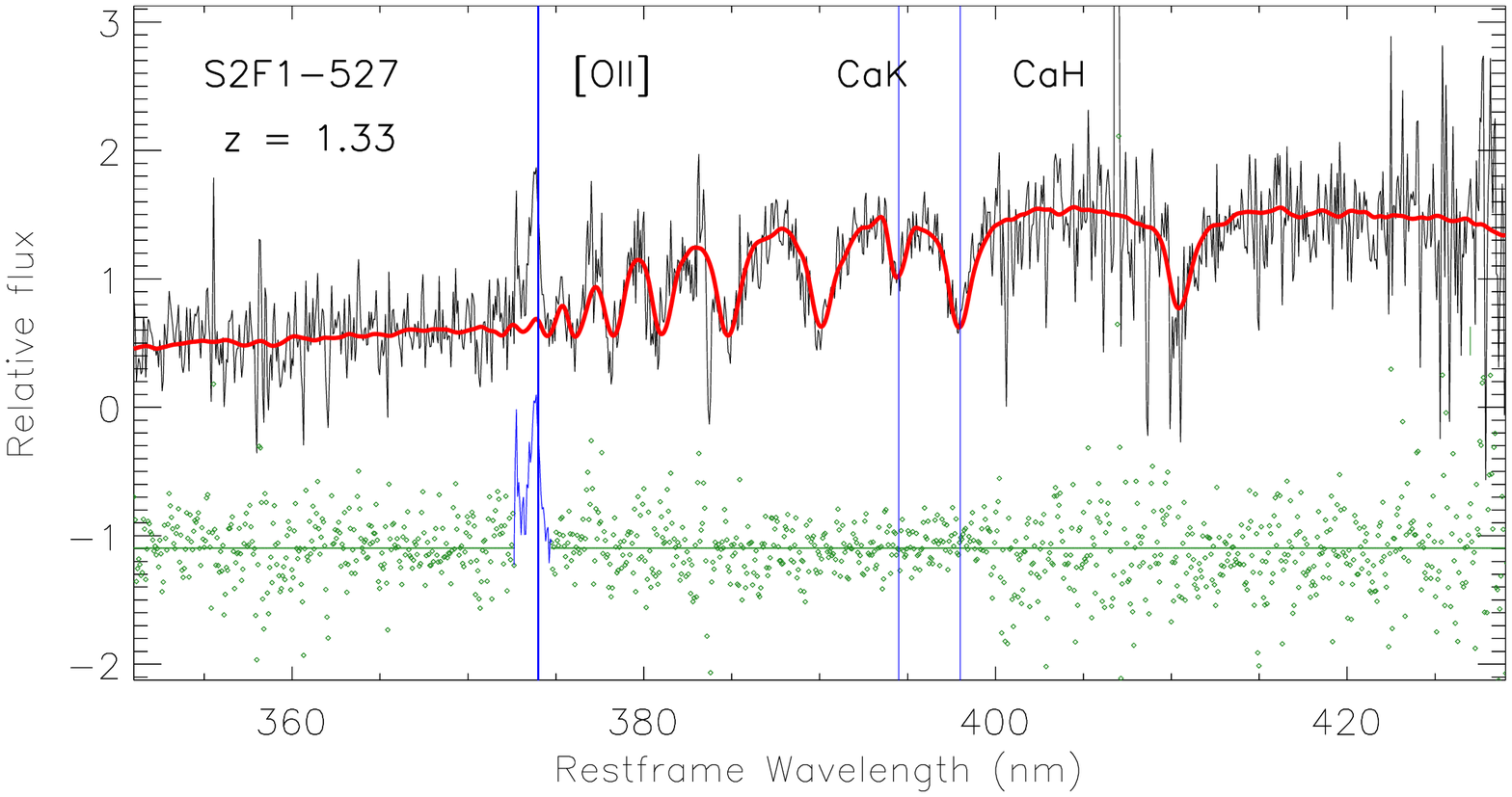} \\ 
        \includegraphics[angle=0,width=2.7cm]{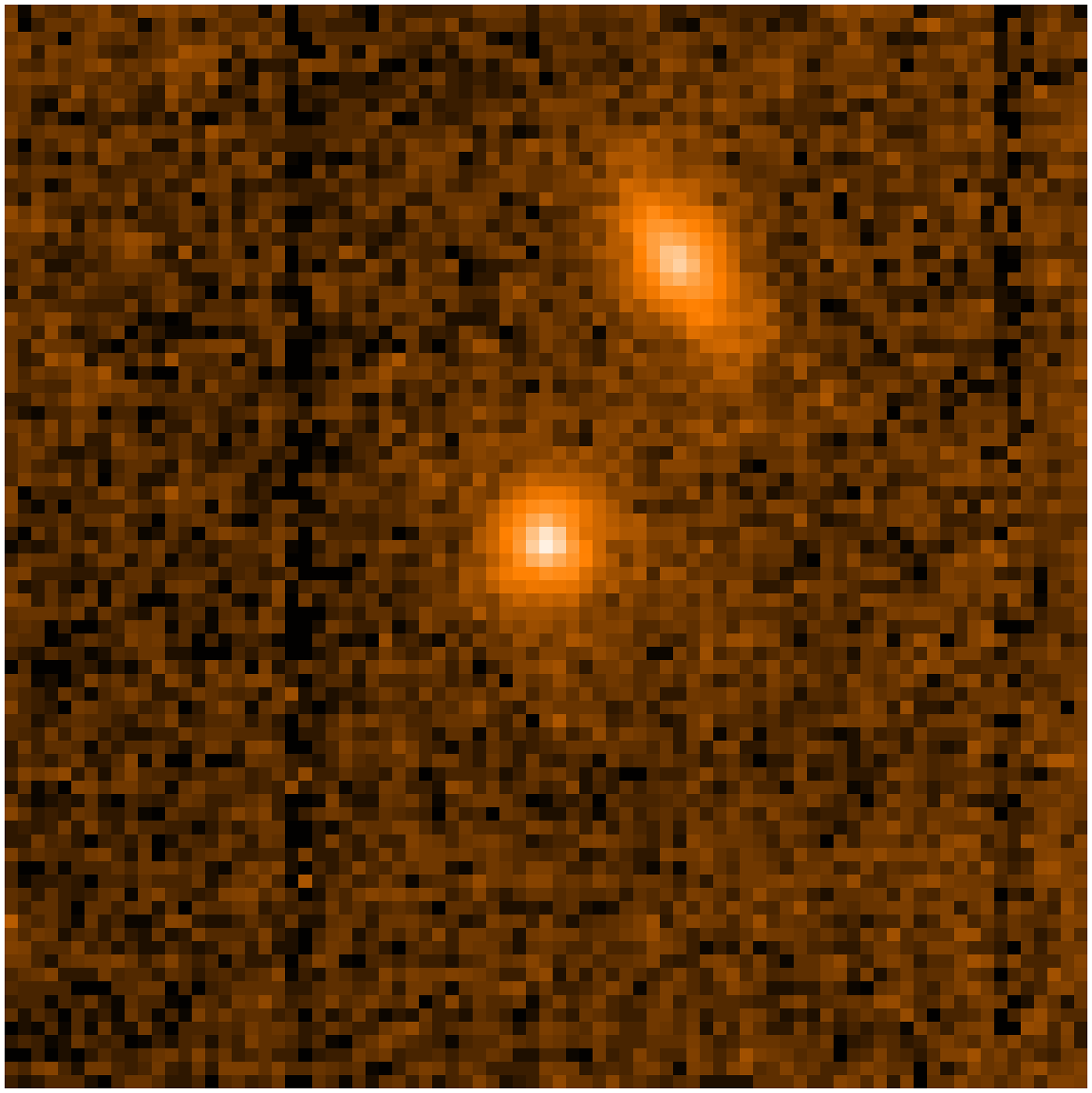}  & \includegraphics[angle=0,width=9.0cm]{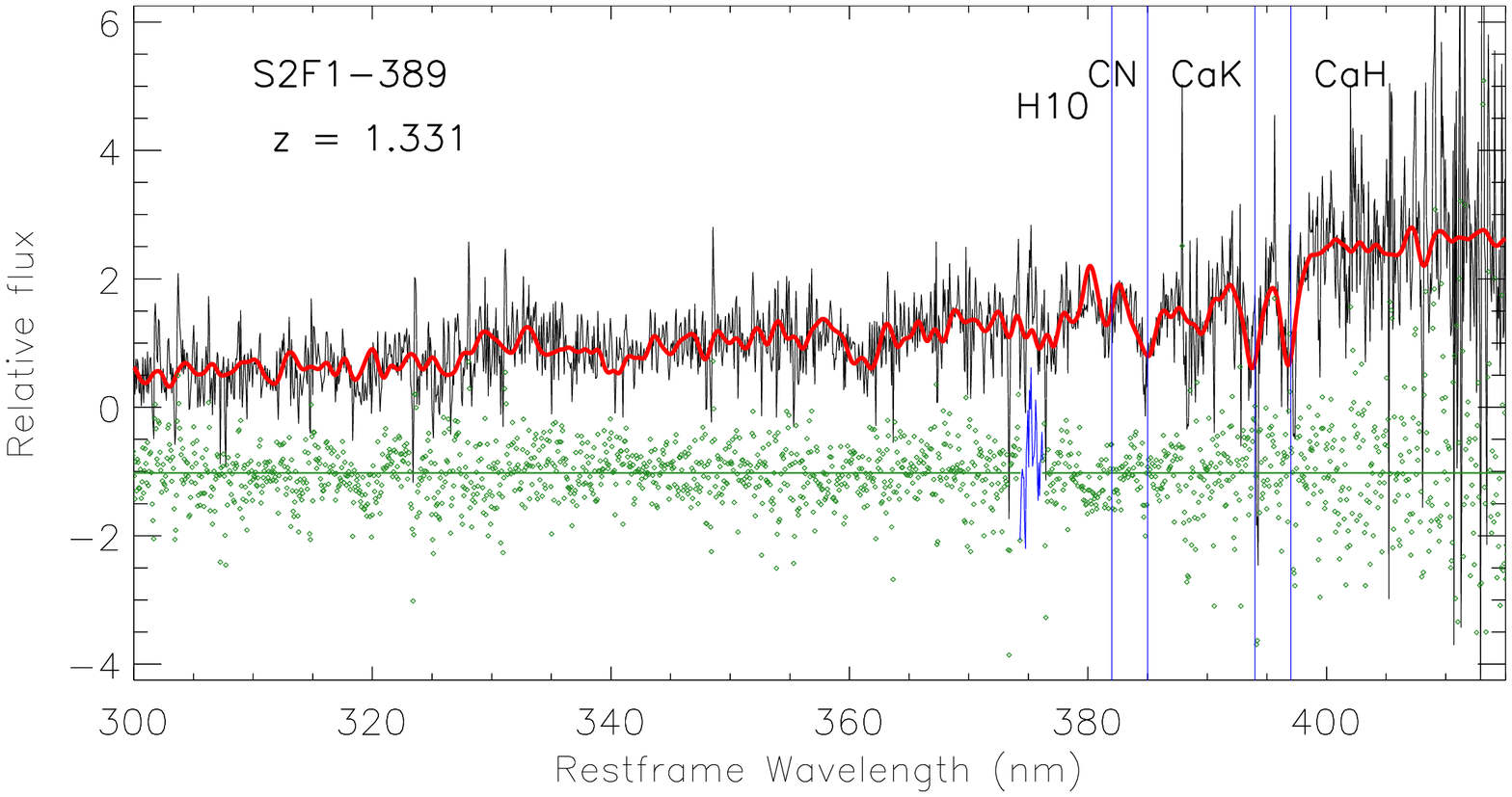} \\
        \includegraphics[angle=0,width=2.7cm]{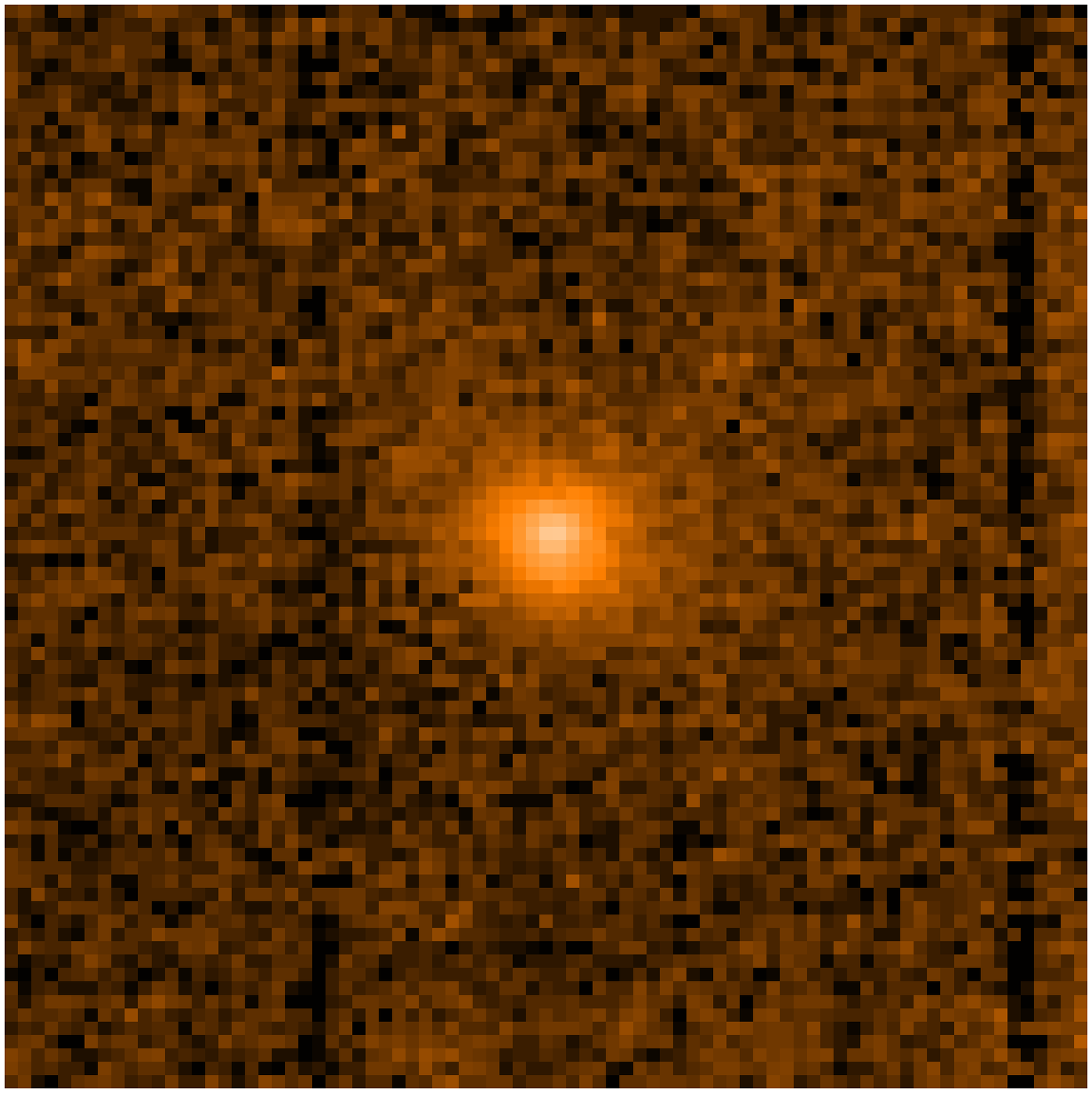}  & \includegraphics[angle=0,width=9.0cm]{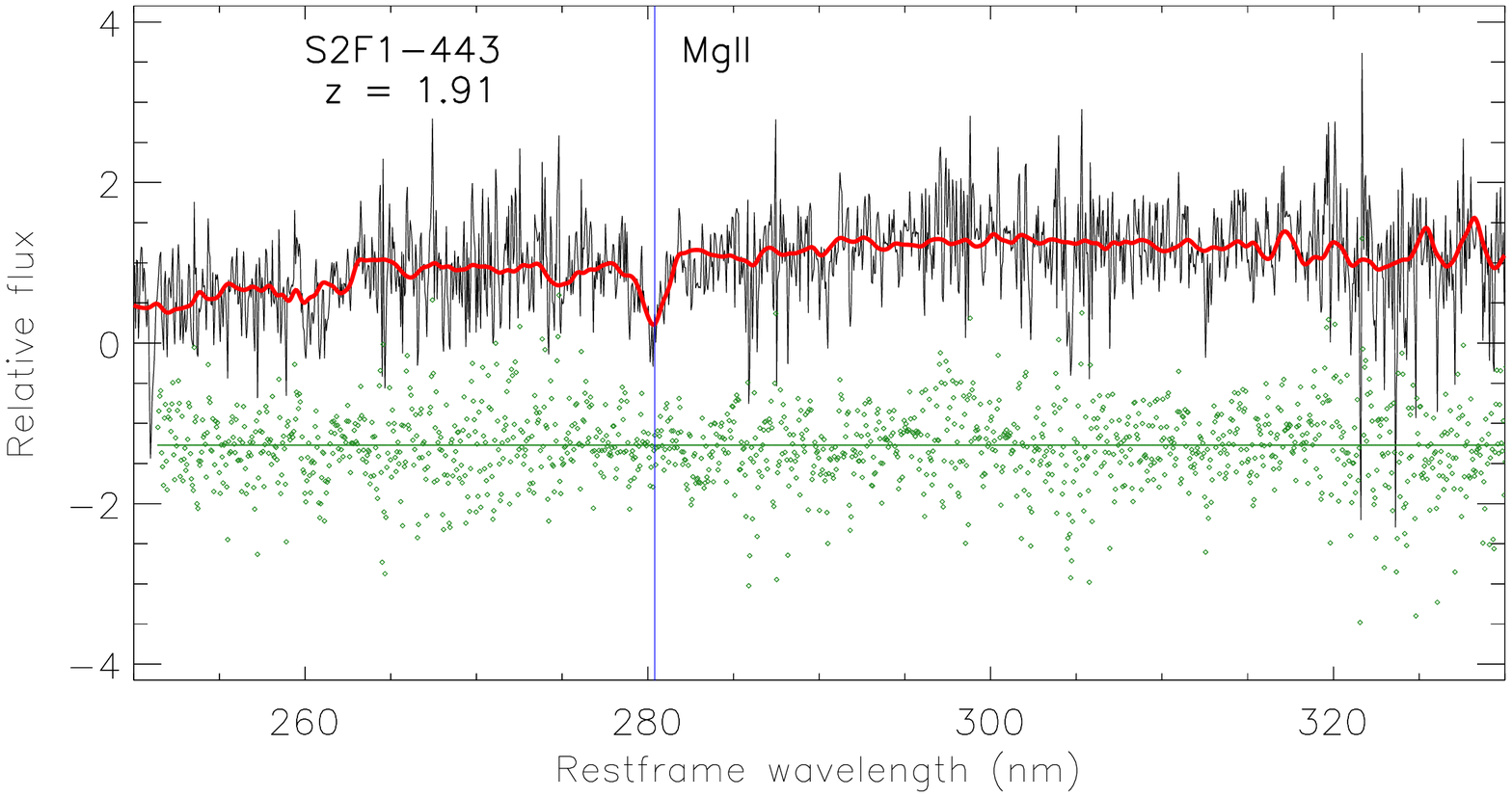} \\
\end{tabular}
\caption{\textit{Left column}: The HST-NIC2/F160W images of the 5 ETGs of our high-z sample 
for which we present here the VLT-FORS2 spectra (each image is 6$\times$6 arcsec). \textit{Right column}
Unsmoothed and not-rebinned spectra of our 5 ETGs (black lines, 8 hours of effective integration time). 
Many optical absorption features (blue lines) are clearly distinguishable from the continuum. 
The best-fit pPXF model is plotted in red. On average, pPXF selects and combines 2-4 star templates to reproduce 
the observed spectrum. Residuals after best-fit model subtraction are shown with green points. 
In the residual tracks, blue lines are the regions automatically excluded by the software pPXF 
as strongly affected by sky emission lines.}
	\label{bestfitspectra}
\end{center}
\end{figure*}

\begin{figure}[h!]
	\includegraphics[angle=0,width=9.5cm]{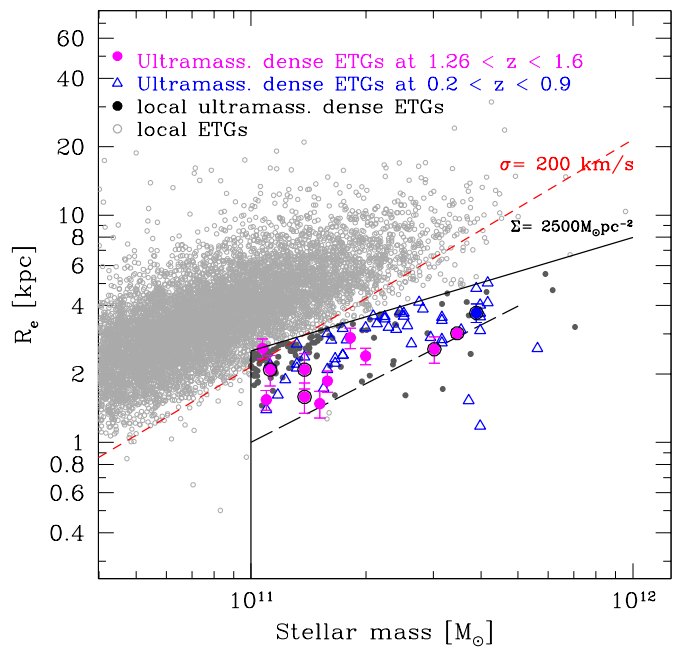} \\
	\caption{The distribution of the ultramassive dense ETGs at 1.2 $< z <$ 1.6 (magenta filled circles),
at 0.3 $< z <$ 0.9 (blue opened squares), and in local universe (dark grey points)
in the R$_{e}$-M$_{\star}$ plane. Solid black lines 
define the cuts in M$_{\star}$  and $\Sigma$ of our samples.
The ETGs from MUNICS sample with available velocity dispersion  (5 presented here and one in \citet{longhetti14})
are black countered. The one at redshift 1.91 (hence not included in the analysis of Sec. 3.3) is indicated with a blue point.
Small open circles are local 
ETGs from \citet{thomas10} in the redshift range 0.063 $ < z <$ 0.1. 
Red line indicates a line of constant $\sigma$
derived as R$_{e}$ = $\sqrt{GM_{\star}/5\sigma}$, i.e. assuming zero dark matter. Long-dashed line indicates 
the lower boundary of the size-mass relation as found by \citet{belli14} and \citet{vanderwel14}.}
	\label{sizemass}
\end{figure}

The remaining four (plus one at z =1.91) ETGs of the total high-z sample 
(Fig. \ref{bestfitspectra}, HST-NIC2/F160W images from Longhetti et al. \citeyear{longhetti07}) 
have been selected among the 7 ultramassive dense ETGs
spectroscopically identified at z$\sim$ 1.5 in the field S2F1 of the 
MUNICS survey \citep{drory01}, thus
belong to the MUNICS sample we have used
to constrain the number density of ultramassive dense ETGs at z $\sim$ 1.4 (see Sec. 2.1). 
Their surface brightness profiles 
are derived by \citet{longhetti07} fitting the HST-NIC2/F160W images with a 2D-psf convoluted 
Sersic law with the software GALFIT \citep{peng02}. 
In Table \ref{table1} we report the effective
radii of the best-fitted Sersic profile for the 5 ETGs. The circularized R$_{e}$ [kpc] are slightly different from those published in 
\citet{longhetti07} due to the more accurate estimates of the redshift we have derived from the
higher resolution VLT-FORS2 spectra.
Stellar masses have been 
re-estimated 
to take into account the updated value of redshift  and the now available
near-IR photometry in the WISE RSR-W1 filter ($\lambda_{eff}$
$\simeq$ 3.4 $\mu$m). 
We have considered a grid of composite stellar
population templates derived from \citet{bruzual03} with solar metallicity Z$_{\odot}$, exponentially declining
star formation history ($\propto$ exp(-t/$\tau$)) with star formation
time scale $\tau$ in the range 0.1-0.6 Gyr and the Chabrier
IMF \citep{chabrier03}. The effect of dust
extinction has been modelled following the Calzetti law \citep{calzetti00}
with A$_{V}$ in the range [0.0-2.0].  The stellar masses M$_{\star}$ associated to 
the best-fit model are reported in Table \ref{table1}. For the estimates of the number density derived in Sec. 2
we have referred to these updated values of R$_{e}$ and M$_{\star}$.

Spectroscopic observations of our 5 ETGs were performed with the ESO
VLT-FORS2 spectrograph in MXU mode during 4 runs in 
2010 
and 2011.
The OG590 filter and the GRIS-600z
grism were adopted to cover the wavelength range 0.6$\mu$m $< \lambda
<$ 1.0$ \mu$m with a sampling of 1.6\,\AA\, per pixel. With a 1`` slit the
resulting spectral resolution is R $\simeq$ 1400 corresponding to a
FWHM $\simeq$ 6.5\AA\, at 9000\,\AA. The total effective integration time on
target galaxies is $\sim$ 8 hours which results in signal-to-noise
(S/N) per pixel $\sim$[8-10] depending on the galaxy. 
Standard reduction
has been performed with IRAF tasks. 
In each frame the sky has been removed by subtracting from each frame the following one
in the dithering observing sequence.
The sky-subtracted frames have been aligned and co-added.
The final stacked spectra have been corrected for the response sensitivity function derived from 
the spectrum of the standard stars (FFEIGE110, GD71) observed in each of the 4 runs.
Finally, we have applied a further correction to the relative flux calibration
related to the x-position of the slit on mask
\citep[][]{lonoce14}.
In Fig. \ref{bestfitspectra} black lines show the 8-hours one dimensional
spectra of our galaxies, not rebinned and unsmoothed.

For the velocity dispersion ($\sigma$) measurements we have adopted the
Penalized Pixel-Fitting software \citep[pPXF]{cappellari04} which derive the $\sigma$ by
fitting the continuum and the absorption features of the spectrum. 
We have derived the velocity dispersions from the
unbinned spectra, weighting each pixel for a quantity proportional to
its S/N, and adopting as templates a set of 33 synthetic stars
selected from the high-resolution (R $\simeq$ 20000) synthetic
spectral library by \citet{munari05}. The stars of the subsample have
temperature 3500K $<$ T $<$ 10000K, 0 $<$ log $g$ $<$ 5, solar abundance
and metallicity and are the same adopted by
\citet{cappellari09} to derive the velocity dispersions of two ETGs at
z$\sim$1.4. 
A detailed description of the velocity dispersion measurements are provided in appendix D, together with 
extensive set of simulations aimed at testing the stability and accuracy of the measures, as well as 
their errors.
In Table \ref{table1} we report the value of the velocity dispersion ($\sigma$) corrected for instrumental 
dispersion ($\sim$ 110 km/s) and their error (see Appendix B). Since data refer to 1'' slit,
the measured $\sigma$ have to be considered within a radius of 0.5
arcsec. In Table 1 we report also $\sigma_{e}$, i.e. the velocity dispersion within the effective radius. We have
scaled the measured $\sigma$ to $\sigma_{e}$ following \citet{cappellari06}.

\begin{table*}\footnotesize
\begin{center}
\caption{Our sample of high-z ETGs. $Column$ $1$: id number, $Column$ $2$: RA, $Column$ $3$: DEC, $Column$ $4$: redshift, 
$Column$ $5$: circularized effective radius in kpc, 
$Column$ $6$: Sersic index, $Column$ $7$: axis ratio, $Column$ $8$: velocity dispersion as measured  within the 1'' 
aperture slit, $Column$ $9$: the range in which varies the offset factor $\sigma_{off}$ to be applied 
to $\sigma$ according to the simulation of Appendix B,
$Column$ $10$: velocity dispersion as measured  within the effective radius, 
$Column$ $11$: stellar mass derived with BC03 code and Chabrier IMF, $Column$ $12$: Vega magnitude in K band \citep{saracco09}}
\begin{tabular}{cccccccccccc}
\hline
\hline

Object & RA & Dec & z &  R$_{e}$ & n &  b/a & $\sigma$  & $\sigma_{off}$ & $\sigma_{e}$ & log(M$_{\star,cha}$/M$_{\odot}$) & K\\
       &    &     &   & (kpc)   &   &      &  (km/s)   & (km/s)         &  (km/s)      &         & \\

\hline
\hline
S2F1-511 & 03:06:34.04 & +00:02:30.9 & 1.267 &  2.09$\pm$0.07 & 3.3 &  0.60 & 269 & [-4 $\div$ 12]   & 281$^{+26}_{-40}$ & 11.05 & 18.1$\pm$0.1\\ 
S2F1-633 & 03:06:35.10 & +00:04:43.6 & 1.297 &  2.59$\pm$0.11 & 4.1 &  0.53 & 434 & [-24 $\div$ -7]  & 447$^{+14}_{-19}$ & 11.48 & 18.2$\pm$0.1\\ 
S2F1-527 & 03:06:43.34 & +00:02:44.8 & 1.331 &  1.57$\pm$0.24 & 3.1 &  0.88 & 226 & [-16 $\div$ 1]   & 240$^{+25}_{-28}$ & 11.14 & 18.3$\pm$0.1\\ 
S2F1-389 & 03:06:28.03 & +00:00:31.6 & 1.406 &  2.10$\pm$0.18 & 4.5 &  0.85 & 224 & [-30 $\div$ -18] & 234$^{+41}_{-41}$ & 11.14 & 18.2$\pm$0.1\\ 
S2F1-443 & 03:06:31.76 & +00:01:13.4 & 1.910 &  3.36$\pm$0.15 & 1.9 &  0.79 & 352 & [-21 $\div$ 8]   & 357$^{+40}_{-64}$ & 11.59 & 18.4$\pm$0.1\\ 
\hline
 \label{table1}
 \end{tabular}
\end{center}
\end{table*}

\subsection{The evolution of structural and dynamical parameters of ultramassive dense ETGs: results}

In the galaxy selection we have 
paid particular attention to the filter in which structural parameters were derived.
The effective radii were derived in the F160W band for all the galaxies at 1.2 $< z <$ 1.6,
in the F814W band for the sample at intermediate redshift range, and in the $r$ band for the local 
SDSS galaxies, in order to sample 
approximately the same restframe band through all the redshift range we probe. 

Figure \ref{sizemass} shows the distribution of ultramassive dense ETGs at 0 $ \lesssim z <$ 1.6 
in the plane defined by effective radius and stellar mass.
The six spheroids belonging to MUNICS sample (5 presented in Sec. 3.2 plus one by \citet{longhetti14}) 
are highlighted with black contours and we have highlighted 
with a blue point the ultramassive ETG of our high-z sample at z = 1.91. 
This galaxy is not included in the comparison with local ETGs.

Given the request of available velocity 
dispersion, the sample we have collected is clearly not complete and does not homogeneously cover 
the region of the size-mass plane occupied by dense galaxies.
However, we have checked that the request of available velocity 
dispersion does not produce a bias against the most compact objects, 
i.e. that we are not missing the smallest ETGs at fixed stellar mass. In the Figure 5 the long-dashed black line
is the lower boundary of the size-mass relation found by \citet{belli14} and \citet{vanderwel14} (they are good  
agreement with each other). We notice that although not complete, our sample is not biased against dense ETGs, 
as expected since velocity dispersion measurements are more successful in dense systems.
\begin{figure*}
\begin{center}
	\includegraphics[angle=0,width=18cm]{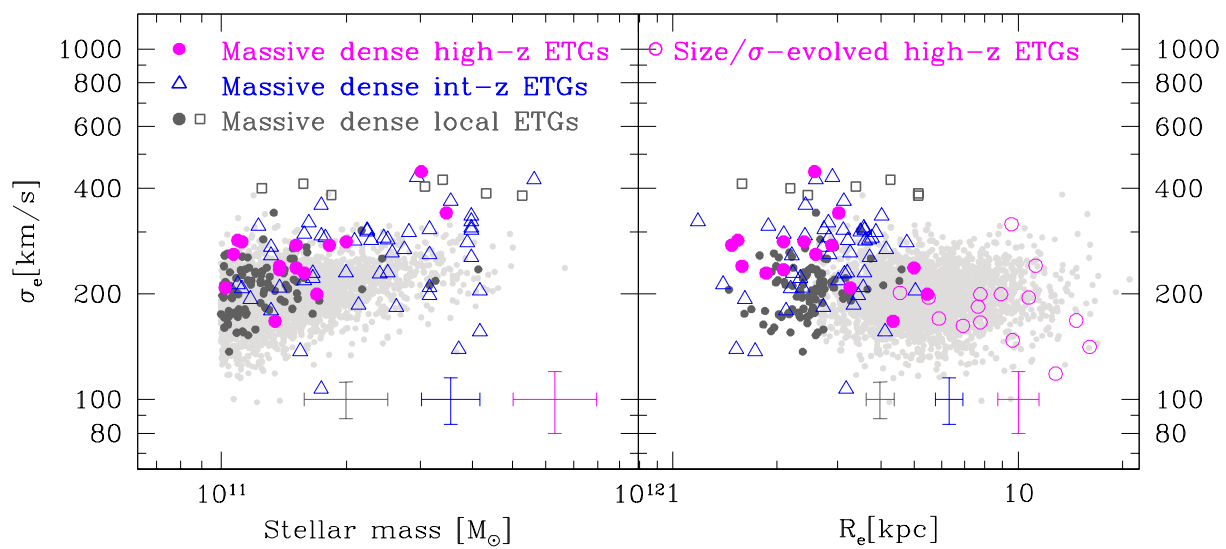} \\
	\caption{Comparison between the structural and dynamical properties of ultramassive dense ETGs over the last $\sim$ 10 Gyr.
 Magenta points are dense ETGs at 1.2 $ \lesssim z <$ 1.6, open blue squares are dense ETGs at 0.3 $ \lesssim z \lesssim$ 0.9, 
 and local ETGs are dark grey small dots. Light grey points are the whole sample of local ETGs with M$_{\star}>$ 10$^{11}$M$_{\odot}$, 
 thus with no cut in mean stellar mass density.
 Grey open squares are the local ultramassive dense galaxies selected from the sample of local galaxies
with velocity dispersion $>$350km/s by \citet{bernardi08}, while open magenta circles are high-z ETGs evolved in size a factor $\sim$ 3 as
 predicted by the relation $<R_{e}> \propto$ (1+z)$^{-1.5}$and in $\sigma$
 according to the merger scenario by \citep{hopkins09}. In each panel, average error bars are reported. 
They are colour coded according to the sample they refer. For what concerns the high-z sample, we have found that on average 
$\delta_{R_{e}}$/R$_{e}$ $\sim$ 11$\%$, $\delta_{\sigma_{e}}$/$\sigma_{e}$ $\sim$ 10$\%$ (see Table 3 and 4), while we have assumed a 
typical uncertainty of 20$\%$ on M$_{\star}$. At intermediate redshift, $\delta_{R_{e}}$/R$_{e}$ $\sim$ 10$\%$, 
$\delta_{\sigma_e}$/$\sigma_e$ $\sim$ 7$\%$, and $\delta_{M_{\star}}$ = 0.10 dex \citep[see][]{saglia10}. In local Universe, 
$\delta_{M_{\star}}$ = 0.07 dex \citep[see][]{salim07}, $\delta_{\sigma_e}$/$\sigma_e$ $\sim$ 10$\%$, 
and we assume $\delta_{R_{e}}$/R$_{e}$ $\sim$ 15$\%$. For the error on local R$_{e}$ we refer to \citet{labarbera10} which
derive the structural parameters of SDSS galaxies on SDSS images using 2DPHOT. In fact, 
the errors on parameters are expected to be mainly driven by the characteristics of the images, and not by the software and/or 
procedure used.}
	\label{comparison}
\end{center}
\end{figure*}

In the 
left and right panels of Fig. \ref{comparison} we plot the velocity dispersion of ultramassive dense ETGs 
a function of stellar mass and effective radius, 
respectively. 
The first thing that we may note is that all the ultramassive dense ETGs at high-z have
an ultramassive dense counterpart in the local Universe with comparable
structural and dynamical properties. 
Also the two most massive (M$_{\star}\gtrsim$ 3$\times$10$^{11}$ M$_{\odot}$) and densest 
($\Sigma> 5000$  M$_{\odot}$ pc$^{-2}$) high-z ETGs, whose parameters are offset with respect
to the others high-z dense ETGs, have a counterpart.
Their counterparts can be found among the local ultramassive dense galaxies belonging 
to the sample of SDSS galaxies with  $\sigma>$350km/s selected by \citet{bernardi08}  (Fig.\ref{comparison} open squares).
Hence, Figure 6 shows that in local Universe there are ultramassive dense ETGs with 
effective radii and velocity dispersions comparable
to those of equally massive and dense ETGs at $<z>$ = 1.4.
At the same time, Fig.\ref{comparison} shows that there are some local galaxies with $\sigma_{e} < $200 km/s 
without an high-z counterpart.
These ETGs populate in Fig. \ref{sizemass}  the region immediately below the line at $\Sigma$ = 2500 M$_{\odot}$pc$^{-2}$
with 10$^{11} <$ M$_{\star} <$ 2$\times$10$^{11}$
where indeed no high-z galaxies fall. 
Actually, the sparse sampling of the high-z sample and its incompleteness, 
prevent us to asses whether the lack of low-$\sigma$ ultramassive dense ETGs at 
high-z is real or not.

Just as exercise, in Fig. \ref{comparison}, with open circles, we show the displacement that
high-z dense ETGs would show in the space (R$_{e}$, M$_{\star}$, $\sigma_{e}$) in
the hypothesis that they evolve in size according to the relation $(1+z)^{-1.5}$
(i.e. a factor $\sim3$ from $z\sim1.4$ to $z=0$ as found e.g. by Cassata et al. (2013),
Cimatti et al. (2012), van der Wel et al. (2014)) and 
in $\sigma$ according to the
merging scenario proposed by Hopkins et al. (2009).
Thus, if all or most of the high-z ultramassive dense ETGs evolve in size (and $\sigma_{e}$) they have to be replaced by other ultramassive
dense ETGs along the time to restore the population seen in the local Universe.

\section{Discussion and Conclusions}

We have investigated the mass accretion history of ultramassive dense 
(M$_{\star} >$ 10$^{11}$M$_{\odot}$, $\Sigma >$ 2500 M$_{\odot}$pc$^{-2}$) ETGs (elliptical and spheroidal
galaxies, visually selected) over the last $\sim$9 Gyr.
We have addressed this topic by tracing the evolution of their comoving number density and by 
comparing the structural (effective radius R$_{e}$ and stellar mass M$_{\star}$) and dynamical 
(velocity dispersion $\sigma_{e}$) parameters of ultramassive dense ETGs from redshift $\sim$ 1.6 to redshift $\sim$ 0.

In order to constrain 
the comoving number density $\rho$ of ultramassive dense ETGs since z = 1.6, we have taken advantage of the
MUNICS and GOODS-South surveys to probe the highest redshift range (z $>$ 1),
while we have used the COSMOS spectroscopic survey
to cover the redshift range from z = 1.0 to z = 0.2.
For the number density of ultramassive
dense ETGs in the local Universe we have referred to the sample of local ETGs selected from the SDSS DR4 
by \citet{thomas10}.
We have found that the comoving number density follows the relation $\rho$ $\propto$(1+z)$^{0.3\pm0.8}$, 
i.e., it decreases at most by 25$\%$ since z=1.6. 

Our results are in good agreement with the other ones presented in literature 
\citep{carollo13, damjanov15, cassata13, valentinuzzi10, bezanson13}, with the only two exceptions concerning the estimates 
provided by \citet{taylor10} at z $\sim$ 0, and by \citet{vanderwel14} at 0.2 $< z <$ 1.5 
(in both cases our number densities are greater than that claimed by the authors). 
In the first case, we have shown that the main source of the observed discrepancy is the 
different definition of R$_{e}$ used  (semi-major axes SM$_{e}$ obtained 
fitting a de Vaucouleur profile in Taylor's paper versus circularized R$_{e}$ obtained 
fitting a Sersic profile in our analysis). In the second case, we argue that the main reason of
the drastic ($\sim$ one order of magnitude) drop observed in the number density by \citet{vanderwel14} can be ascribed
to the procedure they adopted to derive the effective radius at 5000\,\AA\, and the usage of semi-major axis SM$_{e}$.

We have then investigated  whether the ultramassive dense ETGs 
observed at high-z have a local counterpart, i.e. whether they
are similar both in their structure (R$_{e}$ and M$_{\star}$)  and kinematics ($\sigma_{e}$) to the low-z ones. 
To this aim, we have collected an homogeneous sample of ultramassive dense ETGs at 1.2 $ < z <$1.6 with 
available R$_{e}$, M$_{\star}$ 
and  $\sigma_{e}$. 
We have included in our final catalogue only the galaxies with clear elliptical morphology, 
 effective radius derived in the F160W band fitting a Sersic profile and stellar masses estimated with BC03 models and a 
Chabrier IMF. 
The request of available velocity dispersion restricts the sample to 11 ETGs.
For 4 ETGs out of 11 (plus one at z = 1.91) we have presented
new unpublished VLT-FORS2 spectra and the measure of their velocity dispersions.
For the intermediate redshift range (0.2 $< z <$ 0.9) we refer to the sample of ETGs by \citet{saglia10}, 
and to the sample of passive galaxies by \citet{zahid15}, while
as local reference we have adopted the complete sample of local (0.063 $< z <$ 0.1) ETGs selected from the
SDSS DR4 by \citet{thomas10}.
The comparison in the plane [R$_{e}$, M$_{\star}$, $\sigma_{e}$] 
shows that all the ultramassive dense ETGs at z = 1.4, have a local counterpart with similar velocity dispersion,
stellar mass and effective radius.

The two above evidences point toward two simple scenarios. In the last 9-10 Gyr, the vast majority ($\gtrsim$ 70$\%$, according 
to the number density) of ultramassive
dense ETGs passively ages, 
leaving the bulk of the stellar mass, the kinematics, and the structure almost unchanged.
In this case, the observed size growth of ultramassive ETGs has to be ascribed mostly to new born larger ETGs.
The other possibility is that a significant fraction of high-z ultramassive dense ETGs migrates, through inside-out accretion, towards 
less dense systems, sustaining the mean size growth of spheroidal galaxies. Concurrently, the emergence of
new ultramassive dense ETGs at lower redshift maintains the number density almost constant. 
We are not in the position to discriminate among the two possibilities, but some considerations deserve to be discussed.

A conventional approach to distinguish the real descendants from the newly 
formed spheroids (i.e. to discriminate among the two above scenarios) 
is to compare high redshift galaxies with old 
(on the basis of their luminosity weighted age LWA) low-z galaxies.
However, this method does not take into account the evidence that
despite the old global age, spheroidal galaxies
often host very young stars \citep[e.g.][]{labarbera12}. 
This population of young stars although has a negligible weight in term of total stellar mass \citep[e.g.][]{tantalo04},
can significantly lower the LWA. 
The direct consequence of this is that to select local  
descendants through a cut in LWA can exclude part of spheroidal galaxies already in place at z $\sim$ 1.4.
As a further complication, most of the methods used to infer galaxies age are limited by the age-metallicity degeneracy 
and by the possible presence of the dust. 

\begin{figure}
\begin{center}
	\includegraphics[angle=0,width=9cm]{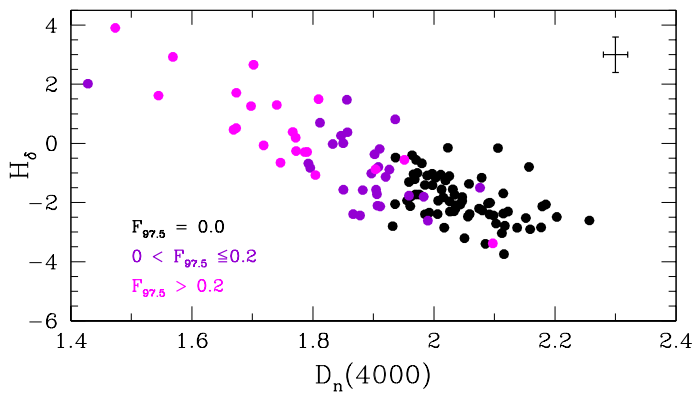} \\
	\caption{The H$_{\delta}$ of ultramassive dense local ETGs versus their D$_{n}$(4000) 
colour coded on the basis of the 97.5 percentile value of F, where F is the fraction of stellar mass formed in the last 2 Gyr.}
	\label{agealpha}
\end{center}
\end{figure}

Considering these aspects, we have decided to investigate the distribution of local ultramassive
dense ETGs in the D$_{n}$(4000) - H$_{\delta}$ plane. 
Actually, the D$_{n}$(4000) index progressively increases with the age of the stellar population, while the strong absorption 
H$_{\delta}$ line arises in galaxies that have experienced a burst of star formation in the past 1-2 Gyr. 
Although both indices are metallicity sensitive, their combination is not so and, moreover,  
is largely insensitive to the dust-attenuation effects.
In Fig. \ref{agealpha} we have reported the D$_{n}$(4000) and H$_{\delta}$ values for our sample of 124 ultramassive dense local ETGs.
For the spectral indices we have referred to the database provided by \citet{kauffmann03} for SDSS galaxies. 
The Figure shows that just $\sim$ 70$\%$ (60$\%$)of local ultramassive dense ETGs have D$_{n}$(4000) $\gtrsim$ 1.9 (1.95) 
(thus age $ \gtrsim$ 9Gyr), 
so, on the basis of its age, only two out of three local ETGs should be considered a descendant of high-z compact ETGs. 
However, for each galaxy, \citet{kauffmann03} best-fitted the observed values of D$_{n}$(4000) and H$_{\delta}$
with a grid of composite stellar population (CSP) models both with typical declining and bursty star formation history.
From the best-fit model, they derived the fraction $F$ of stellar mass formed in the last 2 Gyr, 
and provide both its median value
and the 2.5 and 97.5 percentile of its likelihood distribution.
Actually, all the galaxies in Fig. \ref{agealpha} have a median value of F = 0.
However, in Fig. \ref{agealpha} we have colour coded the galaxies on the basis of the 97.5 percentile value of F (F$_{97.5}$).
The reason of this choice is that more than the exact value of $F$, which can be susceptible 
to the input models grid and prior adopted in the fit, we are interested to investigate whether there is 
a non null probability to have a secondary (minor) event of star formation, 
that can drift the age estimate toward lower value.
Interestingly, the wast majority of the galaxies with D$_{n}$(4000) $\gtrsim$ 1.9 (and less than 
10$\%$ for galaxies with D$_{n}$(4000) $\gtrsim$ 1.95), F$_{97.5}$ = 0.
For the remaining galaxies of the sample with D$_{n}$(4000) $<$ 1.9, 
 their F$_{97.5}$ value is $>$  0, 
thus there is a non null probability that they have experienced a burst of star formation in the last 2 Gyr.
In this context, detailed studies on spectral features of passive galaxies \citep{lonoce14} have shown that
fitting the D$_{n}$(4000) with a single CSP model with a declining star formation history  
returns age values that can be significantly (even $\gtrsim$ 50$\%$) lower than the 
values obtained with a SED fitting or obtained assuming a composite model made by an old component forming the bulk of the stellar mass, and
a young stellar population contributing for just few $\%$ to the total stellar mass amount.
If this is the case, the galaxies with D$_{n}$(4000) $<$ 1.9 could be still considered descendant of high-z ETGs
that experienced secondary and minor events of star formation at later epochs \citep[see also][]{thomas10}.
These events can have a drastic impact on the estimate of galaxy age,
but do not alter significantly the stellar mass. Actually, Fig. \ref{agealpha} shows that the probability that a galaxy 
increases its stellar mass more than 20$\%$  in secondary events of star formation is extremely low.

This analysis, is not intended to be quantitative, but just to show that the treatment of the progenitor bias is really
complex and that the mere selection of local descendants on the basis of the galaxy age obtained through the photometry fit
or even through the more robust D$_{n}$(4000) can provide at most a lower limit to the true number of local descendants.
A more detailed analysis of the whole ensemble of the spectral features that are age and ``star-formation history'' sensitive 
(e.g. H$_{\delta}$; H+K, D$_{n}$(4000)) is mandatory both in local and at intermediate redshift range, to track the real mass assembly 
history of ETGs.

Although not conclusive, the analysis of spectral features of local ETGs have shown that there is a non null probability
that all the local dense ETGs are high-z descendants. Moreover, 
in the last years studies on the properties (number densities, masses, sizes, SFRs) 
of submillimeter galaxies (SMGs) have strengthen the possible evolutionary connection between these objects and dense high-z elliptical, 
with the first being the progenitors of the second ones \citep[e.g.][]{barro13, barro14, toft14}. Nonetheless, these highly 
star forming galaxies are extremely rare at z$<$ 1. If galaxies with, e.g., 
D$_{n}$(4000) $<$ 1.9 are genuinely young ultradense systems (and not old galaxies which 
have experienced recent event of star formation), we should find the way to form them. The lack of progenitors
for young dense ETGs, strengthen the hypothesis that 
the local compact massive galaxies are descendants of the high redshift ones.

Thus, more than one observational evidence would point toward the first of the two proposed scenarios
according to which most ($\gtrsim$ 75$\%$) of the ultramassive dense ETGs observed at z$\sim$ 1.4 have already completed 
their assembly. Although residual event of star-formation may occur,
these cannot significantly modify neither the mass, nor the velocity dispersion, nor the typical dimension
of the ultramassive dense galaxies already formed at z $\sim$ 1.4. 
As far as the remaining 25$\%$ of dense ETGs possibly missing in the local universe, 
stochastic merger events could have transformed them in ''non dense`` systems ($\Sigma \leq$ 2500M$_{\odot}$pc$^{-2}$),
contributing to the increase of the mean size and of the number density of the whole population
of ultramassive ETGs at z $<$ 1.5.
Numerically, their contribution to the total number density of 
local ultramassive ETGs is $\sim$ 2$\times$10$^{-5}$ gal/Mpc${3}$ (see Fig. 2).
Actually, the number density of passive M$_{\star}>$10$^{11}$M$_{\odot}$ galaxies increases 
from $\sim$ 10$^{-4}$ to 5$\times$10$^{-4}$ gal/Mpc$^{-4}$ from redshift 1.5 to z $\sim$ 0 \citep{ilbert10, pozzetti10, brammer11, 
moustakas13, muzzin13}.
Therefore, if the fraction of newly formed ultramassive dense ETGs is trascurable, 
the eventual contribution of the size-evolved ultramassive dense ETGs to the mean size growth in 
term of number density is $\lesssim$ 5$\%$.

\begin{acknowledgements}
We warmly thank the anonymous referee for his/her comments and suggestions
which, in our opinion, have really improved the whole manuscripts.
This work has received financial support from Prin-INAF
1.05.09.01.05 and is based on observations made with ESO Very Large Telescope 
under programme ID 085.A-0135A.
\end{acknowledgements}

\nocite{}
\bibliographystyle{aa}
\bibliography{paper_gargiulo_aea_final}

\begin{appendix}
 \appendix
\section{Testing the reliability of R$_{e}$ and M$_{\star}$ for local SDSS ETGs}
\subsection[]{Effective radius of local SDSS ETGs}

The effective radius for SDSS galaxies has been derived by many authors adopting different methods 
(e.g.  1D fit with the NYU-VAC pipeline by \citet{blanton05}, 2D fit with GALFIT by \citet{simard11} and \citet{guo09})
and it is not straightforward to asses which analysis provides the more accurate estimate of the parameter.
In fact, all these works have estimated the R$_{e}$ on SDSS images 
(PSF-FWHM = 1.4`` in $r$ band, pixel size = 0.396''/px, i.e. 
$\sim$ 0.56 kpc/px at z = 0.075), 
and thus the central region 
of the light profile (much sensitive to the Sersic index) of a typical ultramassive dense ETG 
is sampled by just 3-4 px in the redshift range we probe.
Based on this evidences, to test the accuracy of the R$_{e}$ we have used in this work, we have compared the estimate
of the effective radius derived by \citet{blanton05} on SDSS images, with those obtained using HST optical images 
(FWHM $\simeq$ 0.12``, pixel size 0.05``/px).
We have used the Canadian Astronomy Data 
Centre\footnote{(http://www2.cadc-ccda.hia-iha.nrc-cnrc.gc.ca/en/search/\\?collection=CFHTMEGAPIPE$\&$noexec=true$\symbol{35}$queryFormTab)} 
to find ACS-HST images in the F775W/F814W/F850LP/  filter for the 124 ultramassive dense ETGs of our sample,
but, unfortunately, none of them has HST optical counterpart.

\begin{figure}
\begin{center}
	\includegraphics[angle=0,width=9cm]{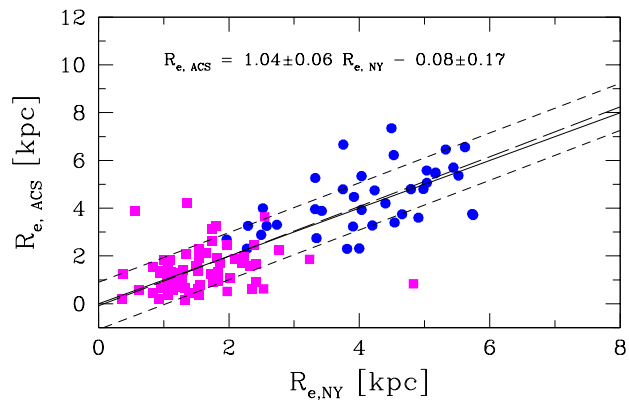} \\
	\caption{Effective radius measured on F814W/HST images for galaxies at 0.1$ < z <$0.4 (blue points) 
and for galaxies in Coma cluster (magenta squares) compared with the estimates obtained 
on SDSS images by \citet{blanton05} in r-band, for galaxies at  0.1$ < z <$0.4, and in i-band images for Coma galaxies.
Solid line is the 1:1 correlation, long-dashed line is the best fit relation obtained with a sigma clipping algorithm, 
and dashed lines indicate the 1$\sigma$ deviation from the best fit.}
	\label{hstsdss}
\end{center}
\end{figure}

Given this, we have collected from literature a sample of galaxies included in the Blanton's catalogue
with R$_{e}$ estimated also on HST images.
We firstly refer to the COSMOS catalogue described in section 2.2. 
In figure \ref{hstsdss} the R$_{e}$ of galaxies at 0.1 $< z <$ 0.4 measured on HST/ACS-F814W images ($\sim$ r-band restframe) 
by \citet{sargent07} is reported versus the R$_{e}$ (r band) taken from 
Blanton's release (blue points). The match between the two releases do not include galaxies with R$_{e}$ smaller than 2 kpc, 
hence, in order to investigate the reliability of Blanton estimates even at lower effective radii, we integrate the sample
with galaxies taken from the ACS-Coma Cluster Survey \citep{carter08}.
The effective radius of Coma galaxies were measured on the F814W images (exposure time $\sim$ 1400s) with GALFIT \citep{hoyos11}
and in Fig. \ref{hstsdss} is compared with the i-band SDSS R$_{e}$.
The range of R$_{e}$ covered by the two samples (1kpc $ \lesssim$ R$_{e}$ $\lesssim $ 6 kpc) is
very similar that covered by our sample of ultramassive dense local ETGs.
The Figure shows that in the range of radius of our ultramassive dense ETGs (1kpc $ \lesssim$ R$_{e}$ $\lesssim $ 6 kpc), the estimates
by Blanton agree with that measured on HST images (R$_{e, HTS}$ = 1.03$\pm$0.08R$_{e, NY}$ + 0.06$\pm$0.24).

\subsection[]{Stellar mass of local SDSS ETGs}

SDSS galaxies are provided with different estimates of stellar masses. 
In our analysis we have decided to adopt the stellar masses derived through the SED fitting for consistency 
with high and intermediate redshift stellar mass estimates.
In order to show that our conclusions are not affected 
by our choice of M$_{\star}$, 
in this Appendix we compare two different estimates of this quantity in order to asses the consistency
of the measures, at least in the range of values we are interested in.

\begin{figure*}
        \includegraphics[angle=0,width=18cm]{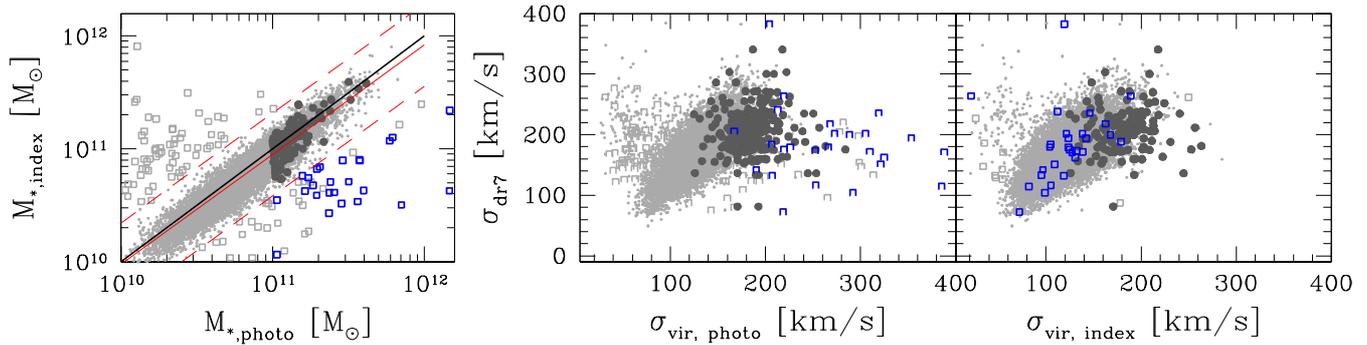} \\
        \caption{\textit{Left panel}: the stellar mass derived through 
the fit of the multiband photometry (used in this paper, M$_{\star, photo}$) versus that 
obtained fitting the spectral indices \citep[M$_{\star, index}$,][]{kauffmann03} for local ETGs at 0.05 $< z <$ 0.1 
(light grey points). Black solid line is the 1:1 correlation, while red line is the best-fit relation 
and red-dotted curves are the 3$\sigma$ lines. 
All the point at $>$ 3$\sigma$ are marked as open squares. Among these outliers, blue ones are those having 
M$_{\star} >$ 10$^{11}$M$_{\odot}$, and $\Sigma >$2500M$_{\odot}$pc$^{-2}$, 
i.e. nominally included in local sample of ultramassive dense local ETGs. Dark grey points are the remaining part of the 
ultramassive dense local ETGs. 
\textit{Central panel}: the measured velocity dispersion ($\sigma_{dr7}$)  versus the velocity dispersion derived from the 
virial theorem assuming as mass the M$_{\star, photo}$ values ($\sigma_{vir, photo}$).   
\textit{Right panel}: the same of central panel but with $\sigma_{vir}$ derived 
assuming the M$_{\star, index}$ values.}
        \label{checkstime}
\end{figure*}

In the left panel of Fig. \ref{checkstime} we plot
the stellar mass derived through 
the fit of the multiband photometry (used in this paper, M$_{\star, photo}$) versus that 
obtained fitting the spectral indices \citep[M$_{\star, index}$,][]{kauffmann03}.

The figure shows that for 
the bulk of local ETGs with M$_{\star, photo} > $10$^{11}$M$_{\odot}$ the two estimates agree well, 
but few percent of them significantly deviate from the main correlation. 
We have identified these outliers, in figure represented by squares, as those galaxies at 3 sigma from 
the best-fit relation derived with a sigma clipping algorithm.
We notice that a significant fraction of outliers below the relation has M$_{\star, photo} >$ 10$^{11}$M$_{\odot}$, 
while M$_{\star, index} <$ 10$^{11}$M$_{\odot}$. Among these ultramassive outliers, we have highlighted with open blue squares 
those with $\Sigma >$2500M$_{\odot}$pc$^{-2}$, i.e. those included in local sample of ultramassive dense local ETGs.
To discriminate for these ultramassive outliers the stellar mass estimate more realistic, in the  
central  panel we plot the the velocity dispersion derived from the virial theorem ($\sigma_{vir}$)
 \begin{equation}
  \sigma_{vir} = \sqrt{G M_{\star, photo}/\beta R_{e}},
 \label{viriale}
 \end{equation}
 where G is the gravitational constant and $\beta$ = (8.87 - 0.831$\times$n + 0.0241$\times$n$^{2}$) following 
 \citet{cappellari06}, versus the measured one ($\sigma_{dr7}$). For the velocity dispersions of local ETGs 
we have referred to the DR7 $\sigma$ 
(see \citet{thomas13} for more tests on the reliability of DR7 $\sigma$ estimates).

The majority of ultramassive outliers deviate from the general trend and in particular at fixed $\sigma_{dr7}$,
$\sigma_{vir}$ should be lower. This suggests that 
the M$_{\star, photo}$ for these galaxies are overestimated. 
As further confirmation, in the bottom panel we plot $\sigma_{dr7}$ versus $\sigma_{vir}$ but
with the last quantity derived from the M$_{\star, index}$ estimates.
We observe that in this case, the ultramassive outliers follow the main relation.

In fact, most of the ultramassive outliers despite having M$_{\star, photo}>$10$^{11}$M$_{\odot}$  
have M$_{\star}<$10$^{11}$M$_{\odot}$, and thus we will exclude them from our sample of local ultramassive dense ETGs. 

\section{Insights on the impact of R$_{e}$ definition on the samples of local dense galaxies.}

In Sec. 2.5.1, we have shown that the number of local massive dense galaxies is drastically dependent on 
the R$_{e}$ used. In particular, we have shown that in the spectroscopic SDSS-DR7 release, the number of galaxies  with 
M$_{\star}>$10$^{11}$M$_{\odot}$, log(R$_{e}$/kpc) $<$ 0.56$\times$(log(M$_{\star}$/M$_{\odot}$) - 9.84) - 0.3, and 
$^{0.1}$(u -r) > 2.5 is $\sim$ 10 if semi-major axis obtained fitting a de Vaucouleur profile (SM$_{e,dV}$) is used
and rises up to $\sim$190 if circularized de Vaucouleur one (R$_{e,dV}$) is used instead. 
Hereafter, for sake of convinience, we refer to this last sample as ''circularized sample``. This implies that circularization 
(hence the axis ratio b/a) has a huge impact on the selection of the sample.
In order to better visualize the impact of the b/a ratio on the selection of the samples, 
we have plotted in the size-mass plane the galaxies of the circularized sample with reliable stellar mass (see Sec. 2.5.1) 
using for each galaxy its semi-major axis SM$_{e,dV}$.
We have color coded the galaxies according to their b/a ratio. The 
solid line indicates the relation log(R$_{e, dV}$/kpc) = 0.56$\times$(log(M$_{\star}$/M$_{\odot}$) - 9.84) - 0.3.
\begin{figure*}
\includegraphics[angle=0,width=18.5cm]{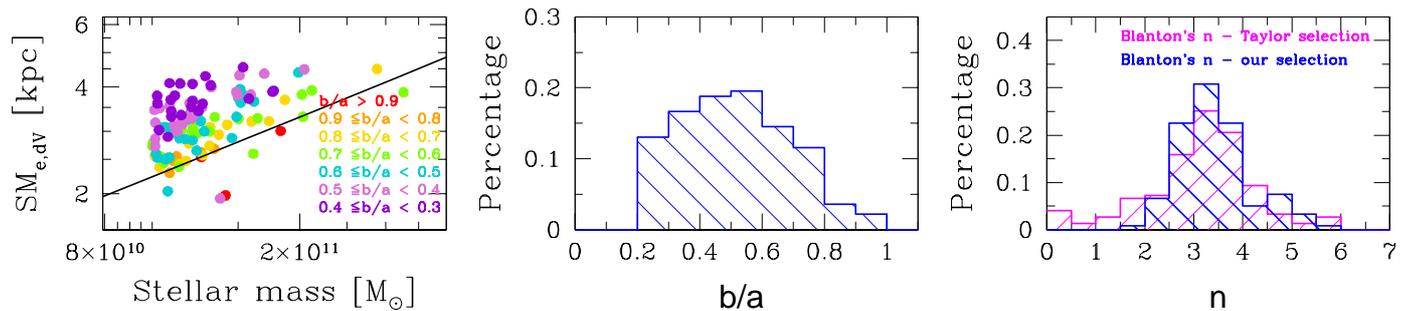}
        \caption{\textit{Left panel}: The sample of local dense galaxies, selected using the Taylor et al.'s cuts and 
circularized de Vaucouleur radius R$_{e, dV}$, 
plotted in the size-mass plane using their semi-major de Vaucouleur axis SM$_{e,dV}$. 
Galaxies are color coded according to their
b/a ratio, as declared in the labels. Solid line indicates the 
log(R$_{e, dV}$/kpc) = 0.56$\times$(log(M$_{\star}$/M$_{\odot}$) - 9.84) - 0.3 relation. 
\textit{Central panel}: The distribution of the axis ratio b/a for the same galaxies in the left-panel. 
Histogram counts are normalized to unity.
\textit{Right panel}: The distribution of the Sersic index n of local dense galaxies selected using Taylor et al.'s cuts,
and circularized Sersic radius (blue histogram). Histogram counts are normalized to unity.
Magenta histogram is the distribution of the Sersic index for our sample 
of ultramassive dense ETGs.}
  \label{b2a}
\end{figure*}
As expected, the Figure shows that $\sim$ 10 galaxies have 
log(R$_{e,dV_{nc}}$/kpc) $<$ 0.56$\times$(log(M$_{\star}$/M$_{\odot}$) - 9.84) - 0.3, while all the others 
lie above the line, with a b/a ratio that progressively decreases with the distance from the solid line, down to b/a $\simeq$ 0.3.
In the central panel of Fig. \ref{b2a} we have reported the distribution of the axis ratio b/a for the same sample of galaxies. 
Half of the sample has b/a $<$ 0.5, 
and this is the main reason of the drastic difference between the samples selected using circularized R$_{e}$, and 
semi-major axis SM$_{e}$.
The leftmost panel of Fig. \ref{b2a} also show that only the galaxies with b/a $\gtrsim$ 0.8-0.9 
are all included in both samples, and this 
is the main reason why even if our ultramassive dense ETGs are expected to be mostly roundish systems, 
i.e. objects for which their semi-major axies should be similar to their R$_{e}$, 
Taylor et al. do not selected them in their sample. In fact, we have checked that
only the $\sim$ 5$\%$ (6 galaxies) of ETGs in our sample has b/a $>$ 0.8, 
and are exactly those satisfying the Taylor's conditions.

Finally, in Sec. 2.5.1 we have found that the number of local galaxies with 
M$_{\star}>$10$^{11}$M$_{\odot}$, log(R$_{e}$/kpc) $<$ 0.56$\times$(log(M$_{\star}$/M$_{\odot}$) - 9.84) - 0.3, 
$^{0.1}$(u -r) > 2.5 still increases if circularized Sersic R$_{e}$ are used 
(in particular, just the $\sim$ 45$\%$ of the galaxies are included in the
sample selected using circularized de Vaucouleur R$_{e}$). This result, as claimed in Sec. 2.5.1, is expected, 
since dense galaxies are known to have Serisc index lower than 4. In the right panel of Fig. \ref{b2a} we have reported
the distribution of the Sersic index $n$ for the local dense galaxies selected using the Taylor et al.'s cut and 
circularized Sersic R$_{e}$: more than 80$\%$ of
the galaxies has $n$ $<$ 4. There is also a tail of galaxies with Sersic index lower than 2.
This is mainly due to the fact that galaxies are selected with a color cut, and it is known that there is a non 
negligible fraction of red disk. Actually, as comparison, 
the magenta histogram in the right panel of Fig. \ref{b2a} shows the distribution of 
the Sersic index for our sample of ultramassive dense ETGs. It is nice to observe how the tail at $n <$ 2
disappears (we remind that no selection on the Sersic index is imposed \textit{a priori} to the local sample of ETGs, 
as well as to no other sample used in our analysis), strengthening the robustness of the visual classification
in selecting spheroidal galaxies.

\section[]{The effect of the extrapolation on the size measurements}

In Section 2.5.3, we have shown that the extrapolation of the size of a galaxy via the Eq. 2 by \citet{vanderwel14}, has an impact 
on the number of dense galaxies.
In this Appendix, we want to deepen this aspect.
\begin{figure}
\begin{center}
	\includegraphics[angle=0,width=9cm]{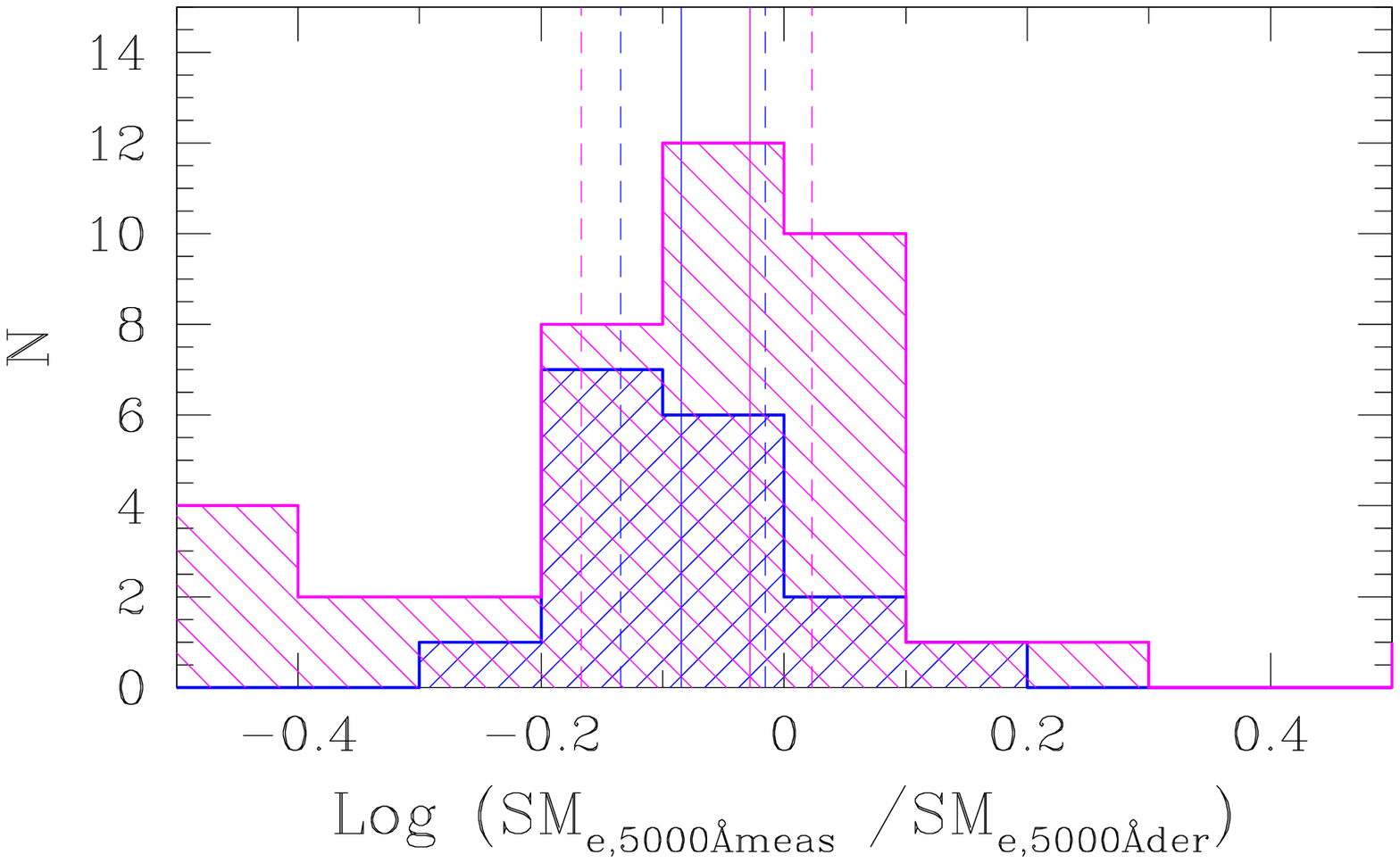} \\
	\caption{The distribution of the SM$_{e,5000\AA meas}$/SM$_{e,5000\AA der}$ ratio as function of the redshift.
SMR$_{e,5000\AA meas}$ is the semi-major axis at 5000\AA restframe measured directly on images, 
while SM$_{e,5000\AA dev}$ is the same quantity, but inferred from F125W-band images using the Eq. 2 \citet{vanderwel14}.
Blue histogram refers to ETGs at 0.5 $ < z <$ 0.7, while magenta one to ETGs at 0.7 $ < z <$ 0.9. Solid lines are the median 
values of the distributions and dotted ones are the 25 and 75 percentile.}
	\label{vdwcol}
\end{center}
\end{figure}
In Figure \ref{vdwcol}, for the two bins of redshift 0.5 $< z <$0.7, and 0.7 $< z <$0.9, 
we have reported the distribution of Log (SM$_{e,5000\AA meas}$/SM$_{e,5000\AA der}$) for a sample of elliptical galaxies, 
where SM$_{e,5000\AA meas}$ is the semi-major axis at 5000\,$\AA$ restframe measured directly on images, 
and SM$_{e,5000\AA der}$ is the one derived from SM$_{e,F125W}$ using the Eq. 2 by \citet{vanderwel14}.
For the galaxies at 0.5 $< z <$0.7, we refer to a subsample of COSMOS ellipticals. 
In details, for the SM$_{e,5000\AA meas}$ values we refer to the measurements by \citet{sargent07}  
on the F814W-band images and discussed in Sec. 2.2, while for the SM$_{e,F125W}$ values we have considered the estimates by
\citet{vanderwel12}. We have crossmatched the two catalogues, and have selected only elliptical (ZEST-type = 1) galaxies  
with M$_{\star} >$ 10$^{10}$M$_{\odot}$.
For the  galaxies at 0.7 $< z <$0.9, we have considered a sample of elliptical galaxies in the GOODS-South field.
For the SM$_{e,5000\AA meas}$ values
we have used  our sample of ETGs on GOODS-South \citep{tamburri14}
with SM$_{e}$ derived on F850LP images (briefly described in Sec. 2.1), and for the SM$_{e,F125W}$ measurements
we have considered the catalogue by \citet{vanderwel12}.
At 0.5 $< z <$0.7, SM$_{e,5000\AA der}$ overestimates the SM$_{e,5000\AA meas}$ by a factor -0.08 dex (median value),
and in particular for more than 80$\%$ of the galaxies SM$_{e,5000\AA meas}$/SM$_{e,5000\AA der}$ $<$ 1. 
At 0.7 $< z <$0.9, where the correction is extrapolated
over a shorter wavelength range, this fraction lowers to $\sim$65$\%$, and the median value is -0.02.
The two samples of galaxies at 0.5 $< z <$0.7 and 0.7 $< z <$0.9 do not have exactly the same distribution 
in stellar mass and this can alter the relative comparison of their  SM$_{e,5000\AA meas}$/SM$_{e,5000\AA der}$ values.
However, we have checked that the ratio SM$_{e,F814W}$/SM$_{e,F125W}$ and SM$_{e,F850LP}$/SM$_{e,F125W}$ does not
depend on stellar mass, as stated by \citet{vanderwel14}.
Although test on more statistically significant samples would be necessary to be conclusive, 
the Fig. \ref{vdwcol} shows that an $average$ variation of SM$_{e}$ with $\lambda$ 
(as assumed in the Eq. 2 by van der Wel) 
could not account for the variety of stellar populations in elliptical
galaxies at high redshift \citep[see e.g.][]{guo11, gargiulo12}, and this should be taken in mind 
in the interpretation and comparison of the achived results.

\section[]{Velocity dispersion measurements at z$\sim$1.4}

Fig. \ref{bestfitspectra} shows the FORS2 spectra 
of the five galaxies of our high-z sample. As expected from the redshift estimates based on the continuum shape provided by AMICI 
spectra in the framework of the MUNICS survey, for the four galaxies at z $\lesssim$ 1.4 (ID 511, 633, 527, 389) we have 
detected many absorption features (e.g. CN(3833\AA), the CaII K and H
(3933+3968\AA),  blue lines) besides the 4000\AA\, break. 
We notice that the ETG 527, whose elliptical morphology is confirmed by the HST image and 
surface brightness profile (see Fig. 1), clearly shows the O$_{2}$ emission line at 3727\,\AA.
We exclude that the presence of this emission line is due to AGN activity. In facts, the 5 ETGs of our sample
were observed with XMM observations and only the source 443 has an X-ray detection \citep{severgnini05}.
For galaxy 443 we have detected, as expected, the MgII absorption feature. 
However, we notice that 
the detection of this absorption line fixes the redshift of this galaxy at z=1.91 slightly higher than the value 
 originally estimated from the AMICI continuum (z =1.70$\pm$0.05). 
In Table \ref{table1} we report the values of the redshifts for the five galaxies of our sample.

The tool pPXF we have adopted to derive the velocity dispersion describes the line of sight velocity
distribution (LOSVD) as a Gauss+Hermite polynomial to take into
account possible deviation of the line
profile from a pure Gaussian shape. 
Following the
prescription of the pPXF documentation, we have run simulations to define
the value of the BIAS parameter, a penalty term which biases the
solution towards a Gaussian shape if the higher moments of Hermite
polynomial are not constrained from the data, and have found that for our galaxies it ranges between
0.1 and 0.2, in agreement with \citet{cappellari11}.

Once fixed the BIAS level, we have measured the velocity dispersion by means of a weighted 
fit and adopting as templates a set of synthetic stars (see Sec. 3.1).
To check for possible dependence of the $\sigma$ estimate on the 
selected templates,
we have repeated the measure adopting
a set of BC03 models with solar metallicity, $\tau$ = 0.1 Gyr and
varying age. We have found $\sigma$ values perfectly comparable
(differences $<$ 5$\%$) with those obtained with Munari's stars.  For
clarity, the values reported in this paper and used in the analysis
refer to $\sigma$ estimates derived adopting the set of stars as
templates.

We have included in the fitting function both additive and multiplicative
polynomials to take into account possible residual 
template mismatch, and eventual sky subtraction and spectral calibration errors. We
have verified the stability of our
measures varying the degree (up to 6) of the two set of polynomials and have found
that the detected variations in the estimates of 
velocity dispersion are lower than 3$\%$, well within 1$\sigma$ error bar.

Four out of five ETGs of our sample have spectra covering the restframe wavelength range $\sim$ [0.3 - 0.43]$\mu$m, thus
including the 4000 \AA\, break and many absorption features among which Ca H+K. 
In the past, several analysis have reported that the inclusion of the H+K features lead 
to an overestimate of the $\sigma$ value, 
up to a factor of $\sim$20$\%$ \citep{kormendy82a, kormendy82b}. These lines are intrinsically 
broad ($\sim$ 5 \AA), and in addition, the CaH line at 3968\AA\, is blended with H$\epsilon$ at 3969\AA. 
Moreover, error in the fitting the steep continuum gradient at 4000\AA\, could produce a template mismatch.
All these factors can invalidate the measured dispersions. Nonetheless the real impact of these feature on the 
$\sigma$ estimates is not quantifiable a priori \citep[e.g.][]{treu01, vandesande13}, thus
we have organized a set of simulations to test the reliability of our
velocity dispersion measurements.

\begin{figure}
\includegraphics[angle=0,width=9.0cm]{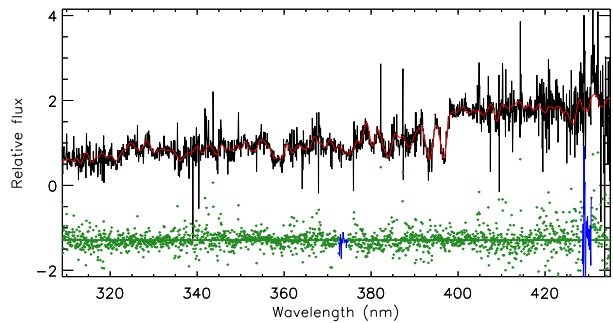} \\
\caption{An example of the galaxy spectrum template (black line) we have constructed for the real galaxy S2F1-633. 
Red line is the best-fit model, while green points are the residuals.}
\label{fakespectrum}
\end{figure}

\begin{figure*}
\begin{center}
  \begin{tabular}{ccc}
 \includegraphics[angle=0,width=6.0cm]{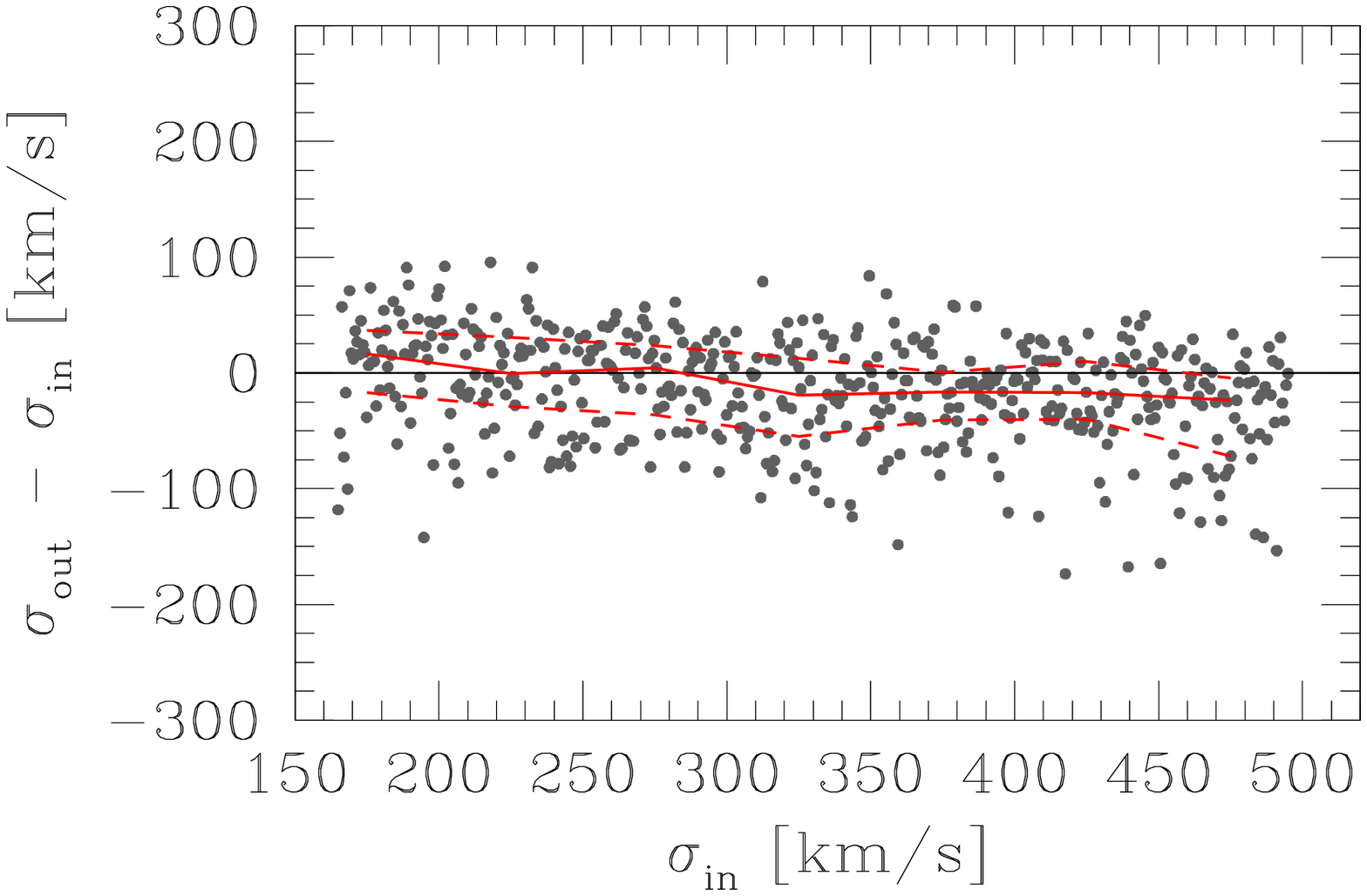} & 
\includegraphics[angle=0,width=6.0cm]{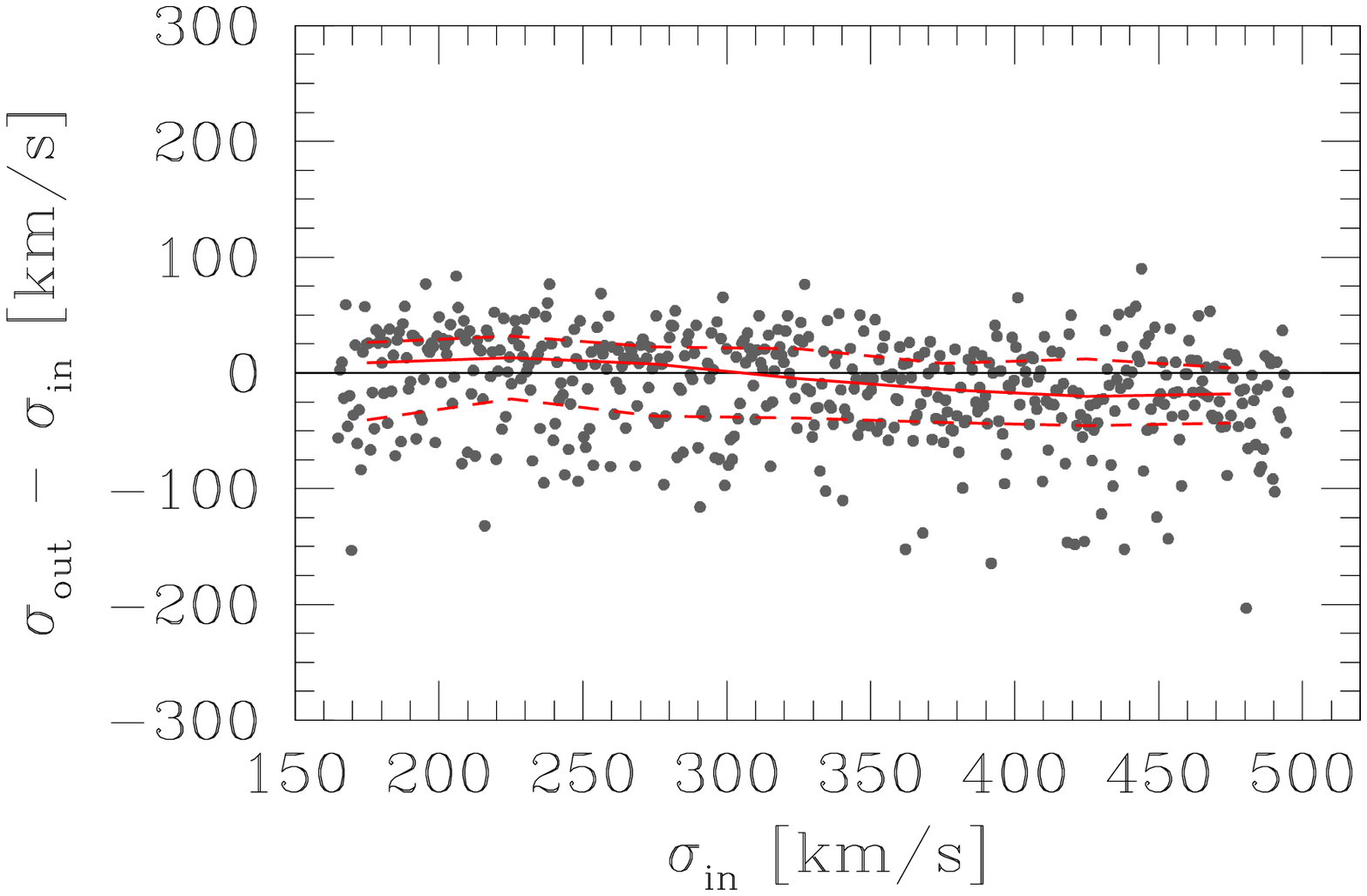} & \includegraphics[angle=0,width=6.0cm]{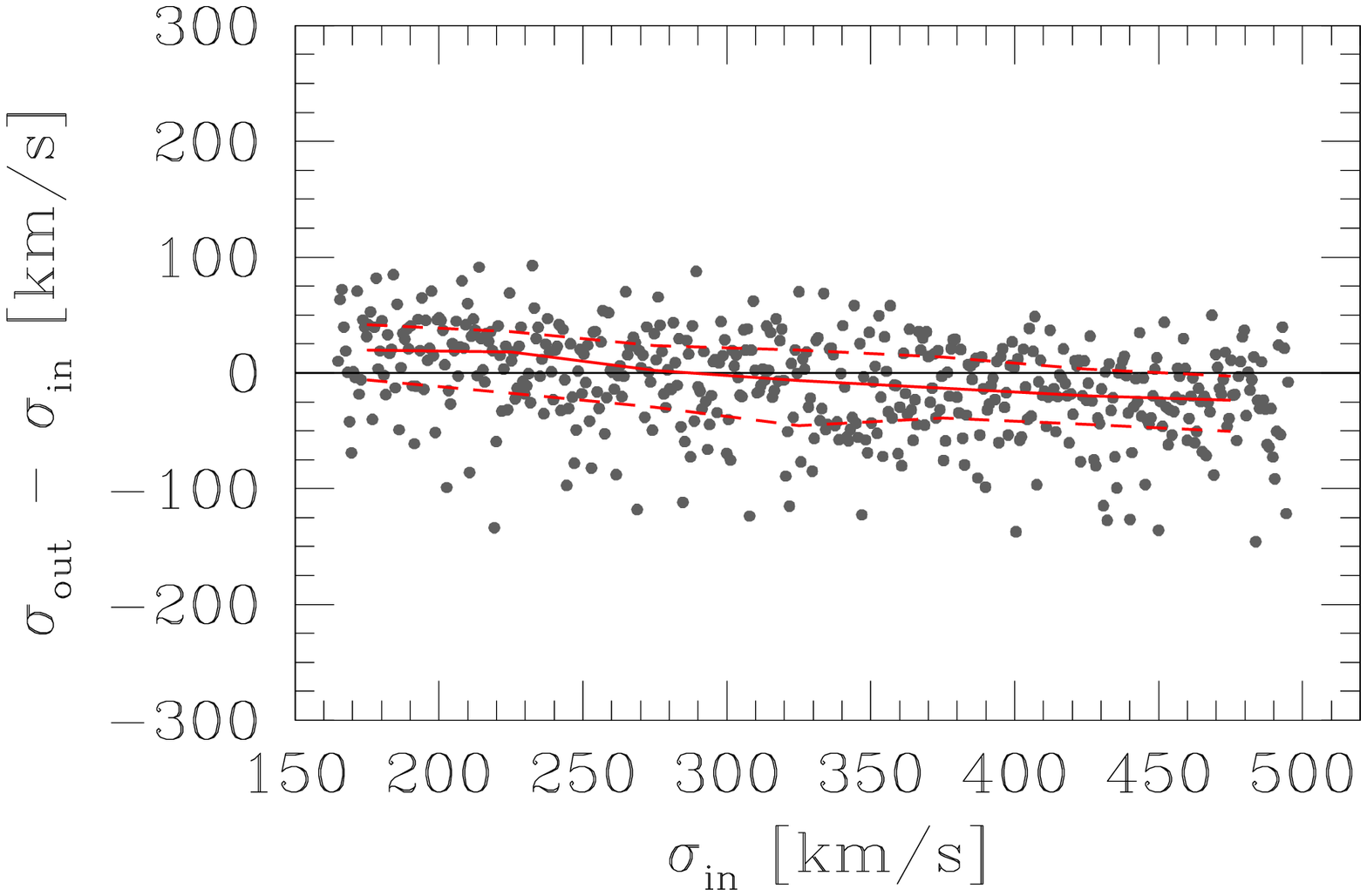} \\
 \includegraphics[angle=0,width=6.0cm]{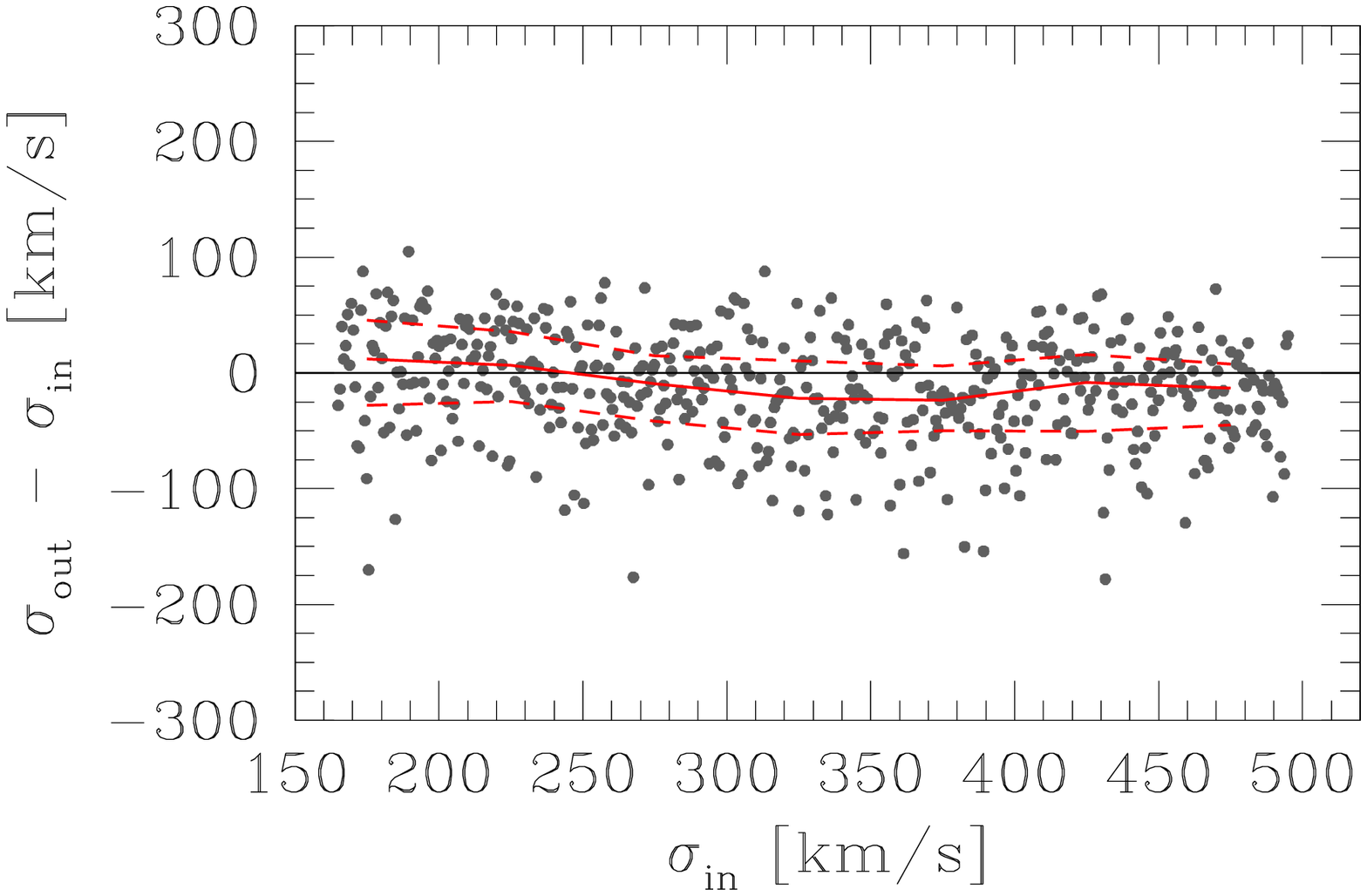} & \includegraphics[angle=0,width=6.0cm]{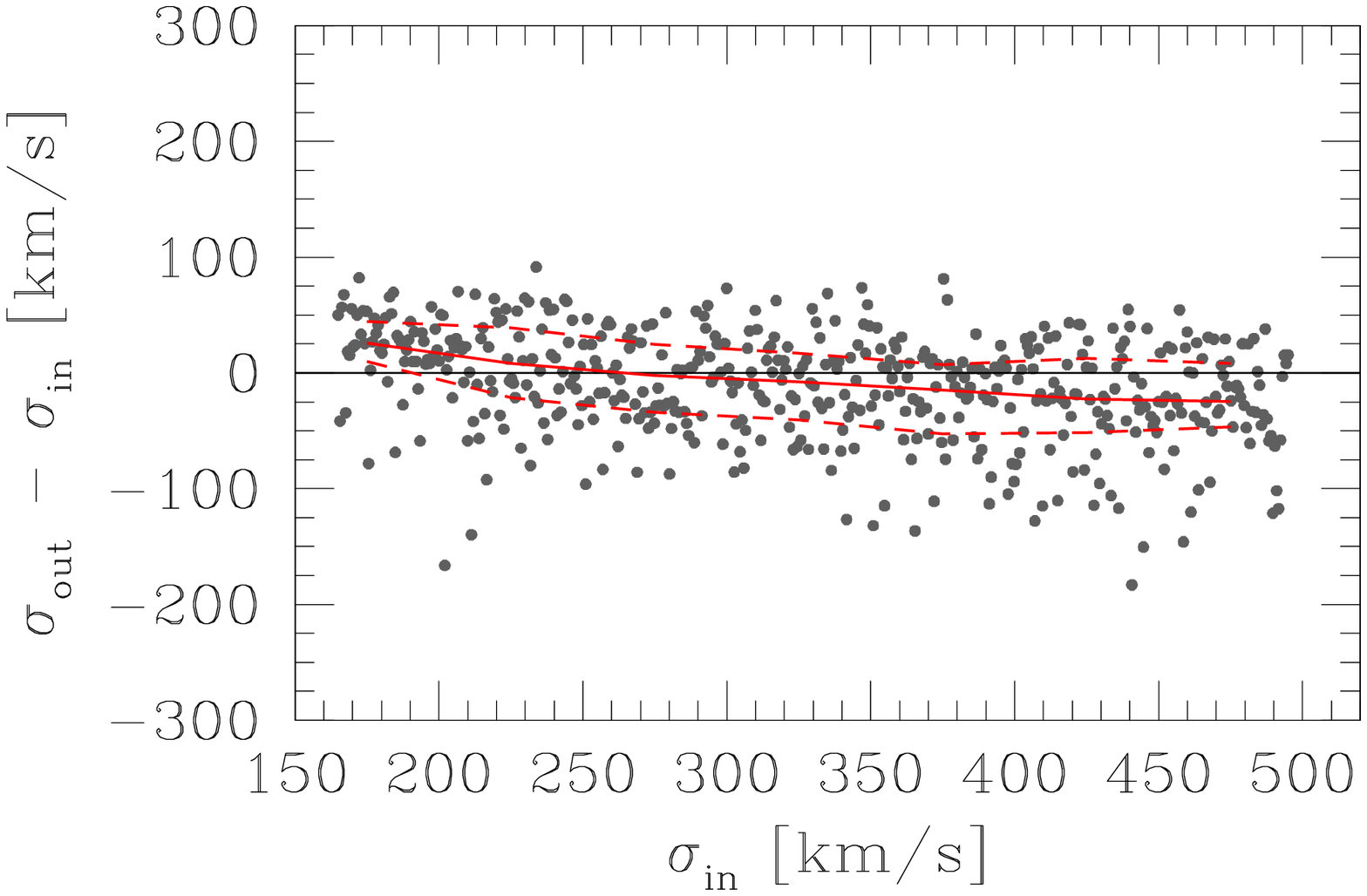} & 
\includegraphics[angle=0,width=6.0cm]{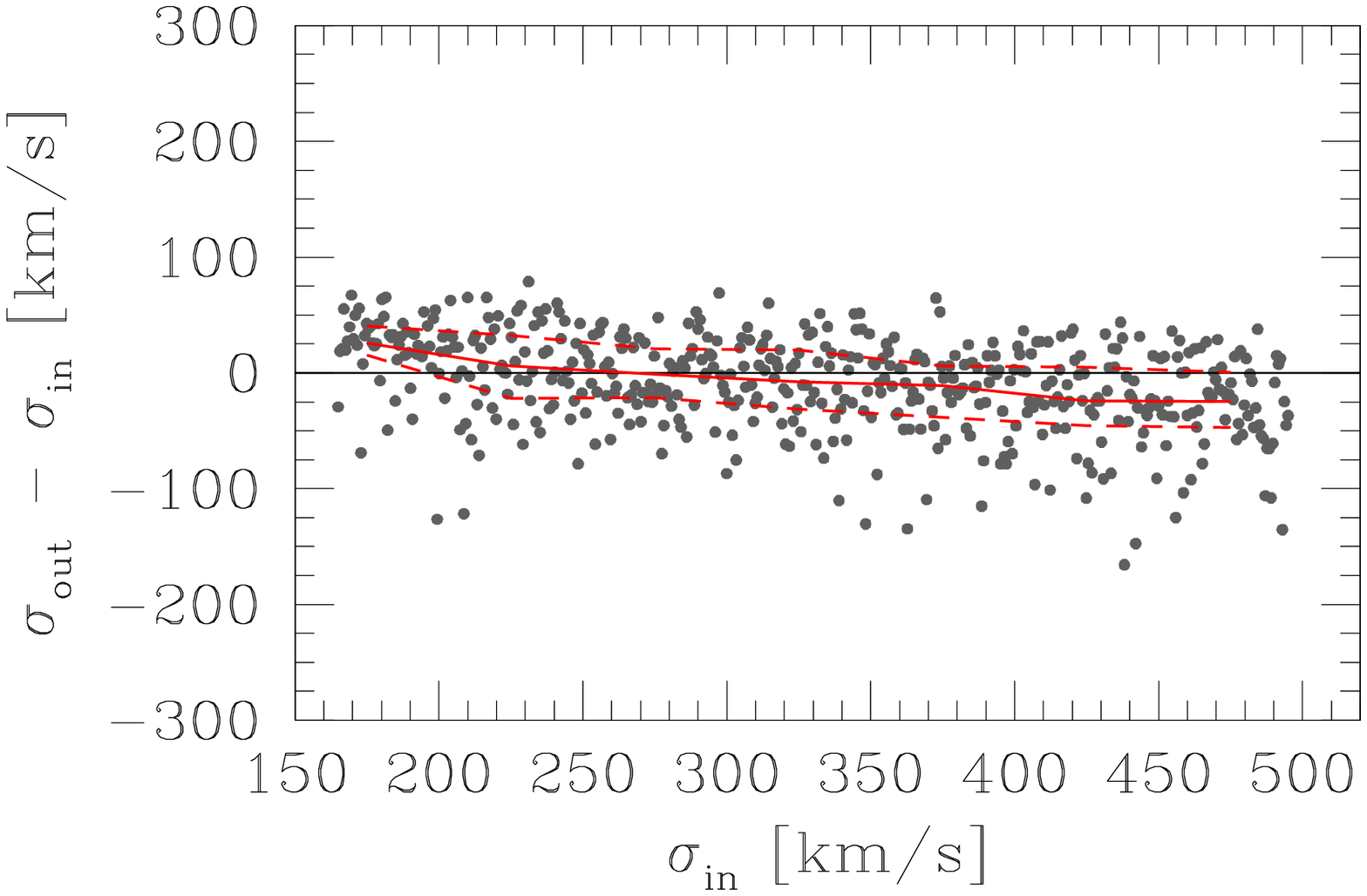} \\
 \end{tabular}
\end{center}
\caption{\textit{Top row}: The difference between the velocity dispersion measured ($\sigma_{out}$) and that provided as
input ($\sigma_{in}$) in the model as function of $\sigma_{in}$ for galaxy S2F1-511. 
The template adopted in the simulation have solar metallicity and age 2, 4, 5 Gyr from left to right. 
It is evident that the age of stellar population do not affect our measurements.
\textit{Bottom row}: Same of top row for the template with age 3 Gyr and sub-solar, solar and super-solar metallicity 
from left to right. The three panels show that our uncertainty on stellar metallicity does not compromise our $\sigma$
estimates. Red solid line is the median of the distribution, while dashed lines are the 
25th and the 75th percentile.  }
\label{simulage1}
\end{figure*}

\begin{figure*}
\begin{center}
  \begin{tabular}{ccc}
 \includegraphics[angle=0,width=6.0cm]{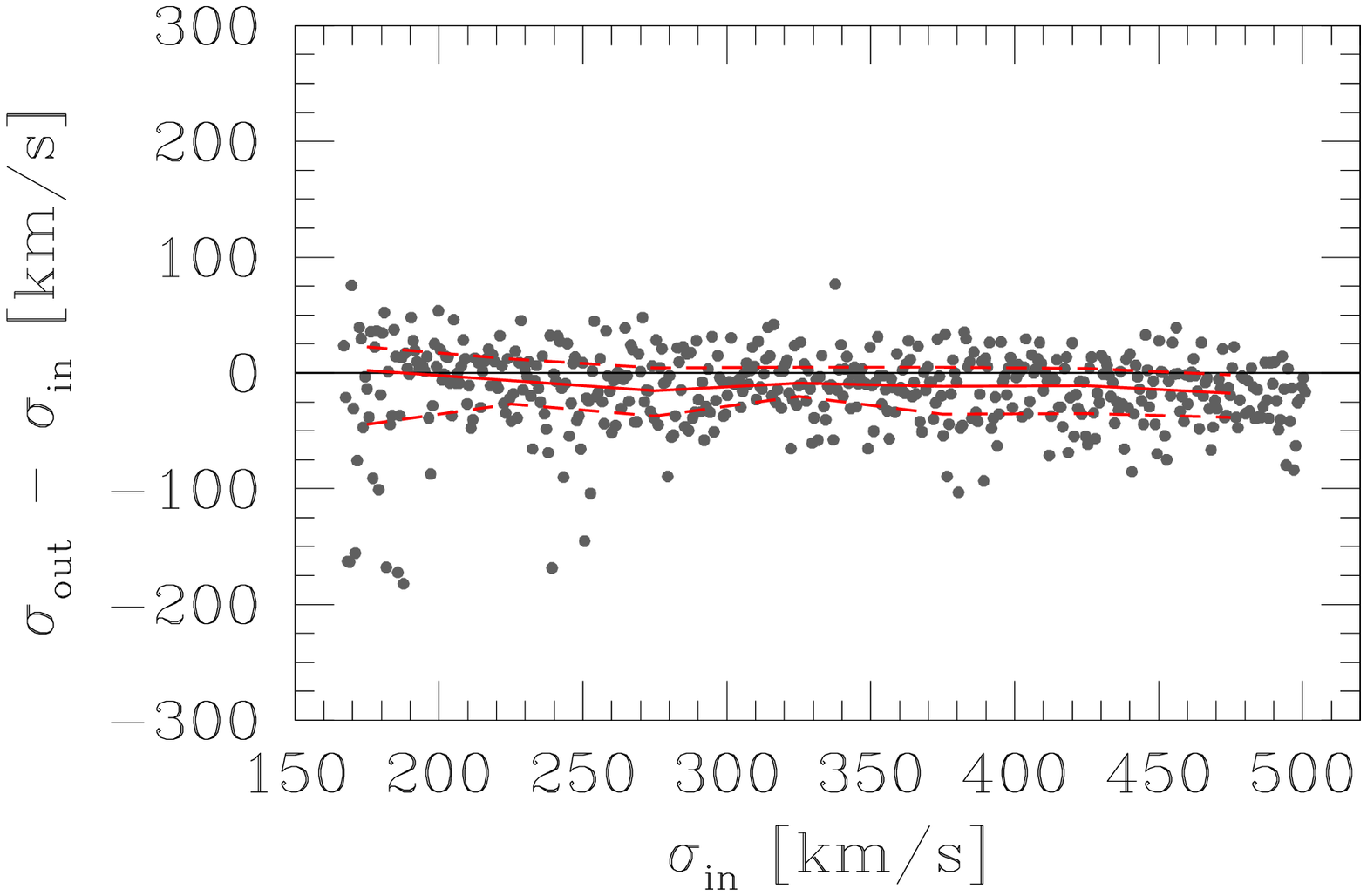} & 
\includegraphics[angle=0,width=6.0cm]{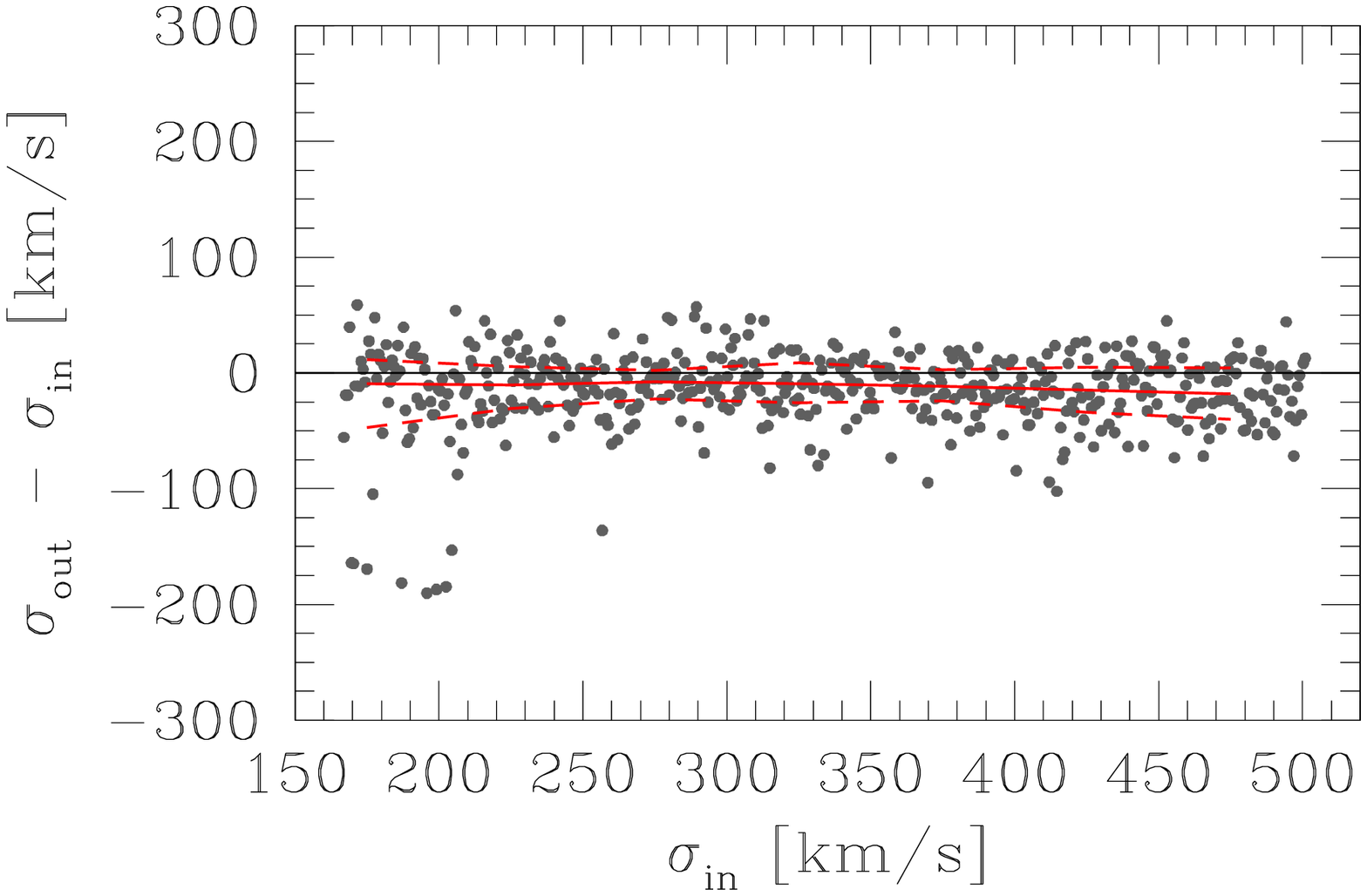} & \includegraphics[angle=0,width=6.0cm]{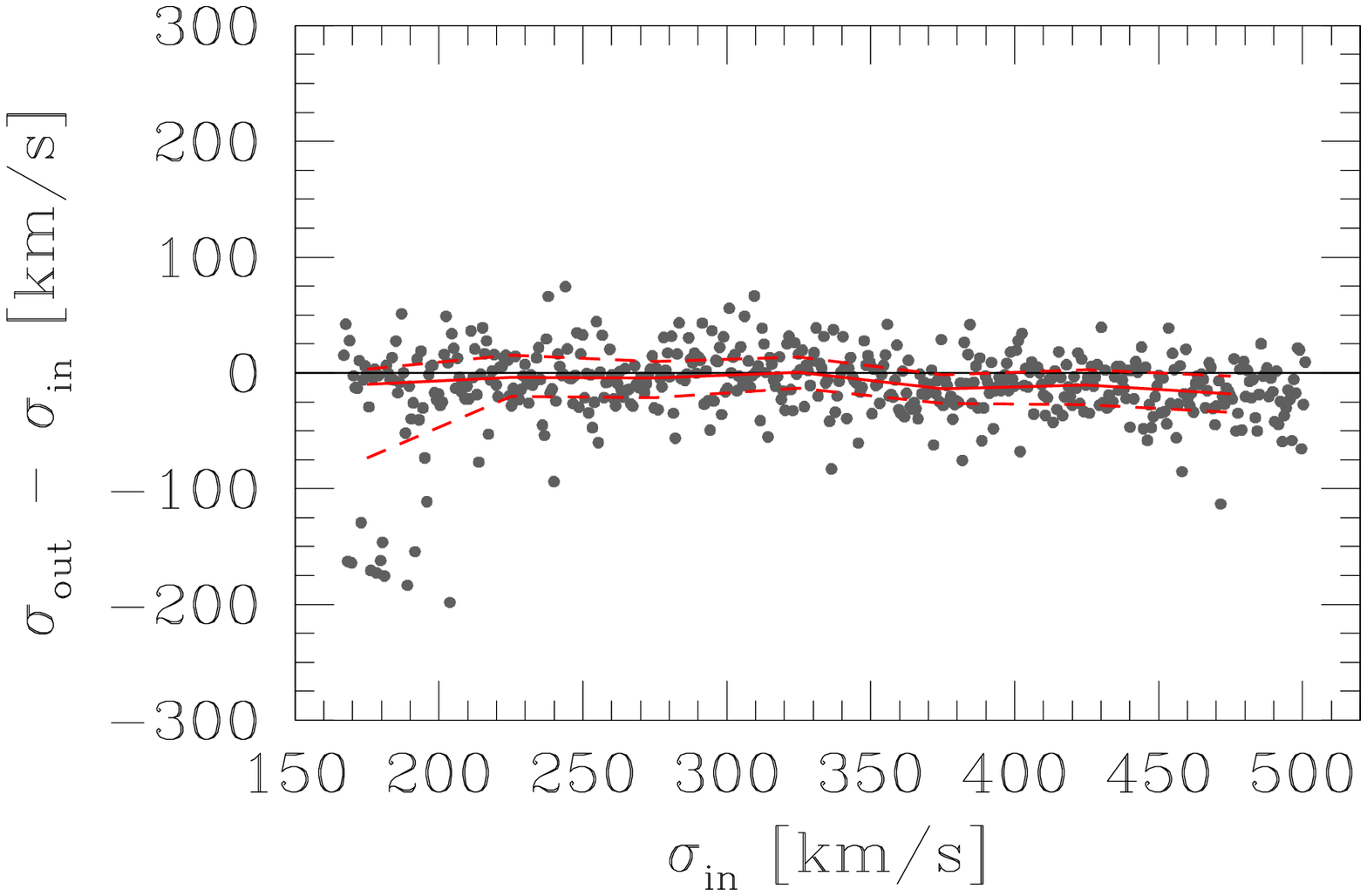} \\
 \includegraphics[angle=0,width=6.0cm]{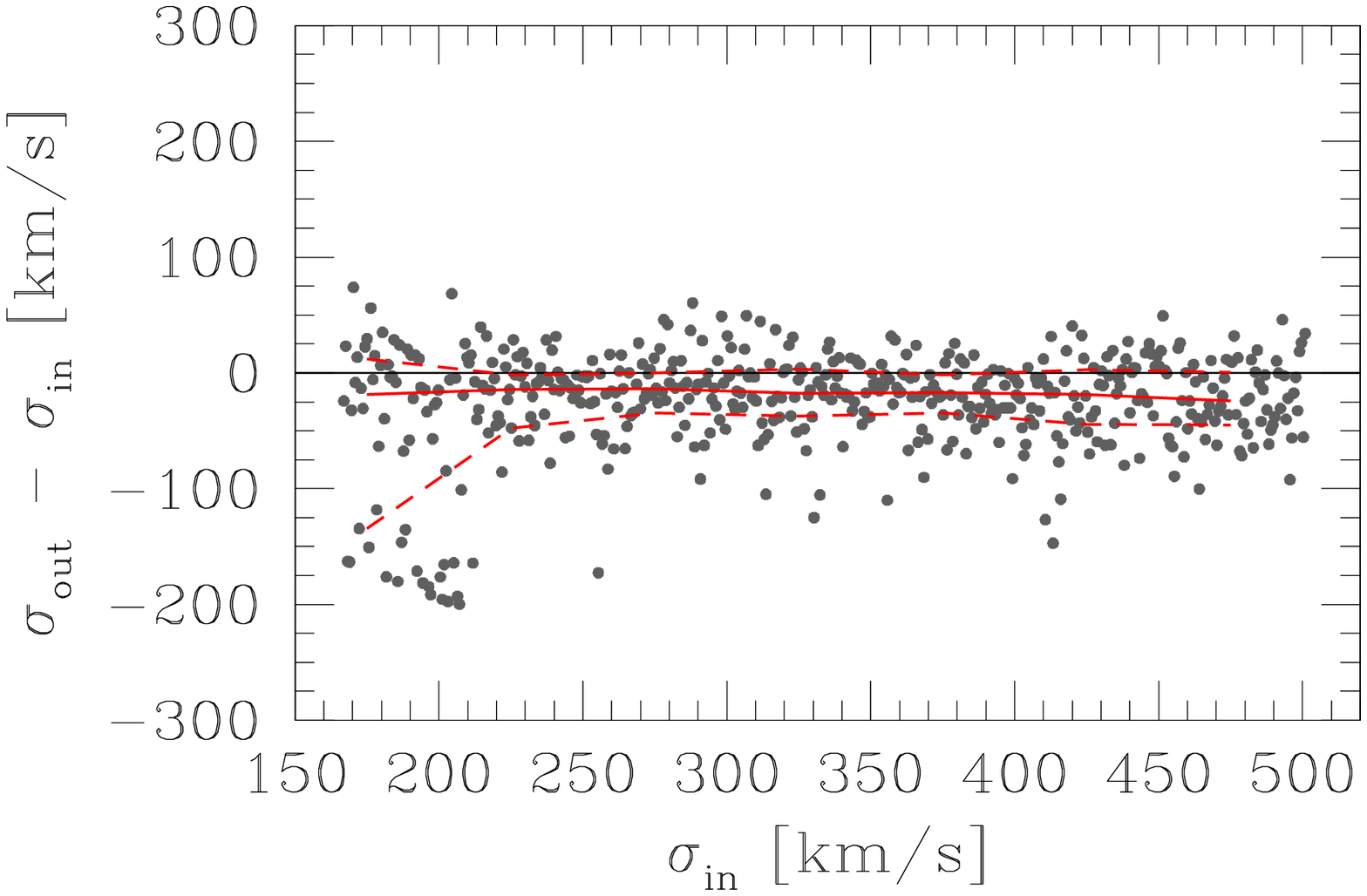} & \includegraphics[angle=0,width=6.0cm]{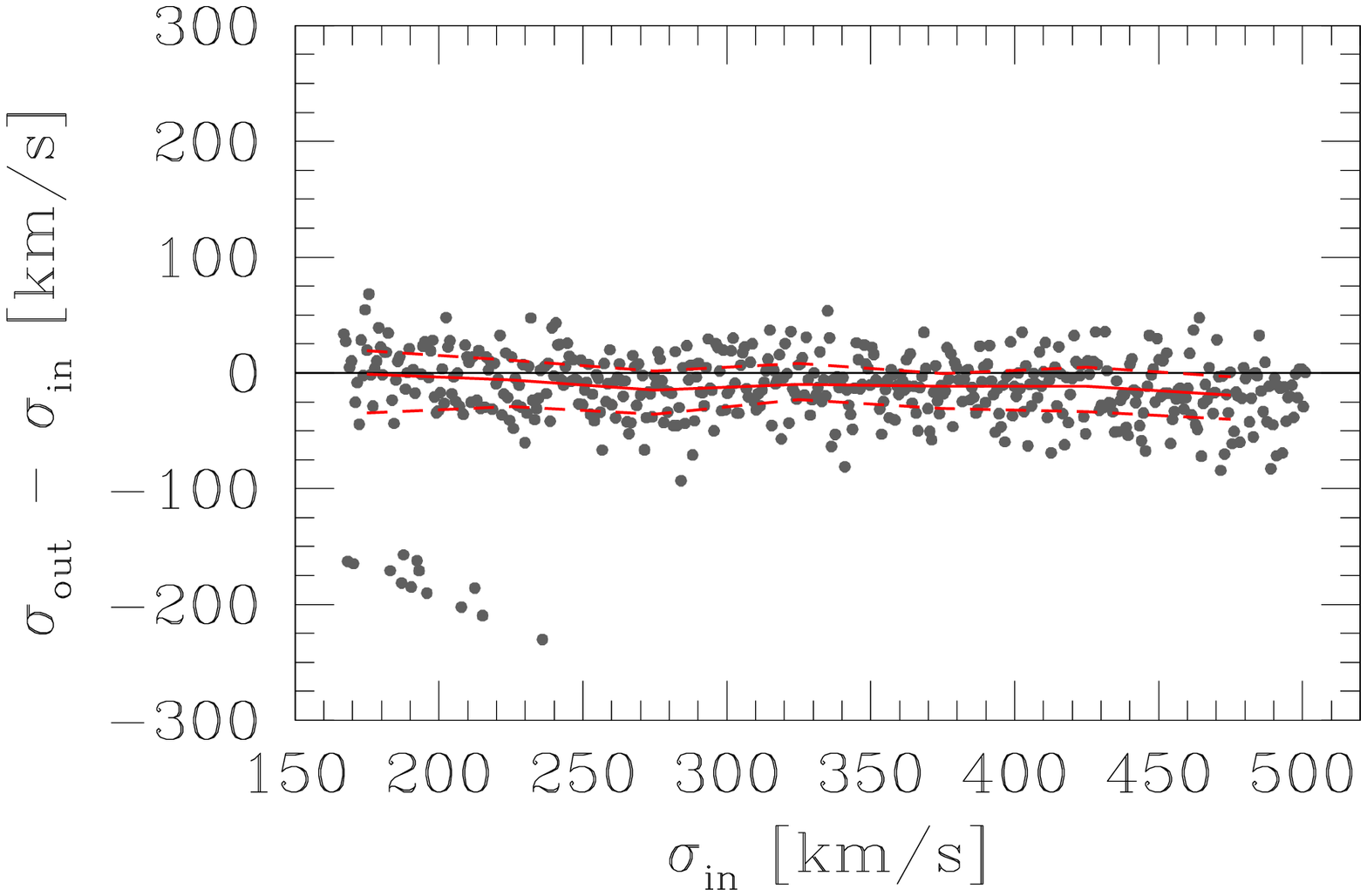} & 
\includegraphics[angle=0,width=6.0cm]{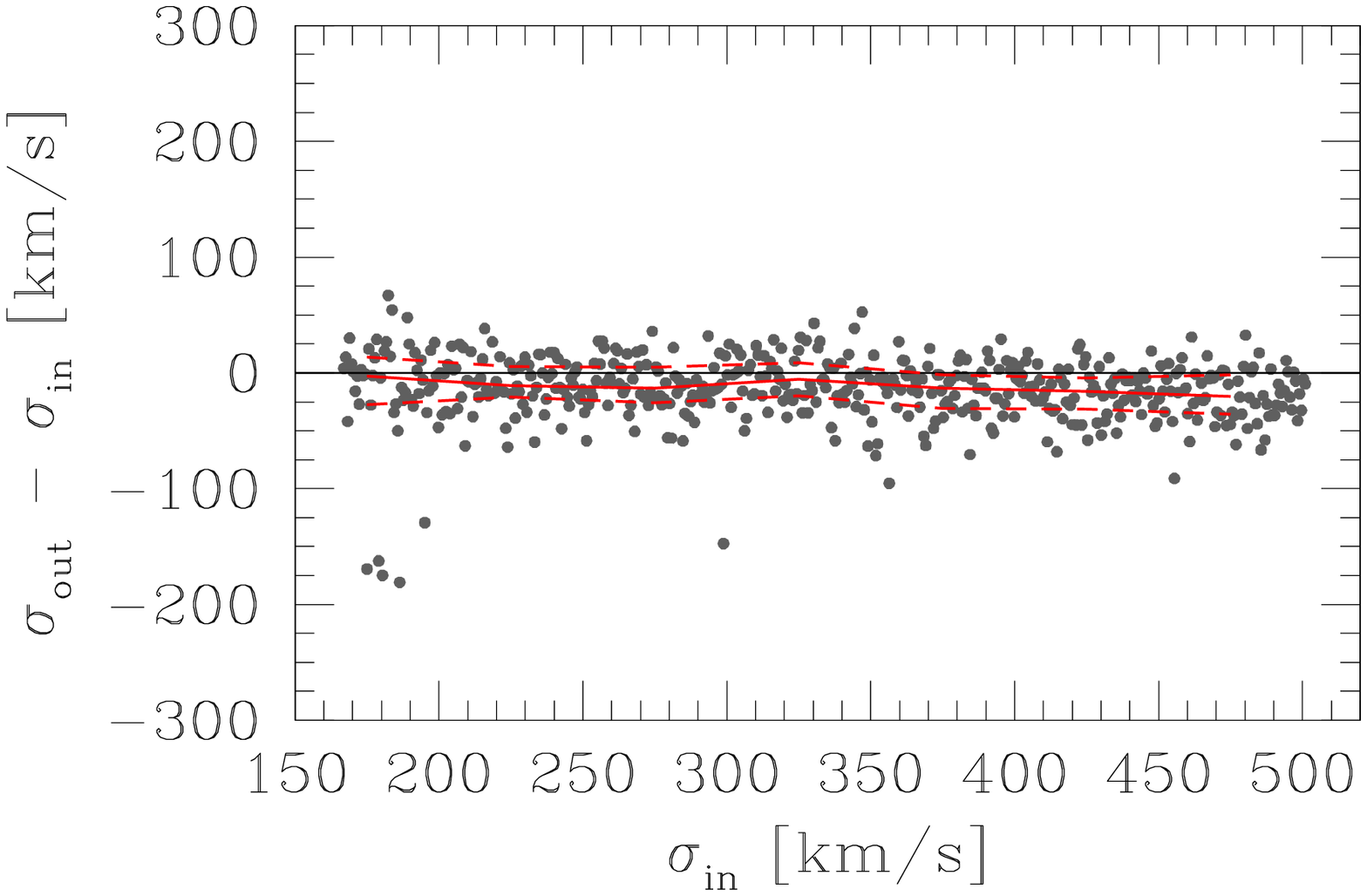} \\
 \end{tabular}
\end{center}
\caption{The same as Fig. \ref{simulage1} for galaxy S2F1-633.}
\label{simulage2}
\end{figure*}

\begin{figure*}
\begin{center}
  \begin{tabular}{ccc}
 \includegraphics[angle=0,width=6.0cm]{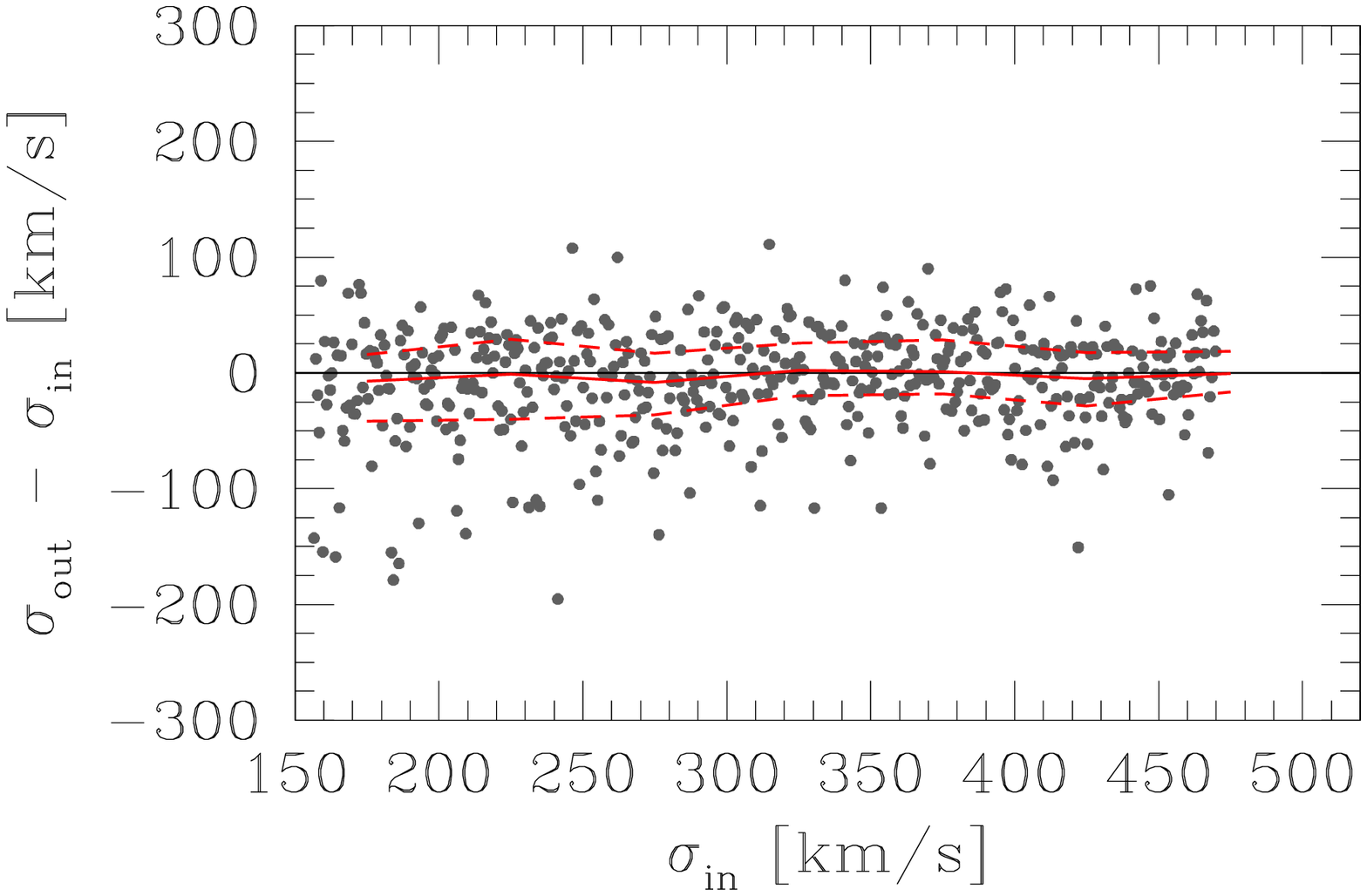} & 
\includegraphics[angle=0,width=6.0cm]{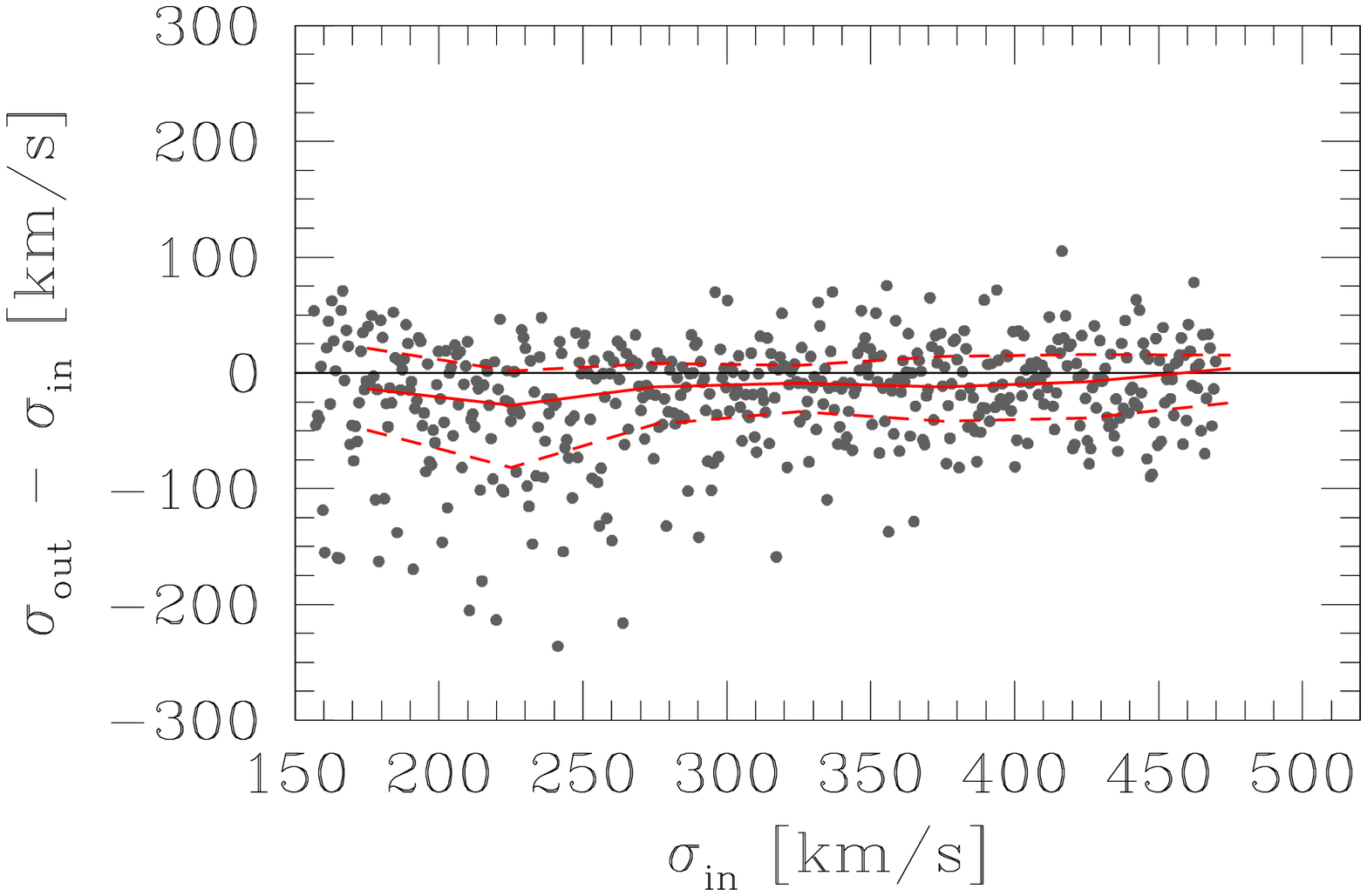} & \includegraphics[angle=0,width=6.0cm]{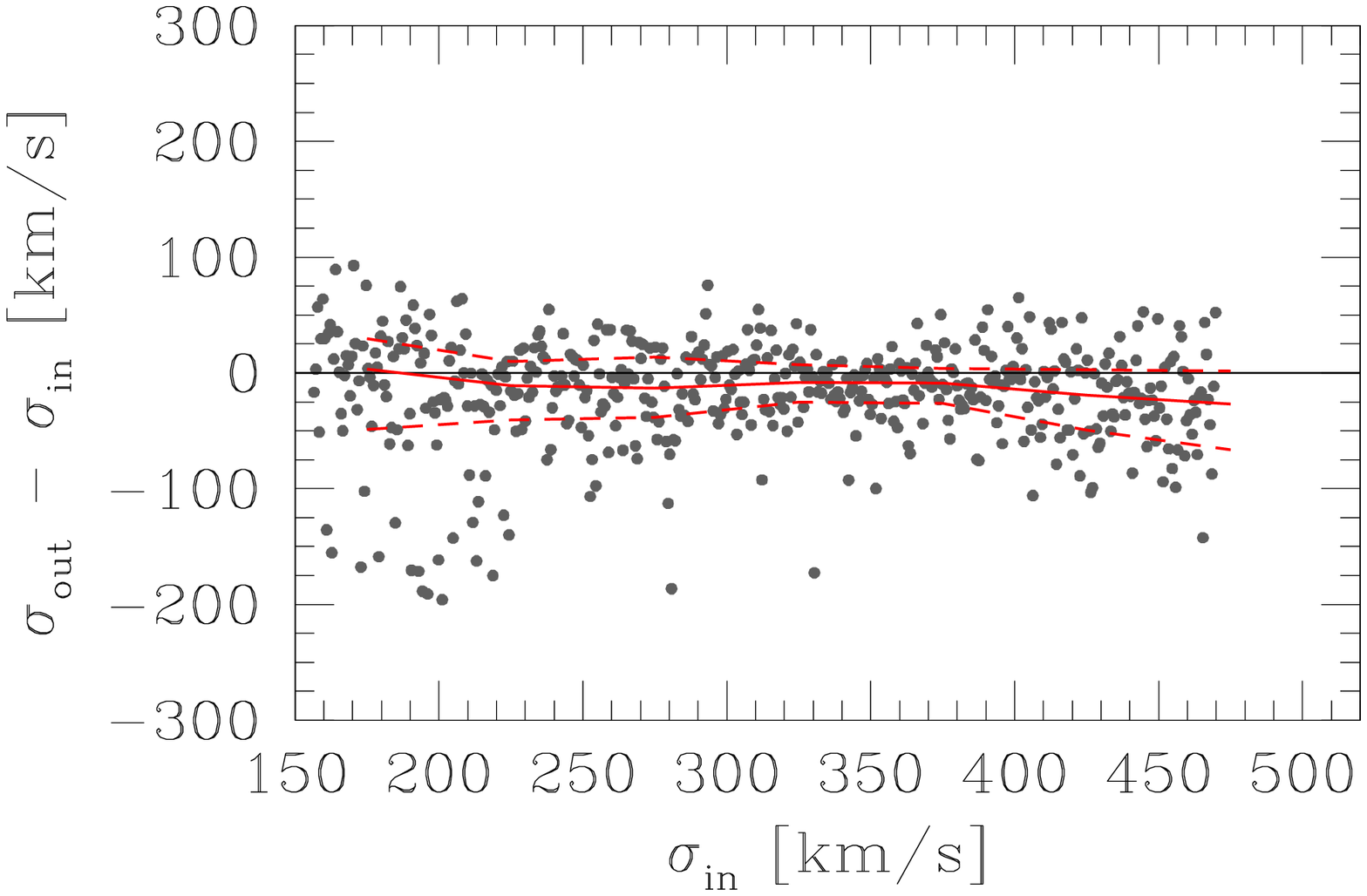} \\
 \includegraphics[angle=0,width=6.0cm]{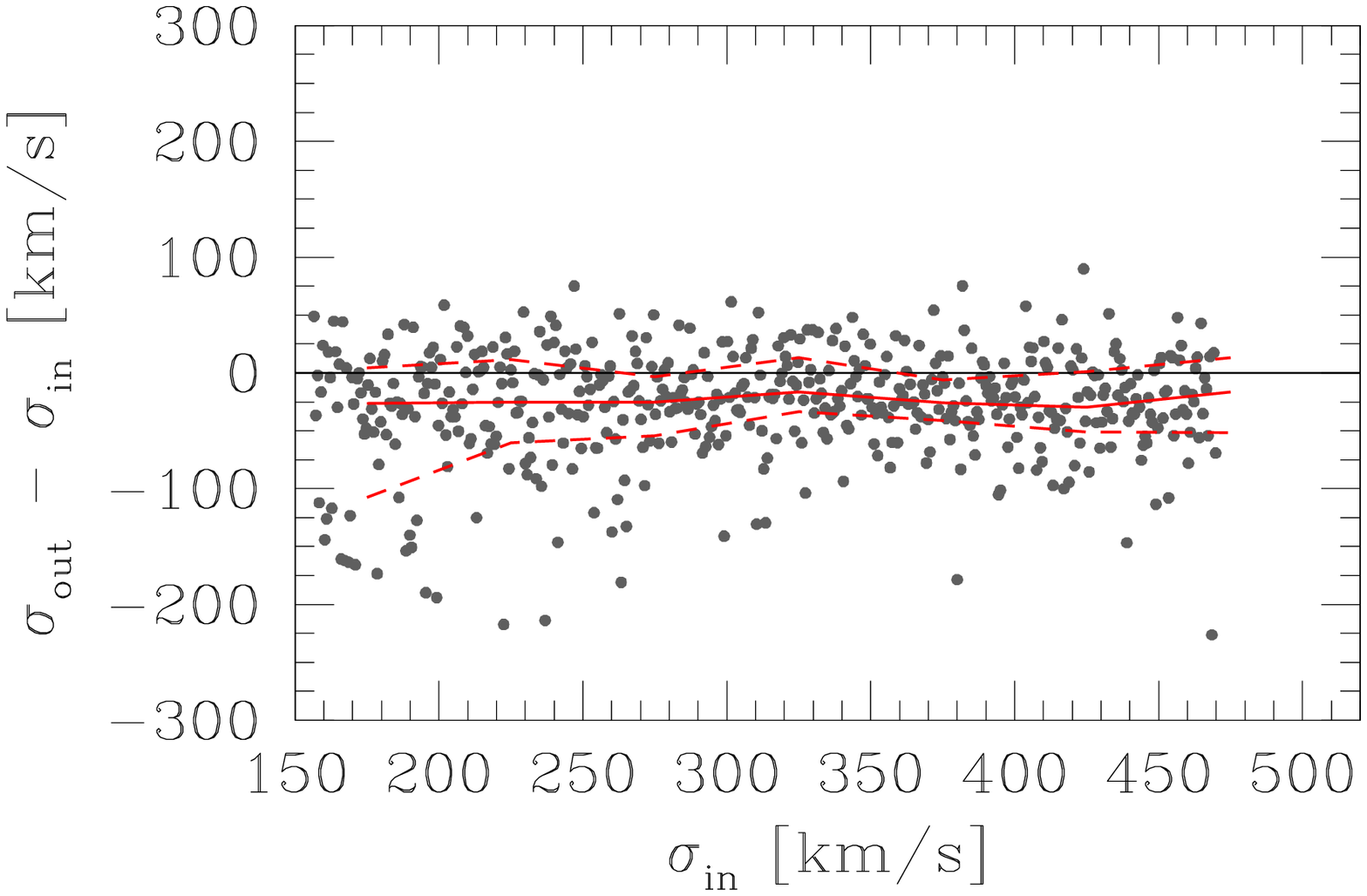} & \includegraphics[angle=0,width=6.0cm]{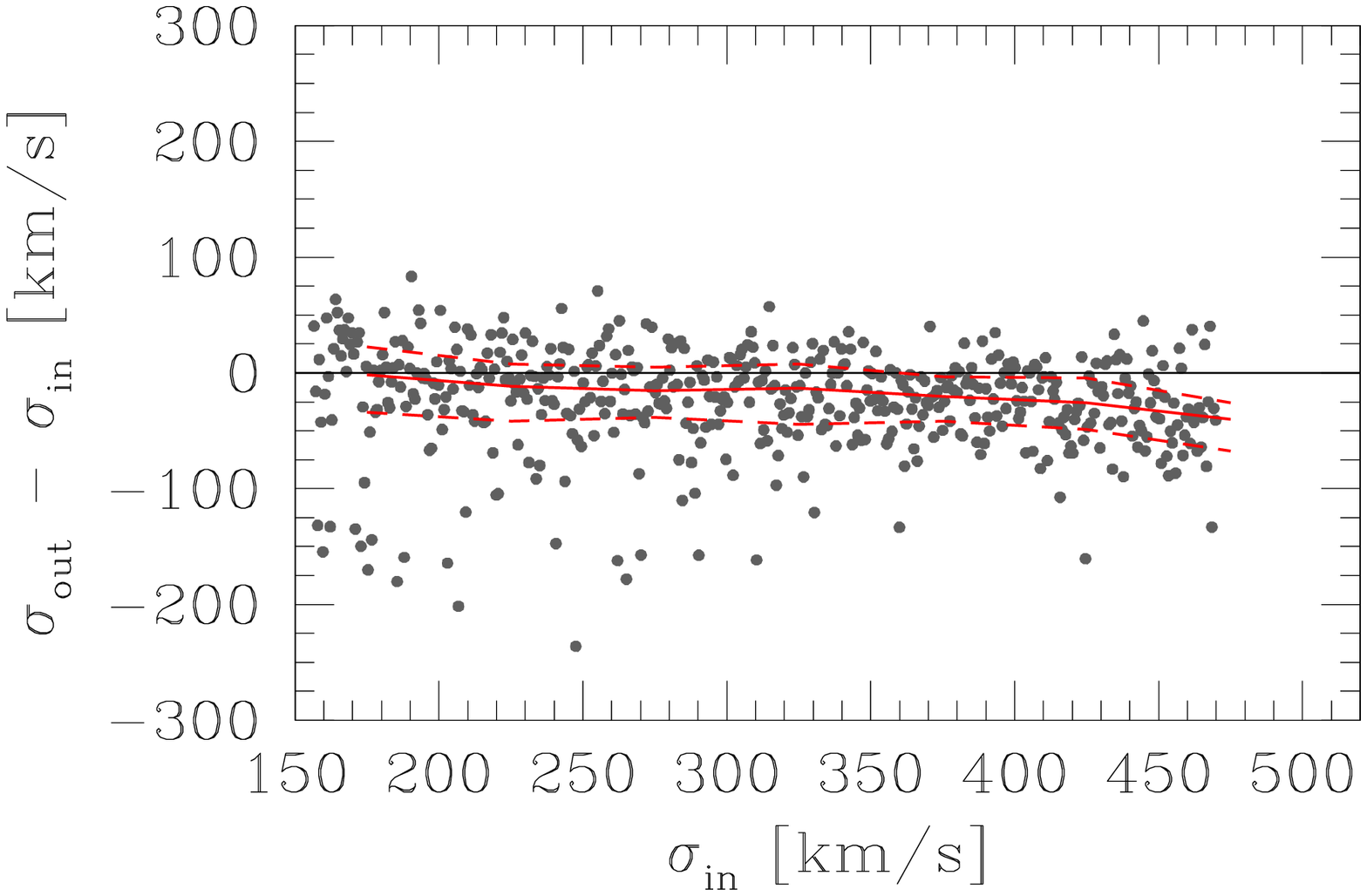} & 
\includegraphics[angle=0,width=6.0cm]{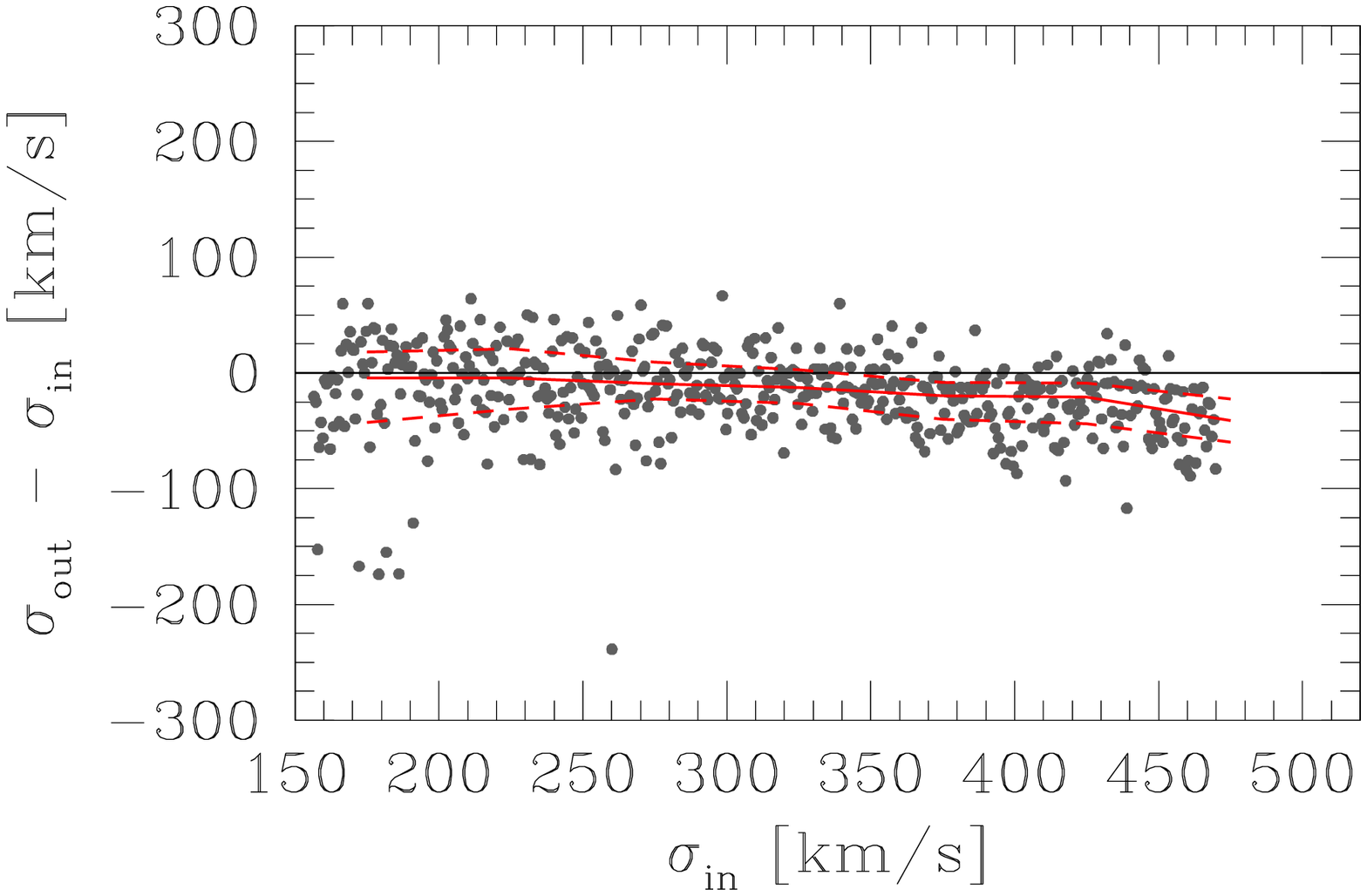} \\
 \end{tabular}
\end{center}
\caption{The same as Fig. \ref{simulage1} for galaxy S2F1-527. Given the spectrum and the SED fitting of this ETG, 
which both suggest a very young age for its stellar population, in the simulation of this galaxy we have adopted as galaxy template 
simple stellar population templates with solar metallicity and age
0.5, 1, 2 Gyr (upper panels, from left to right) and stellar population templates with an age of 3 Gyr and sub-solar, solar, and super-solar 
metallicity (lower panels, from left to right).}
\label{simulage3}
\end{figure*}

\begin{figure*}
\begin{center}
  \begin{tabular}{ccc}
 \includegraphics[angle=0,width=6.0cm]{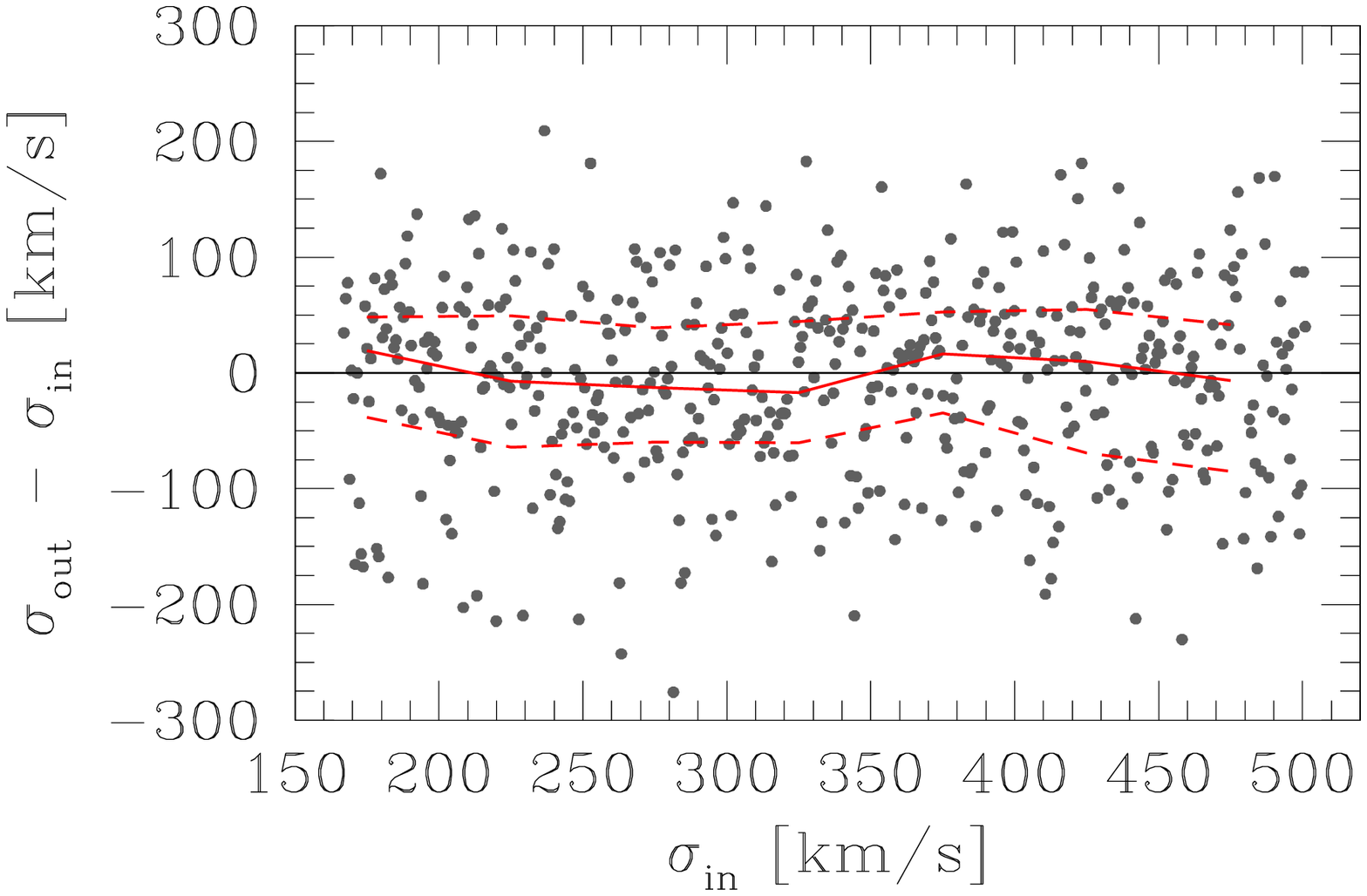} & 
\includegraphics[angle=0,width=6.0cm]{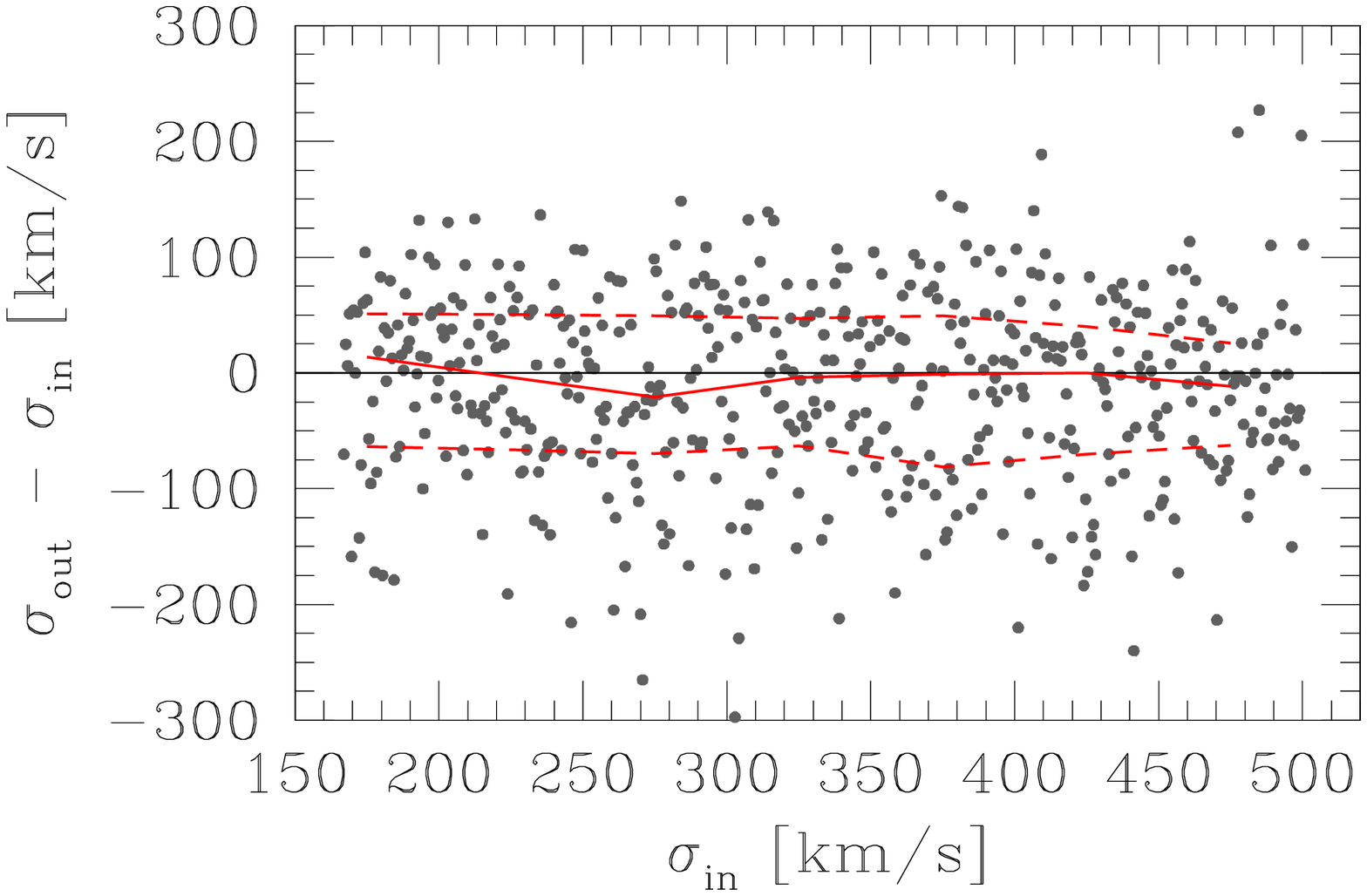} & \includegraphics[angle=0,width=6.0cm]{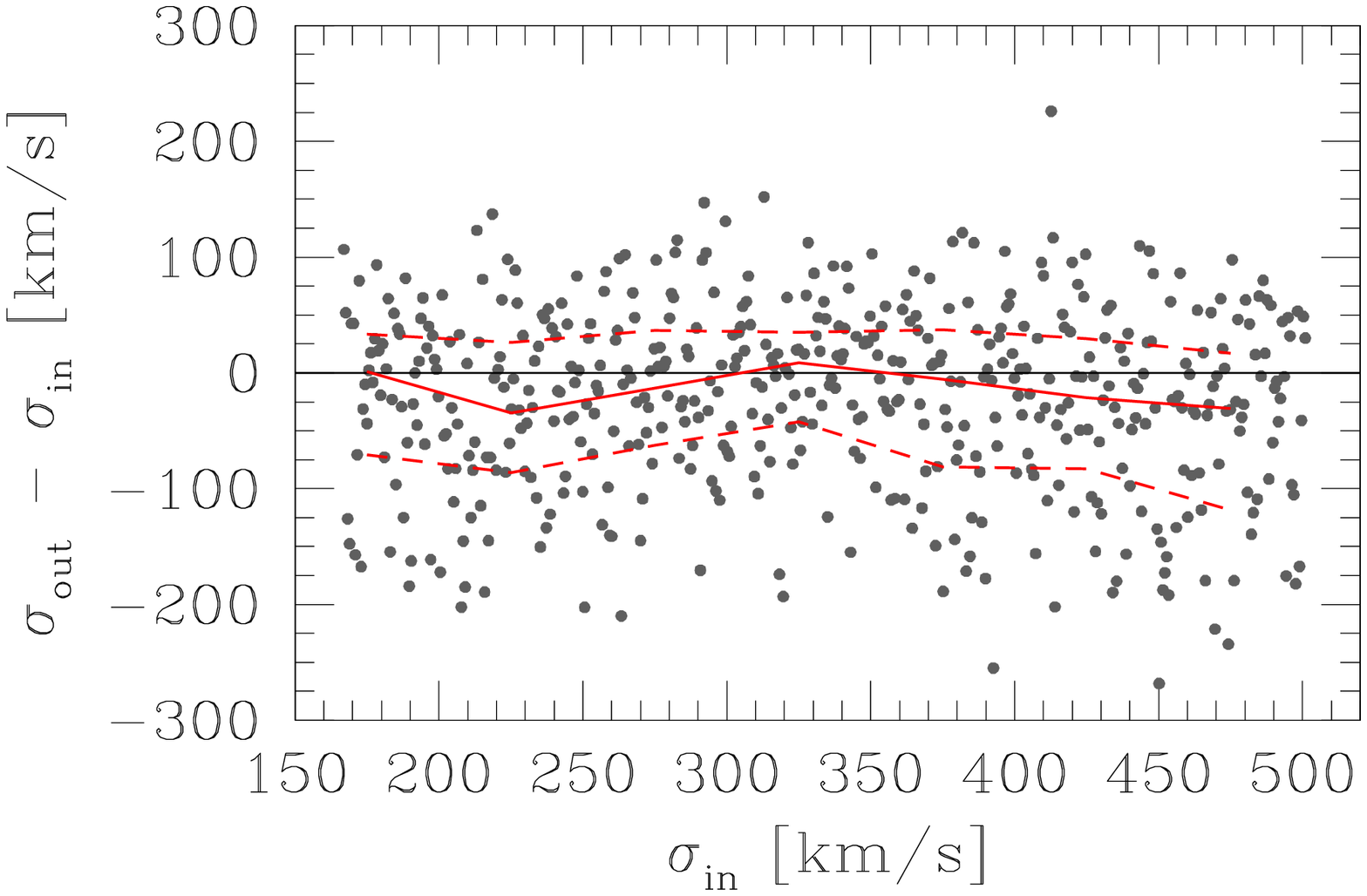} \\
 \includegraphics[angle=0,width=6.0cm]{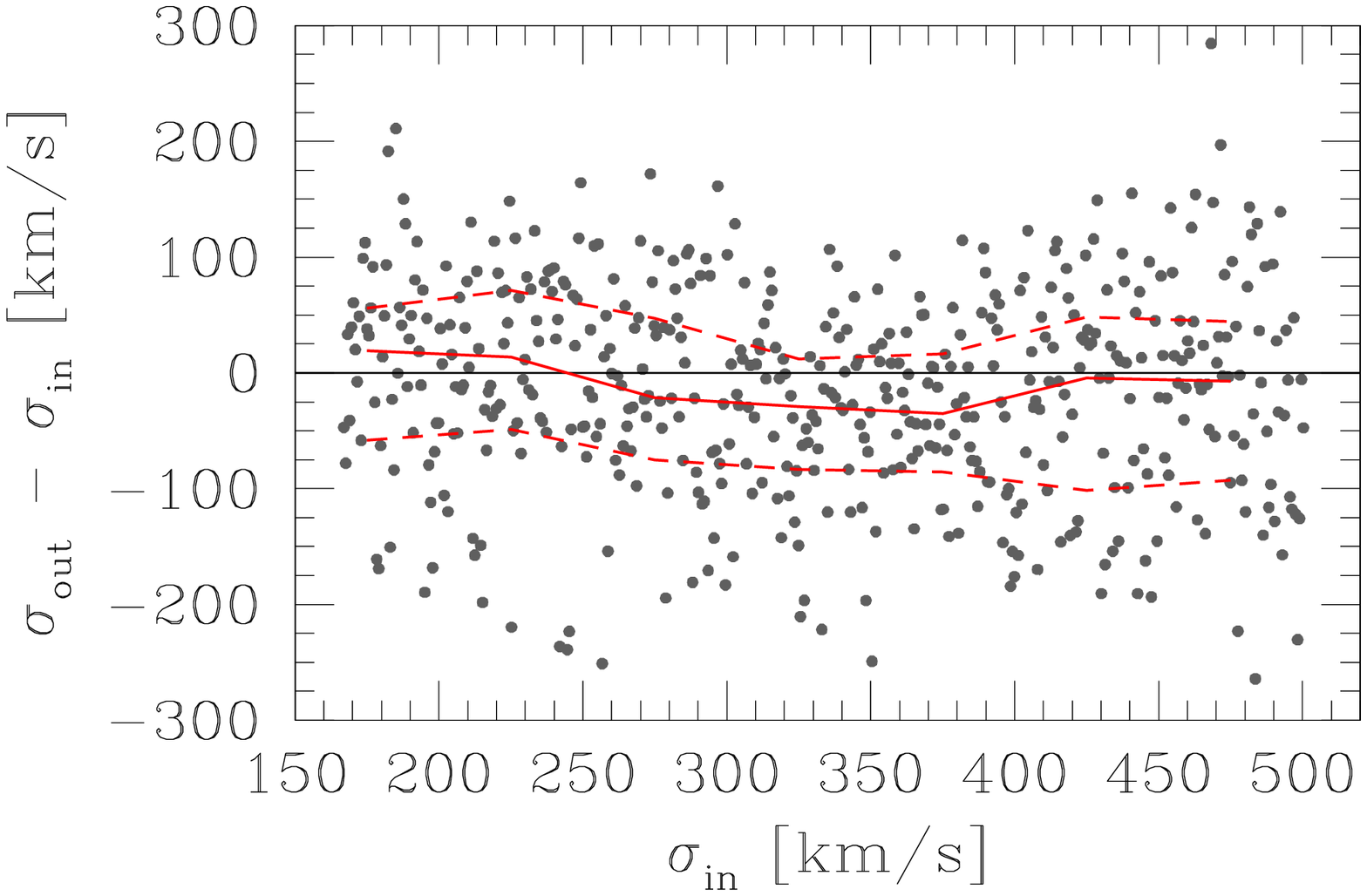} & \includegraphics[angle=0,width=6.0cm]{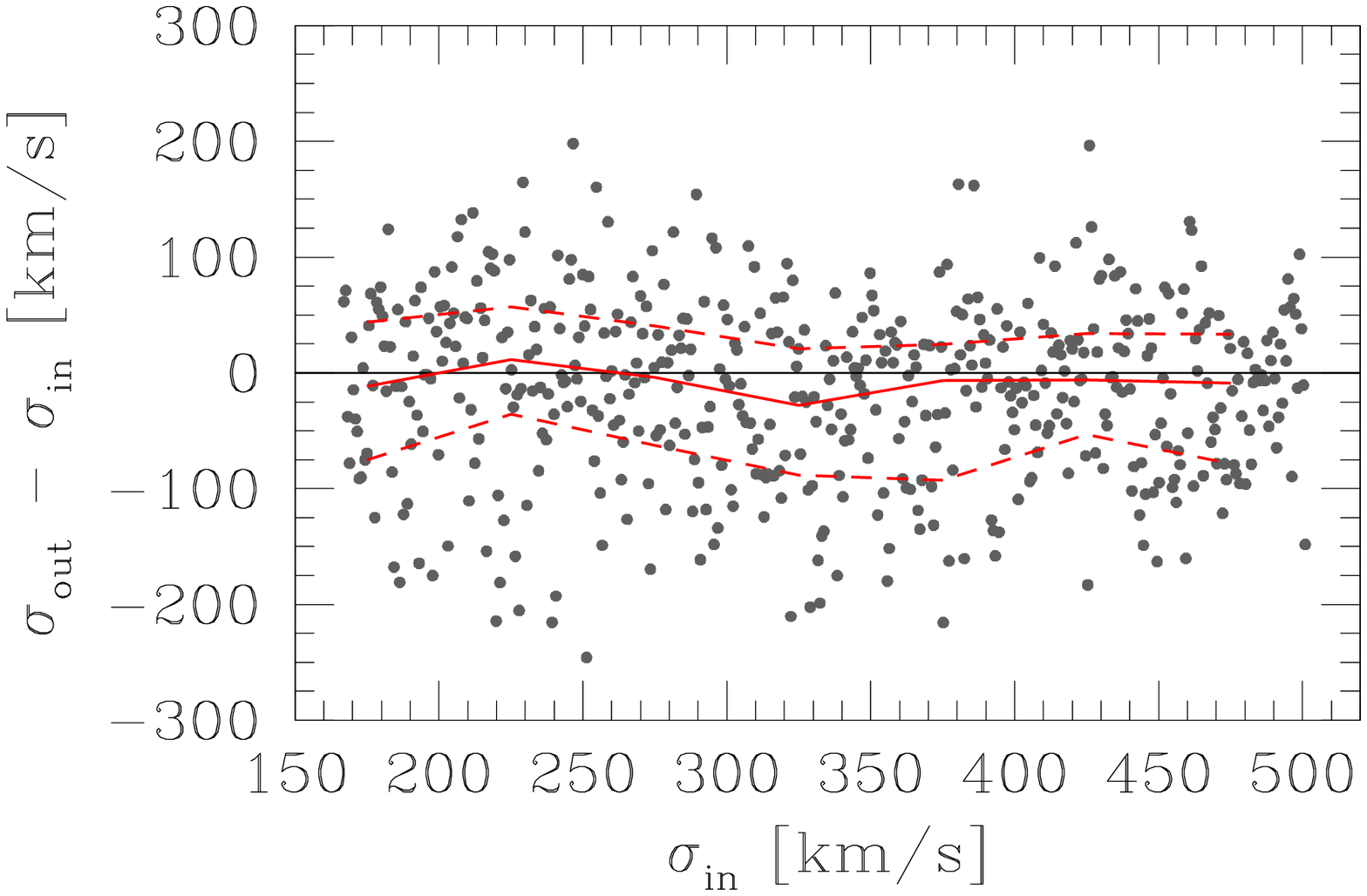} & 
\includegraphics[angle=0,width=6.0cm]{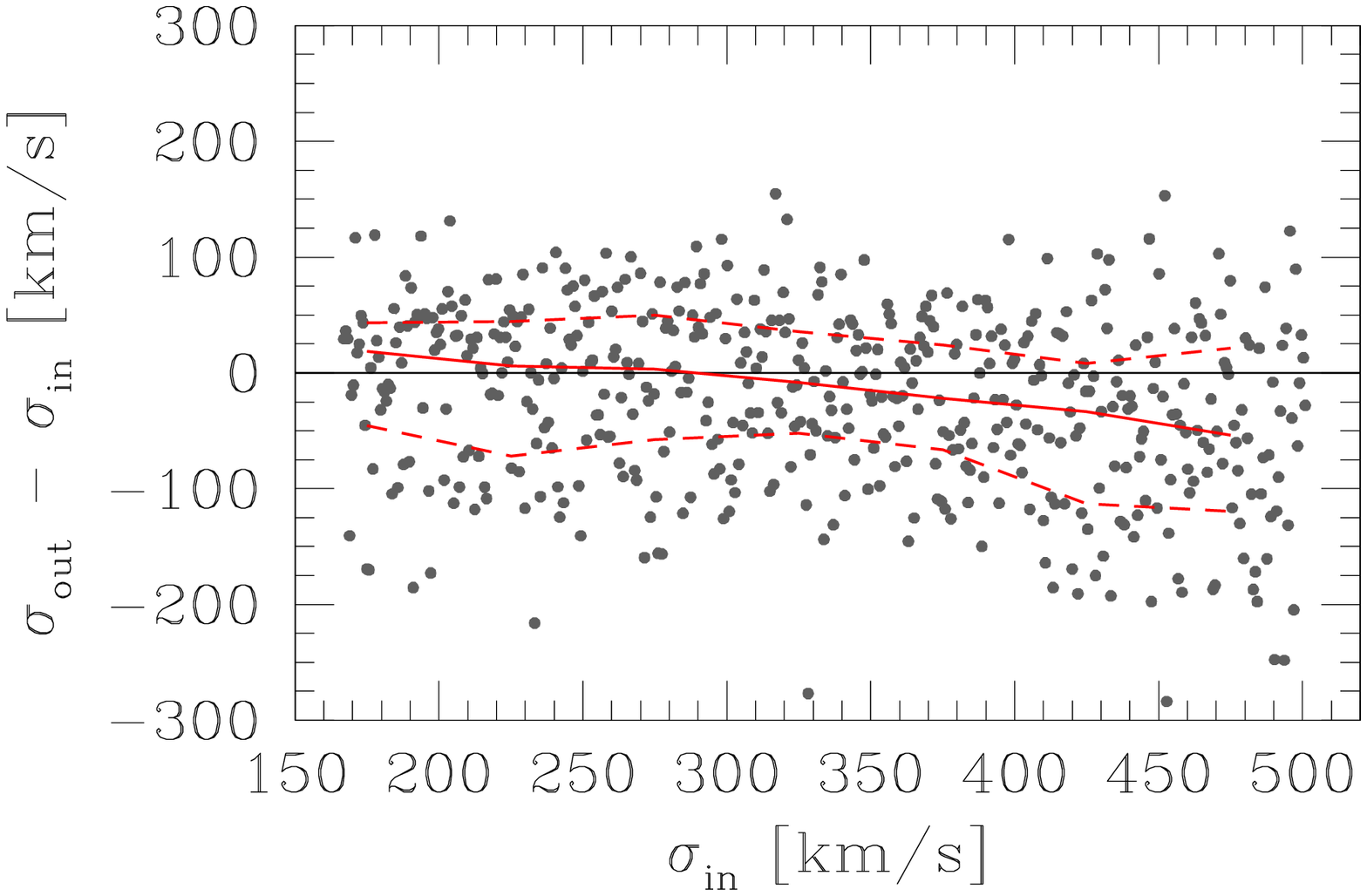} \\

 \end{tabular}
\end{center}
\caption{The same as Fig. \ref{simulage1} for galaxy S2F1-389.}
\label{simulage4}
\end{figure*}

\begin{figure*}
\begin{center}
  \begin{tabular}{ccc}
 \includegraphics[angle=0,width=6.0cm]{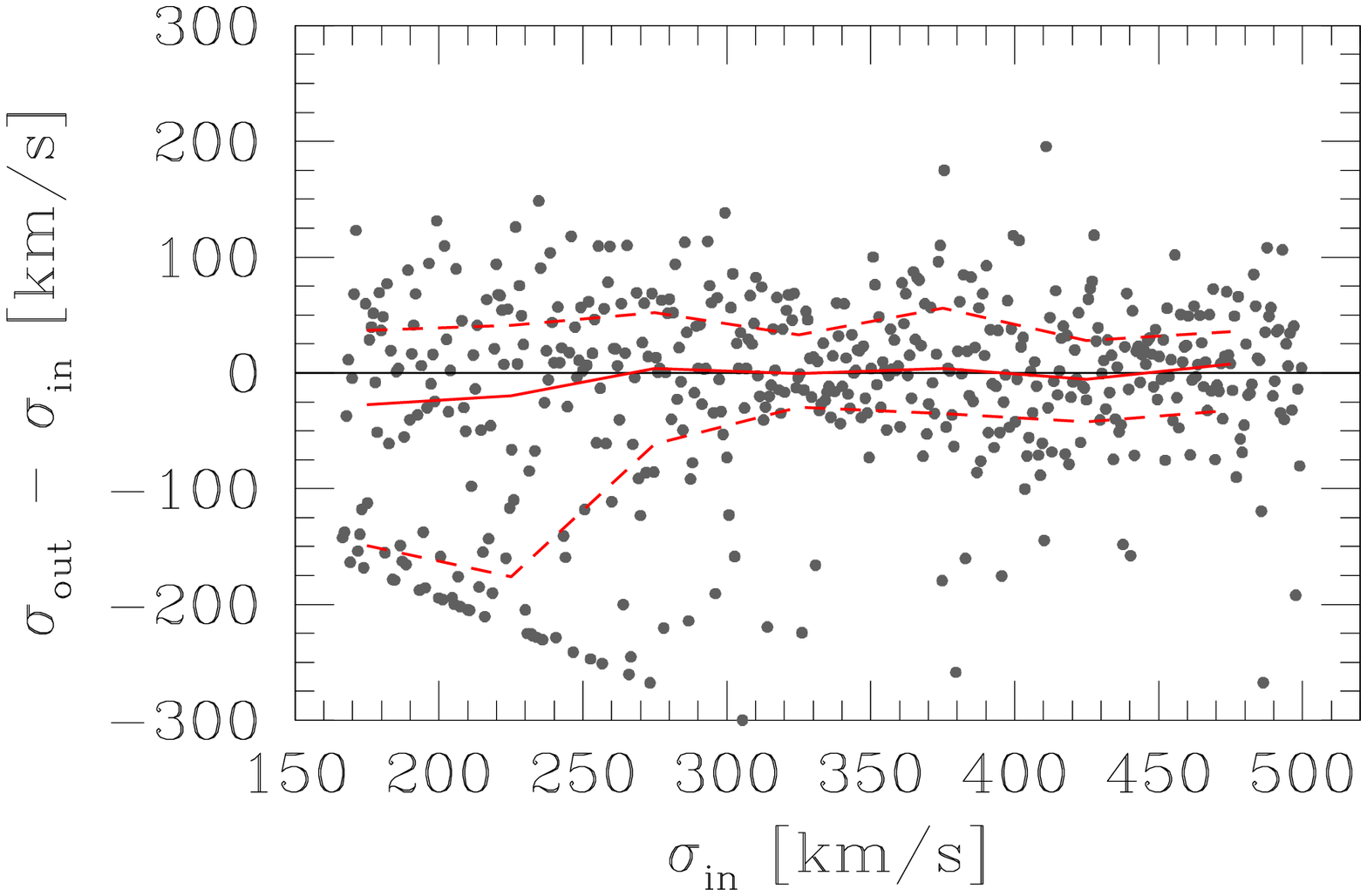} & 
\includegraphics[angle=0,width=6.0cm]{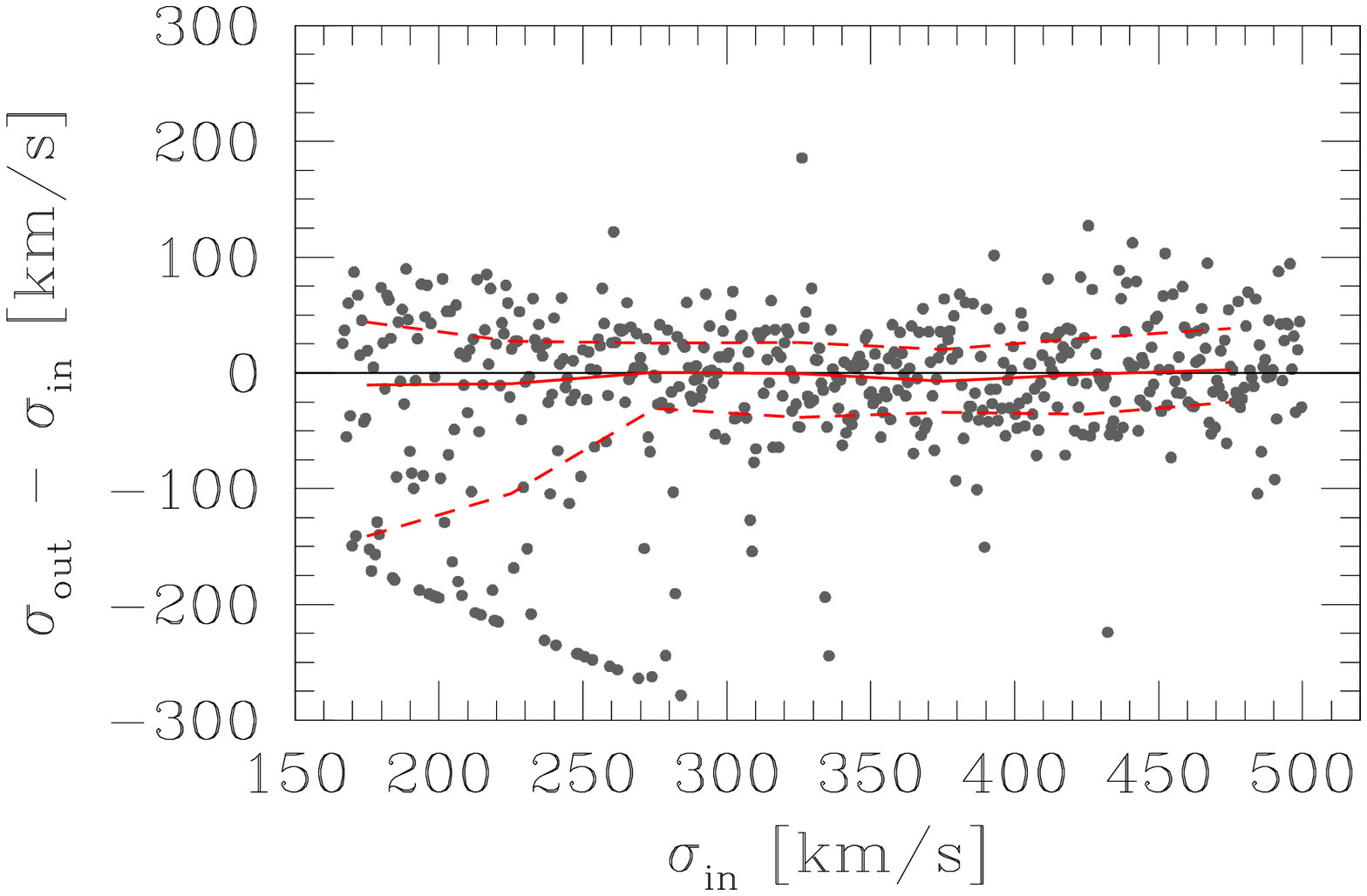} & \includegraphics[angle=0,width=6.0cm]{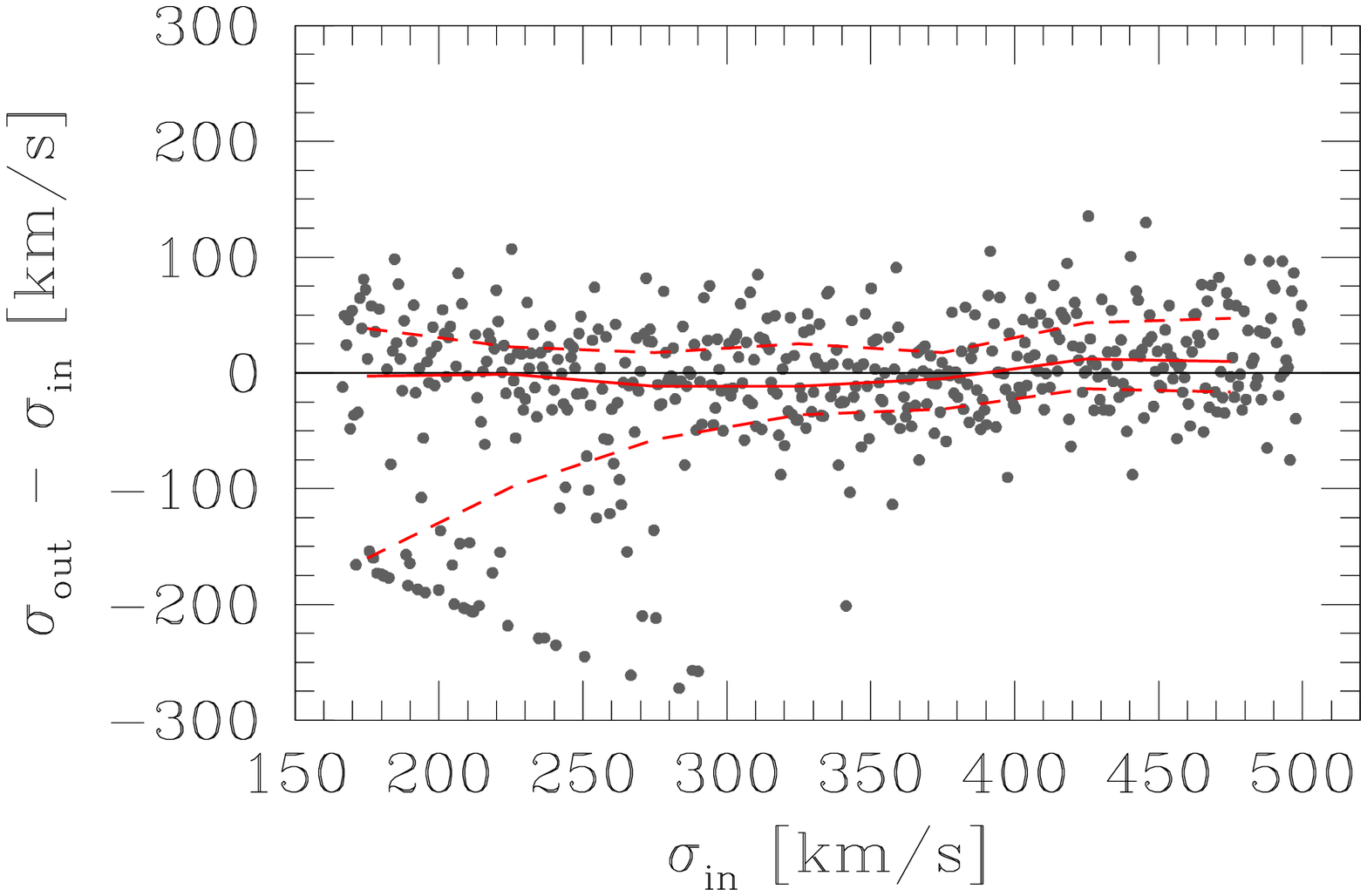} \\
 \includegraphics[angle=0,width=6.0cm]{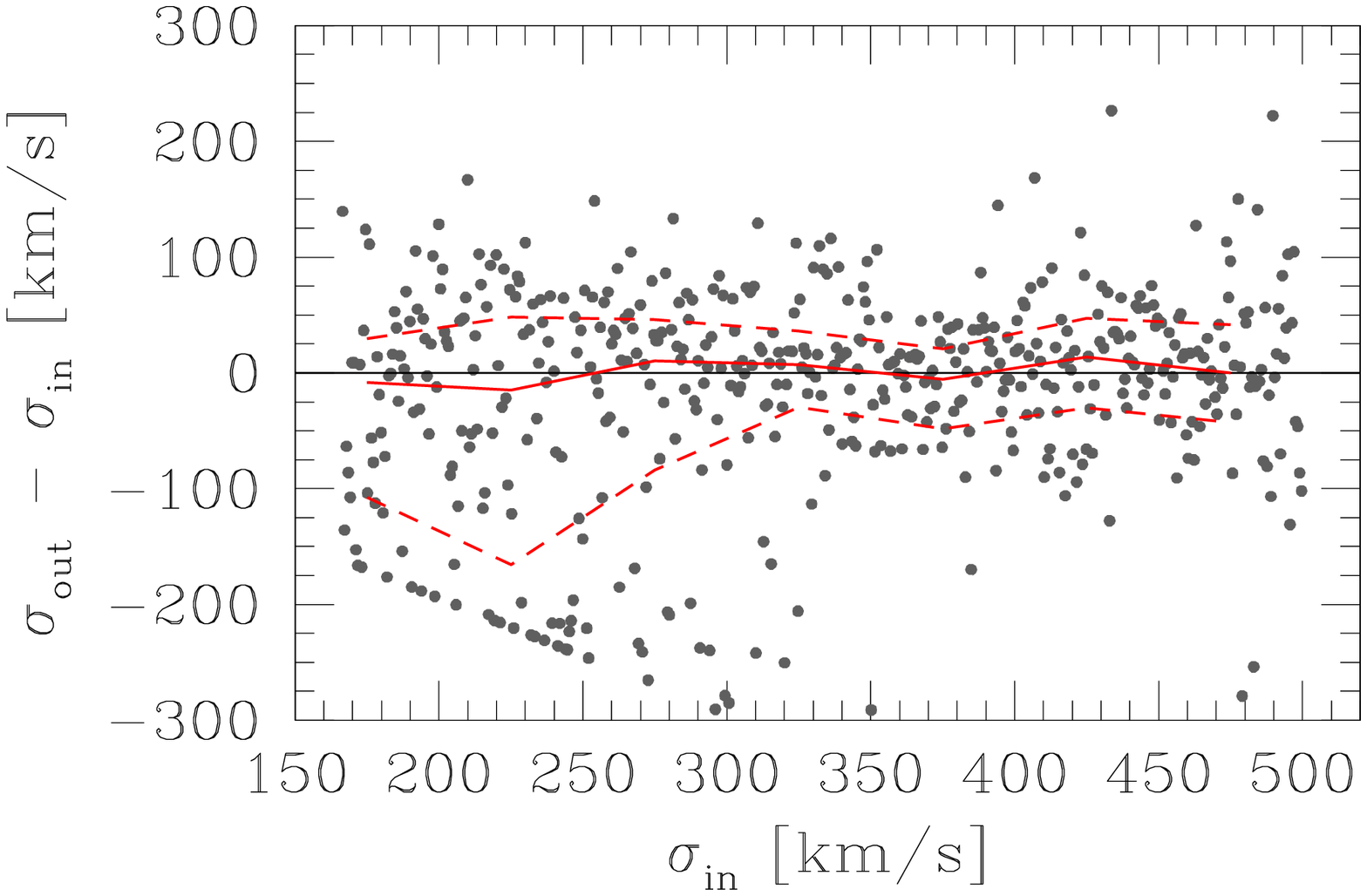} & \includegraphics[angle=0,width=6.0cm]{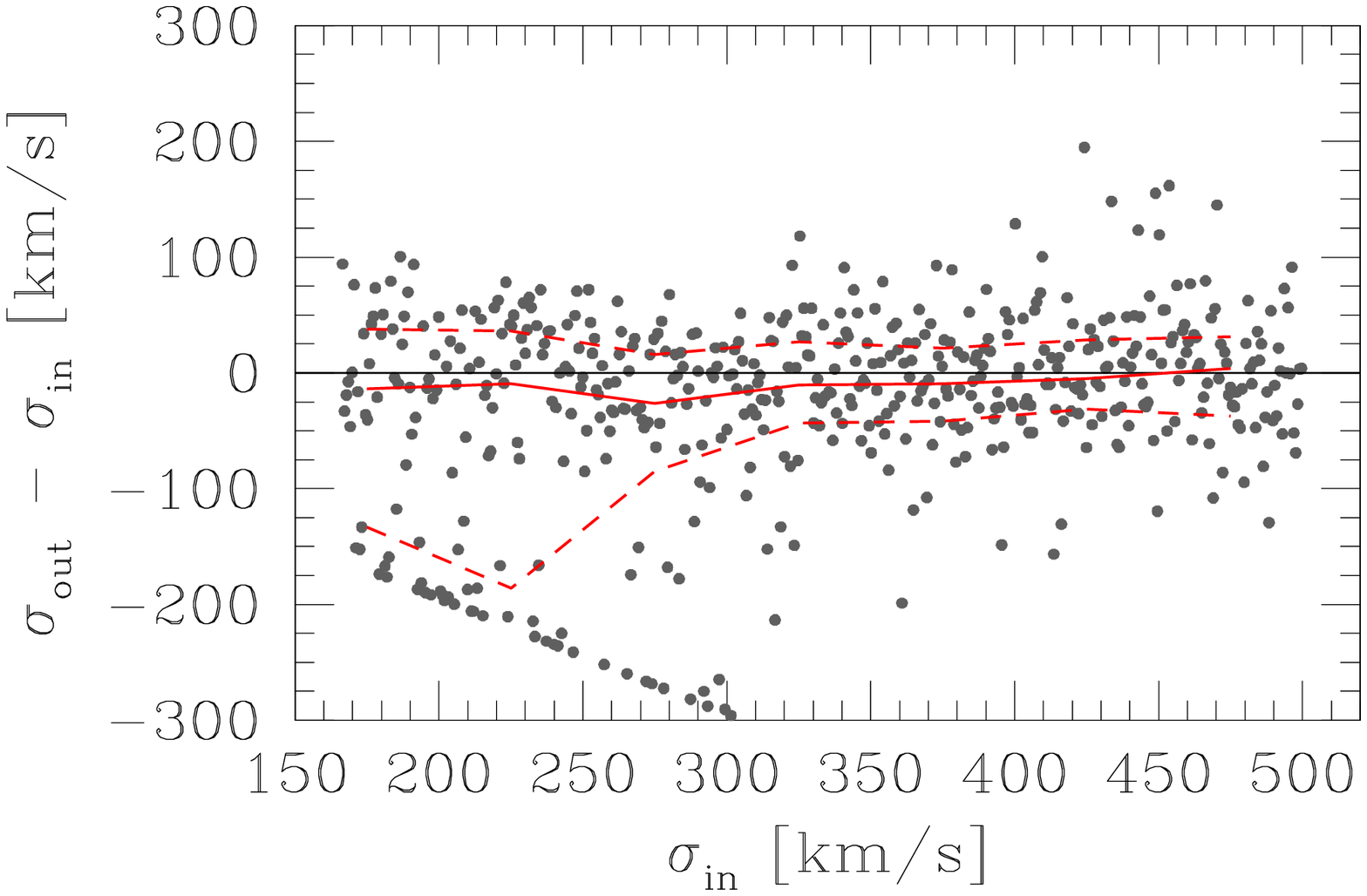} & 
\includegraphics[angle=0,width=6.0cm]{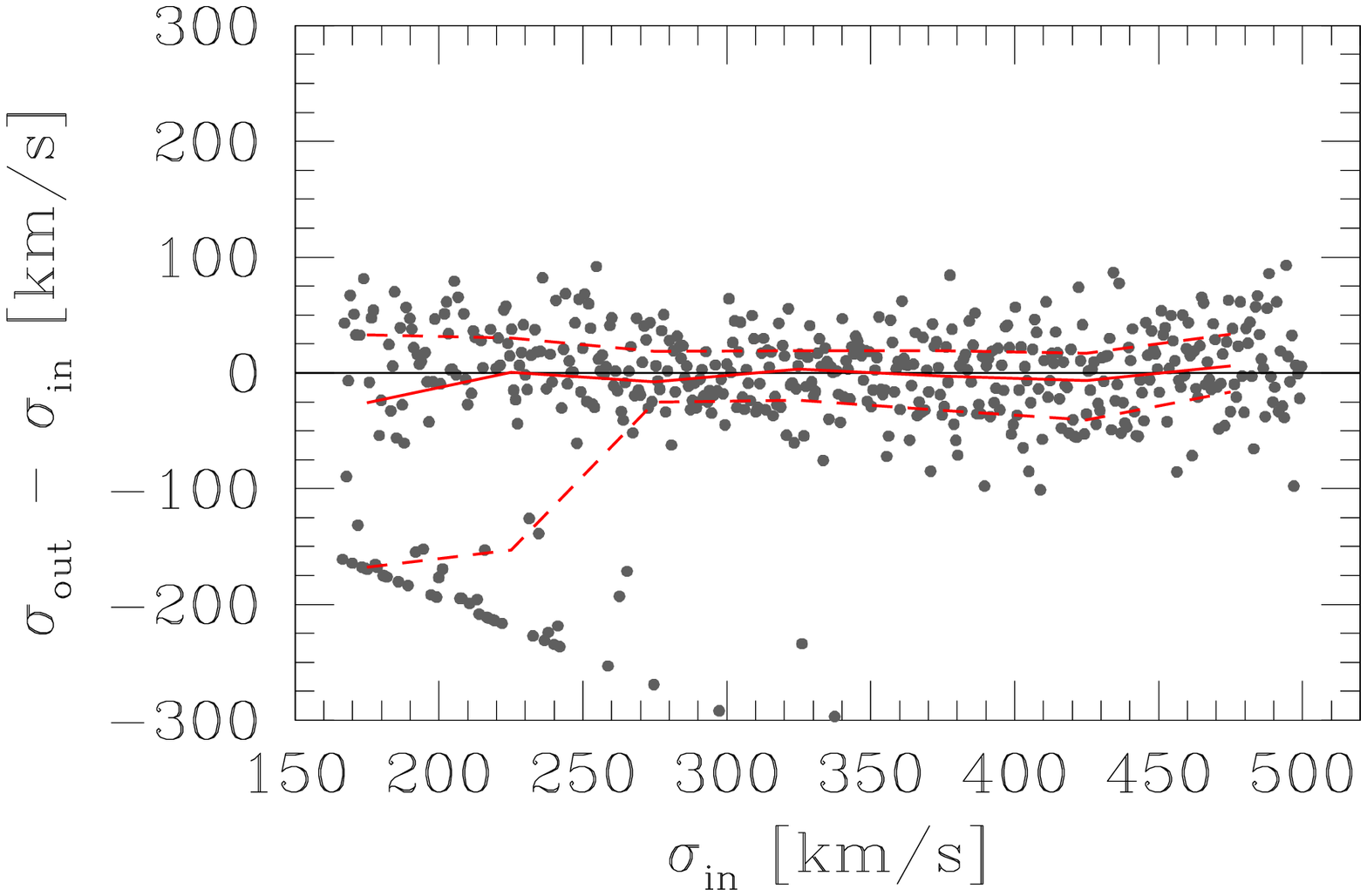} \\

 \end{tabular}
\end{center}
\caption{The same as Fig. \ref{simulage1} for galaxy S2F1-443.}
\label{simulage5}
\end{figure*}

Briefly, for each galaxy of our sample, we have
created four model galaxy spectra starting from a simple stellar population model from \citet{maraston11}
with solar metallicity and age 2, 3, 4, 5 Gyr. 
We have redshifted the spectra to the galaxy redshift, degraded them to the
dispersion of real spectrum ($\sim$ 55 km/s per pixel) and added to them, pixel-by-pixel, 
a fake noise randomly extracted by the real one. These template spectra have been convoluted
with a Gauss+Hermite LOSVD with the $h_{3}$ and $h_{4}$ Hermite's moment equal to those
measured on real spectrum and velocity dispersion ($\sigma_{in}$) varying in the range $\sim$[200-500] km/s over 500 steps.
We have fixed the line-of sight velocity $V$ to zero. 
As example, in Fig. \ref{fakespectrum}  we report a galaxy spectrum template we have constructed for the galaxy
S2F1-633, with velocity dispersion 400 km/s (black line). 
With the same star templates used on real spectra, we have measured their velocity dispersions ($\sigma_{out}$)
assuming as fitting function a Gauss+Hermite
polynomial. In Fig \ref{simulage1}, \ref{simulage2}, \ref{simulage3}, \ref{simulage4}, \ref{simulage5}
we present the results of the simulations for the galaxy S2F1-511, S2F1-633, S2F1-527, S2F1-389, S2F1-443.

The results we have obtained show that, irrespective of the stellar population properties, 
for all the galaxies of our sample, fitting a 
Gauss+Hermite polynomial over the whole wavelength range covered by our spectra,
with a penalty BIAS $\simeq$ 0.1, is the best choice to recover the intrinsic value of velocity 
dispersions. The same set of simulations for the galaxy 443 shows the feasibility of extracting
a reliable estimate of velocity dispersion from
spectrum covering the blue wavelength range shortward the 4000 \AA\, in agreement with what found by \citet{cappellari09}. 

In facts, the difference  $\sigma_{out}$ - $\sigma_{in}$ shows 
a slight offset which is always lower than the dispersion of 
the real spectrum ($\sim$ 55 km/s per pixel). The exact magnitude of the offset depends on the age and metallicity of the 
stellar population. We verified that the amount of the offset is not reduced excluding from the fit the region around
the Ca H+K features, as done by \citet{kormendy82b}, for this reason we have decided to include this part of the spectrum in our 
fit.

Moreover, our simulations have shown that 
for the $\sigma$ we expect
for our galaxies, the median value of the offset between $\sigma_{in}$ and $\sigma_{out}$ 
($\sigma_{off, median}$ = $\sigma_{out}$ - $\sigma_{in}$) depends on the age and metallicity of
the input template, but is always smaller than the statistical error on measured velocity dispersion.
In Table \ref{table1} we have reported, for each galaxy, the minimum and maximum value
of $\sigma_{off, median}$ we have found in the 6 sets of simulations when $\sigma_{out}$ is equal 
to the value we have measured on real spectrum. From the Table appears that 
the offset is never greater than 30 km/s and can be both positive and negative. Given these
two aspects, and our uncertainty on stellar population properties, mostly on metallicity, 
we have decided to not correct our measures for this offset, 
since we are not able to properly quantify its magnitude. Nonetheless, we note that 
our results would not change if we corrected the velocity dispersion of each galaxy for the maximum value of its offset.

The errors on velocity dispersion reported in Table 1 are the 25th and 75th percentile 
of the distributions of $\sigma_{off}$ estimated at $\sigma_{out}$ equal to the velocity dispersion measured on real spectrum. 
Among the different set of simulations, the quoted errors are extracted from the simulation performed starting from a simple stellar population model 
with solar metallicity and age equal to that obtained from the fit of the observed SED.
However, although the value of the offset is slightly dependent on the stellar population properties, its scatter not.

\end{appendix}

\end{document}